\documentclass[]{aa}  
\usepackage{graphicx}

\usepackage{natbib}
\usepackage[colorlinks=true,linkcolor=blue,citecolor=blue]{hyperref}  
\usepackage{txfonts}
\usepackage{xcolor}
\usepackage{float}
\usepackage{soul}

\begin{document} 

\title{Simulated performance of the molecular mapping for young giant exoplanets with the Medium Resolution Spectrometer of JWST/MIRI}
\subtitle{}

 \author{
 M. Mâlin \inst{1}\and
 A. Boccaletti\inst{1} \and 
 B. Charnay\inst{1} \and 
 F. Kiefer\inst{1} \and  
 B. Bézard\inst{1}
 }

   \institute{LESIA, Observatoire de Paris, Université PSL, CNRS, Sorbonne Université, Université Paris Cité, 5 place Jules Janssen, 92195 Meudon, France.}

   \date{Received ; accepted }
  \abstract
   {Young giant planets are the best targets for characterization with direct imaging. The Medium Resolution Spectrometer (MRS) of the Mid-Infrared Instrument (MIRI) of the recently launched James Webb Space Telescope (\textit{JWST}) will give access to the first spectroscopic data for direct imaging above 5 $\muup$m with unprecedented sensitivity at a spectral resolution up to 3700. This will provide a valuable complement to near-infrared data from ground-based instruments for characterizing these objects.}
   {We aim to evaluate the performance of MIRI/MRS to detect molecules in the atmosphere of exoplanets and to constrain atmospheric parameters using Exo-REM atmospheric models.}
   {The molecular mapping technique, based on cross-correlation with synthetic models, has been introduced recently. This promising detection and characterization method is tested on simulated MIRI/MRS data.}
   {Directly imaged planets can be detected with MIRI/MRS, and we are able to detect molecules (H$_2$O, CO, NH$_3$, CH$_4$, HCN, PH$_3$, CO$_2$) at various angular separation depending on the strength of the molecular features and brightness of the target. We find that the stellar spectral type has a weak impact on the detection level. This method is globally most efficient for planets with temperatures below 1500\,K, for bright targets and angular separation greater than 1$''$. Our parametric study allows us to anticipate the ability to characterize planets that would be detected in the future.}
   {The MIRI/MRS will give access to molecular species not yet detected in exoplanetary atmospheres. The detection of molecules as indicators of the temperature of the planets will make it possible to discriminate between the various hypotheses of the preceding studies, and the derived molecular abundance ratios should bring new constraints on planetary formation scenarios.}
  
  \titlerunning{Molecular mapping analysis of simulated directly imaged planets}             
  \maketitle
%

\section{Introduction}
An important outcome of exoplanet searches in the last decades is the diversity of their orbital and bulk properties. Understanding the mechanisms at play during their formation, or migration history is identified as a promising avenue to account for such a diversity \citep{madhusudhan_toward_2014, mordasini_imprint_2016}. The characterization of exoplanet atmospheres has now become a priority in this field, one goal being precise to put some meaningful constraints on their formation. In particular, according to formation models \citep{oberg_astrochemistry_2021}, measuring molecular abundances,  like the ratios C/O and N/O, are relevant to link the atmospheric properties of giant planets to the locations of the snowlines in a planetary system.

A wide range of methods has been used to explore the properties of exoplanetary atmospheres.
The first molecules were detected with transit spectroscopy \citep{charbonneau_detection_2002}. This successful method enables to observe transmission spectra of the day-night terminator, the thermal emission spectra of the day side, and the phase curve on the orbit  \citep[e.g.][]{deming_infrared_2013}, but is limited to planetary systems whose semi-major axis is less than $\sim$1\,au as it suffers from the low probability that the planet is perfectly aligned with the observer. 
Phase-resolved high-resolution Doppler spectroscopy has proven to be a powerful means to detect molecules in the atmosphere of transiting close-in giant planets. 
For instance, this led to the detection of CO in several hot jupiters like HD\,209458 b by \citet{snellen_orbital_2010}, 
or $\tau$\,boo by \citet{brogi_signature_2012} as well as H$_2$O by \citet{birkby_detection_2013} in  HD\,189733 b.
In the case of long-period planets, direct imaging with coronagraphy can also provide spectral information, although mostly at low to medium resolution.
Owing to the inherent contrast limitation, post-processing methods to attenuate the starlight such as Angular Differential Imaging \citep{marois_angular_2006} or Spectral Differential Imaging \citep{racine_speckle_1999}, were decisive to perform observations of young giant planets being warm and bright in the infrared.
A dozen systems have been characterized by spectroscopy with adaptive optics (AO), for example, 51\,Eri\,b with SPHERE/VLT and GPI \citep{samland_spectral_2017, macintosh_discovery_2015}, and with GRAVITY/VLTI the planet $\beta$\,Pictoris\,b \citep{nowak_peering_2020}.

Considering the best of both worlds, \citet{snellen_combining_2015} proposed to combine high contrast imaging with high resolution spectroscopy, while some similar concepts were already formulated earlier \citep[e.g.][]{sparks_imaging_2002}.
An implementation of this original idea was introduced as the so-called "molecular mapping" technique by \cite{hoeijmakers_medium-resolution_2018} yielding the detection of H$_2$O and CO in $\beta$\,Pic\,b, atmosphere, taking advantage of archival VLT/SINFONI data.
Recently, with the same instrument, \cite{petrus_medium-resolution_2021} used this method to characterize the planet HIP\,65426 b. 
In contrast, no molecular species was found in the planets of PDS\,70 \citep{cugno_molecular_2021}, likely because the planet's atmosphere or surroundings is dominated by dust.
Similarly, \cite{petit_dit_de_la_roche_molecule_2018} and \cite{ruffio_radial_2019} took advantage of cross correlation with molecular templates to characterize the HR\,8799 system with Keck/OSIRIS IFS data.

As of today, all direct observations were obtained in the near IR because of the reduced transmission of the Earth's atmosphere in the mid-IR. The James-Webb Space Telescope ({\it JWST}) is expected to be a game-changer in the characterization of directly imaged exoplanet atmospheres \citep{hinkley_jwst_2022}, in allowing us to explore a relatively new spectral range for wavelengths longer than 5\,$\muup$m where planets emit most of their flux (implying smaller brightness ratio between the star and a planet), and exhibit clear molecular signatures from their atmosphere. Complementary to near IR data, a broader wavelength coverage will help to recover for example the temperature of the planet with higher accuracy, as already shown with the early results of {\it JWST} coronagraphy on HIP\,65426\,b \citep{carter_jwst_2022}.

Starting in 2022, MIRI, one of the science instruments of {\it JWST}, optimized for mid-infrared observations \citep{wright_mid-infrared_2015}, is offering a unique opportunity for exoplanet science. 
MIRI has four observing modes: imaging, coronagraphy, low-resolution spectroscopy (LRS), and medium-resolution spectroscopy (MRS). 
The MRS provides integral field spectroscopy across the wavelength range 4.9 to 28.3\,$\mu$m 
\citep{wells_mid-infrared_2015}, which constitute interesting features for exoplanet atmospheres characterization. 
As one of the early outcomes of {\it JWST} programs, \cite{miles_jwst_2022} illustrated the potential of MIRI MRS on the planetary-mass companion VHS\,1256 b, for which several molecules were detected (CH$_4$, CO, CO$_2$, H$_2$O, K, Na).

Following the work by \cite{patapis_direct_2022}, who demonstrated the potential of molecular mapping for MIRI in two well-known systems, HR\,8799 and GJ\,504, we aim to further explore this concept with the self-consistent atmosphere model Exo-REM \citep{charnay_self-consistent_2018} using a parametric study and extending to other known directly imaged planets. 

The paper is organized as follows: Section \ref{sec:data_simu} presents the data simulation and reduction for the MRS. Section \ref{sec:mm_method} introduces the molecular mapping method we implemented. Section \ref{sec:parametric} provides a parametric study and section \ref{sec:atm_charac_mm} the application to a few known directly imaged systems. Section \ref{sec:atm_charac_GJ504} presents a more in-depth atmospheric study for the target GJ\,504\,b and Section \ref{sec:discussion} discusses the results.
\section{Data simulation and reduction for the MRS}
\label{sec:data_simu}
\subsection{The Medium Resolution Spectrometer of MIRI}
The MRS is one of the four observing modes of MIRI. It is an integral field spectrometer that provides diffraction-limited spectroscopy between 4.9\,$\muup$m and 28.3\,$\muup$m, within a field-of-view (FoV) ranging from {$3.7''\times3.7''$} at the shortest wavelengths to {$7.74''\times7.95''$} at the longest wavelengths. The MRS includes four channels that have co-aligned FoV, observing the wavelength ranges simultaneously. Three observations using each time a different set of gratings are needed to observe the entire wavelength range \cite[SHORT, MEDIUM, LONG,][]{wells_mid-infrared_2015}. The spectral resolution decreases with increasing wavelength.
The parameters of each subchannel (bands) are indicated in Tab. \ref{tab:table_MRS}.
The MRS is spatially undersampled at all wavelengths and mostly in the first channel, therefore dithering is necessary to improve this spatial sampling. Different dithering patterns are possible and depend on each scientific case, mainly the 4-point dither pattern is preferred as it provides robust performance at all wavelengths and adequate point source separation in all channels. 
\begin{table*}[h!]
\centering
\begin{tabular}{c c c c c c}
\hline
\hline
Channel & Band & Wavelength ($\mu$m) $^{1}$ & Resolution (best estimate) $^{1}$ & FoV (arcsec) $^{2}$ & Pixel size (arcsec) $^{2}$ \\
\hline
1 & SHORT (A) & 4.885 – 5.751 & 3300-4000 & 3.70 × 3.70 & 0.196 \\
 & MEDIUM (B) & 5.634 – 6.632 & 3420-3990 & & \\
 & LONG (C) & 6.408 – 7.524 & 3330-3840 & & \\
 \hline
2 & SHORT (A) & 7.477 – 8.765 & 3190 - 3620 & 4.71 x 4.52& 0.196 \\
 & MEDIUM (B) & 8.711 – 10.228 & 3040 - 3530 & &\\
 & LONG (C) & 10.017 – 11.753 & 2890 - 3374 & & \\
\hline
3 & SHORT (A) & 11.481 – 13.441 & 2450 - 3010 & 6.19 x 6.14 & 0.245 \\
 & MEDIUM (B) & 13.319 – 15.592 & 2300 - 2460 & & \\
 & LONG (C) & 15.400 – 18.072 & 2020 - 2790 & & \\
\hline
4 & SHORT (A) & 17.651 – 20.938 & 1400 - 1960 & 7.74 x 7.95 & 0.273 \\
 & MEDIUM (B) & 20.417 – 24.220 & 1660 - 1730 & & \\
 & LONG (C) & 23.884 – 28.329 & 1340 - 1520 & & \\
\hline
\end{tabular}
\caption{MRS instrumental parameters in each MRS band. $^{1}$ \cite{labiano_wavelength_2021}, $^{2}$\cite{wells_mid-infrared_2015}}
\label{tab:table_MRS}
\end{table*}

The minimum integration time of the MRS detector in full frame is $t_{fast}=2.775$\,s. One integration is a ramp composed of several groups ($N_{group}$), and an exposure is made of several integrations ($N_{int}$). A reset is applied after each ramp (overhead $=t_{fast}$). Therefore, a series of multiple exposures ($N_{exp}$) corresponds to an actual integration time of $N_{exp}\times N_{int}\times N_{group}\times t_{fast}$, and an observation time (including overheads) of $N_{exp} ( N_{int}\times N_{group}\times t_{fast}+(N_{int}-1)\times t_{fast})$.

\subsection{MIRISim}
MIRISim is a software to simulate representative MIRI data \citep{klaassen_span_2020} that incorporates the best knowledge of the instrument. The simulation takes into account effects due to the detectors, slicers, distortion, and noise sources. MIRISim outputs are the detector images in the uncalibrated data format that can be used directly in the JWST pipeline. In this work, we used version 2.4.1 \footnote{\href{https://wiki.miricle.org/bin/view/Public/}{https://wiki.miricle.org/bin/view/Public/}}. The simulations are parameterized using three configuration files that define the astronomical scene, the setup of the instrument and the parameters of the simulator itself, as described in the following.

\textbf{Scene.} 
For this study, the scene is composed of a host star and one or several planetary companions. Each object is simulated by attributing a spectrum and its position in the FoV that are calculated based on the known astrometric positions. Low background emission is added.

\textbf{Simulation parameters.}
The number of groups, integrations, and expositions are determined using the Exposure Time Calculator (ETC)\footnote{\href{https://jwst.etc.stsci.edu}{https://jwst.etc.stsci.edu}} in order to avoid saturation on the detector and to obtain the signal-to-noise ratio desired. 
The PSF of the MRS is under-sampled by design, a well-sampled PSF requires that the object is observed in at least two dithered positions that include an offset as explained in \cite{wells_mid-infrared_2015}. We chose to do our simulations using the 4-point dithering pattern.
Of the two possible detector read modes we selected the FAST mode (2.775\,s per frame) which is more appropriate for bright targets.
The grating position, as well as the observing channel, are also specified in this file.

\textbf{Simulator.}
The last configuration file defines the various noise components. We apply Poisson noise (for each object in the scene including the background), bad pixels, dark current, hot pixels, flat-field, gain, and non-linearity. Moreover, we include the effect of fringes, detector drifts, and latency. The cosmic ray environment is set to define a minimum solar environment.
We note that MIRISim is producing excess noise on the integration ramps using the FAST mode, so it is advised to turn off the read noise component.

Channel 4 suffers from a drop in sensitivity \citep{glasse_mid-infrared_2015}. Therefore, we do not expect to achieve the planetary mass regime at such wavelengths, and we intentionally omit wavelengths larger than 18 $\mu$m.

\subsection{JWST Pipeline }
\label{sec:jwst_pipeline}
The steps of the JWST pipeline for the MRS are detailed in \cite{labiano-ortega_miri_2016}. Starting with the detector images simulated with MIRISim, the pipeline is divided into three successive steps: \textit{calwebb$\_$detector1, calwebb$\_$spec2, calwebb$\_$spec3}, each including several intermediate steps that are listed below. In this work, we use version 1.4.0\footnote{\href{https://jwst-pipeline.readthedocs.io/}{https://jwst-pipeline.readthedocs.io/en/latest/}} of the pipeline that is compliant with the version used for MIRIsim.\\
In appendix \ref{sec:appendix_pipeline}, we provided the relevant steps to reduce the simulated data of MIRI/MRS.

\subsection{Background treatment}
The PSF of bright stars we are studying extends across almost all the field of view. It is therefore impossible to define a region where the pipeline could estimate and subtract the background directly from the science image.
To overcome this issue, we simulated a scene with only the background emission, all other MIRISim configuration files (simulation and simulator parameters) being the same as those used for the astrophysical target simulation. This simulated background goes through stage 1 of the pipeline to correct for detector effects and is subtracted from target exposures using the step \textit{background} in stage 2.

\section{Molecular mapping method}
\label{sec:mm_method}
\subsection{Atmospheric models}
The basic concept of molecular mapping relies on the correlation of spectro-imaging data with a model of the exoplanet atmospheres we are trying to detect. In the following, we will use Exo-REM, a self-consistent 1D radiative-equilibrium model. It has been first developed to simulate the atmospheres and spectra of young giant exoplanets 
\citep{baudino_interpreting_2015,charnay_self-consistent_2018} and more recently extended to irradiated planets \citep{blain_1d_2021}. This model has been used to characterize some directly imaged planets at low and medium spectral resolution \cite[e.g.][]{delorme_-depth_2017, bonnefoy_gj_2018, petrus_medium-resolution_2021}. 
The radiative-convective equilibrium is solved by assuming that the net flux (radiative and convective) is conservative. The conservation of the flux over the pressure grid (64 pressure levels) is solved iteratively using a constrained linear inversion method.
The input parameters of the model are the effective temperature of the planet, the acceleration of gravity at 1 bar, and the elemental abundances. 
The model includes non-equilibrium chemistry comparing chemical reaction timescales and vertical mixing, using parametrizations from \cite{zahnle_methane_2014}.
The cloud scheme is detailed in \cite{charnay_self-consistent_2018}; it takes into account micro-physics and simulates the formation of silicate, iron, sulfide, alkali salt, and water clouds. The cloud distribution is computed by taking into account sedimentation and vertical mixing with realistic eddy mixing coefficient Kzz profiles based on the mixing length theory. 
It takes into account Rayleigh scattering from H$_2$, He, and H$_2$O, as well as absorption and scattering by clouds – calculated from extinction coefficient, single scattering albedo, and asymmetry factor interpolated from pre-computed tables for a set of wavelengths and particle radii.
Sources of opacity include the H$_2$–H$_2$, H$_2$–He, H$_2$O-H$_2$O and H$_2$O–air collision-induced absorption, ro-vibrational bands from molecules (H$_2$O, CH$_4$, CO, CO$_2$, NH$_3$, PH$_3$, TiO, VO, H$_2$S, HCN, and FeH), and resonant lines from Na and K. 
Lines lists used in Exo-REM are indicated in \cite{blain_1d_2021}.

In our simulations, the planetary spectra are modeled with Exo-REM.
We built a grid of models using the ranges of parameters provided in Tab. \ref{tab:table_grid_exorem}. 
In particular, we considered clouds of iron and silicates (forsterite), and the particle radii are computed with simple microphysics in the cloud scheme. This method is based on the comparison of the timescales of the main physical processes involved in the formation and growth of cloud particles, which includes a supersaturation factor S, that we fixed at S = 0.03. This model reproduces the L-T transition, with the passage of clouds below the photosphere at the transition. Therefore, for the T-types, the clouds are forming below the photosphere and have a weak impact on spectra. The clouds are calculated in a self-consistent way depending on the condensation curves at each temperature \citep{visscher_atmospheric_2010}.
\begin{table}[h!]
\centering
\begin{tabular}{c c c}
\hline
\hline
Parameters & Values & steps\\
\hline
Temperature (K) & 400 - 2000 & 50\\
$\mathrm{Log} g$ & 3.0 - 5.0 & 0.5\\
C/O & 0.1 - 0.8 & 0.05\\
Metallicity & 0.32 ; 1.0 ; 3.16 ; 10.0\\
\hline
\end{tabular}
\caption{Exo-REM grid models}
\label{tab:table_grid_exorem}
\end{table}

As for the molecular templates, they are computed from the pressure-temperature profile at equilibrium and from the abundance profiles that were previously calculated. The radiative transfer is computed again with all chemical species removed, except the one considered. The clouds are also removed but the collision-induced absorption (H$_2$-H$_2$, H$_2$-He, H$_2$O-H$_2$O) is still included.\\

Stellar spectra are taken from the BT-NextGen online libraries\footnote{\href{http://svo2.cab.inta-csic.es/theory/newov2/index.php}{http://svo2.cab.inta-csic.es/theory/newov2/index.php}}.
For stars cooler than 3000\,K we used BT-Settl models.

\subsection{Subtracting stellar contribution and cross-correlation}
\label{sec:method_sub_stellar}

\begin{figure*}[h]
     \centering
     \includegraphics[width=180mm]{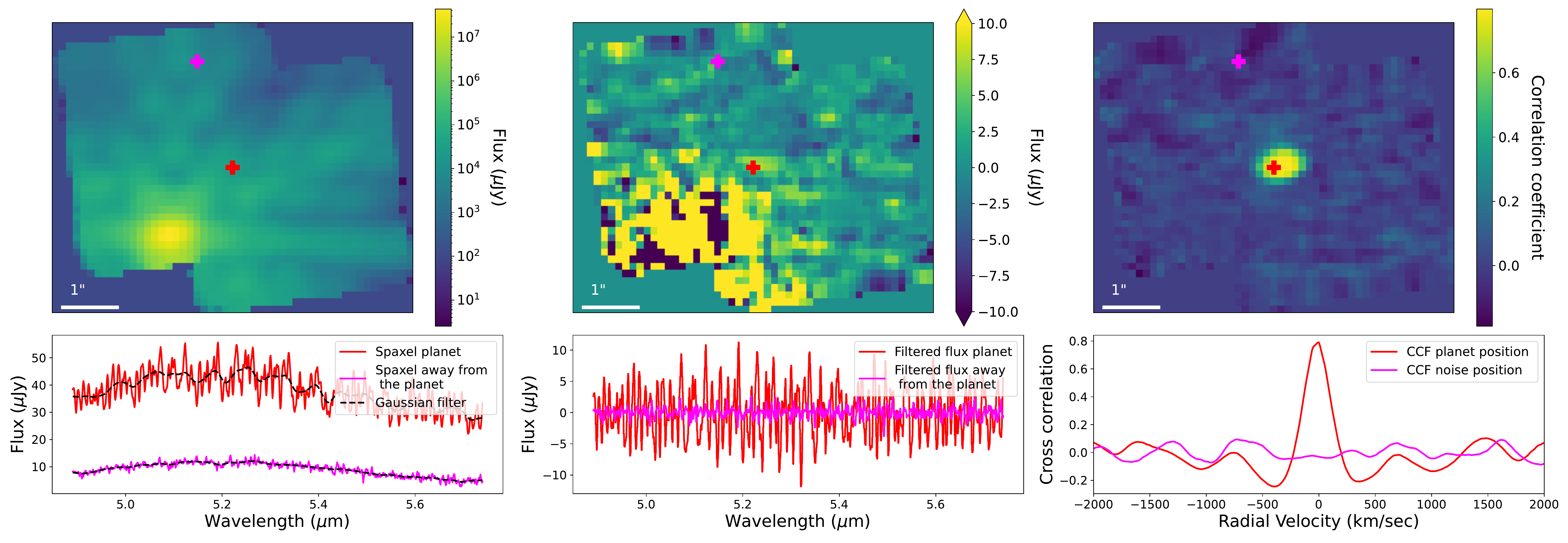}
     \caption{Top: Simulation in channel 1A for a star ($T=6000\,K$) and a planet ($T=1000\,K$) separated by 1.8$''$. From left to right: direct image resulting from four dithered positions, the star being offset in the bottom left corner (sum over the wavelengths of the channel); high frequencies residuals after subtracting the spectral Gaussian filter in each spaxel; correlation map with the very same template spectra as injected into the simulation for $\delta V=0$. The red cross indicates the planet's position, and the pink cross is the arbitrary position chosen away from the planet.\\
     Bottom: Illustration of the molecular mapping technique in two spaxels, one at the position of the planet (red) and the other one at a position away from the planet (pink), for channel 1A ($4.885-5.751\,\muup$m). From left to right: we display the filtering process applied to the model, the combined spectra and the Gaussian filter (black) in the two spaxels (pink and red), the high-frequency component after subtraction, and the cross-correlation function for $\delta V = [-2000;+2000]$\, km/s.}
     \label{fig:method_complete}
\end{figure*}

The stellar contribution in high-angular resolution data is a mixture of the ideal diffraction pattern, and speckles due to optical aberrations, whose intensity scales with the star's spectrum, and the phase-induced chromaticity of speckles. Both scales radially with the wavelength, at first order.
A planet buried in the diffracted halo and a star have very different spectral dependence, hence they can be disentangled \citep{sparks_imaging_2002}.
In the infrared, and in space conditions (no telluric lines), the atmospheric signature of a giant planet would appear as a high spectral frequency as opposed to the star, due to molecular absorptions. Therefore, as a prerequisite to apply the correlation with a model, the stellar contribution can be greatly attenuated by high-pass filtering while preserving the molecular signatures of the planet's spectrum almost intact \citep{ruffio_radial_2019}. 
In our case, we used a Gaussian filter to suppress low frequencies on each spaxel (spectral pixel) of the cube.
We adopt experimentally a filter parameter of $\sigma = 10$ which globally maximizes the detection of the simulated planets in our sample. 
Prior to applying the correlation, the Exo-REM models are degraded to the maximum resolution of the MRS (3700 in the first band 1A) and interpolated on the wavelength values of each MRS channel. The very same high-pass filter is applied to the Exo-REM models. 
Finally, we calculated the cross-correlation function (CCF) between the model and the data (high-pass filtered) for each velocity offset ($\delta V$) between the two spectra. 
Models and data spectra which are provided at a constant $\delta\lambda$ in MIRISim, are converted to velocity and re-interpolated to get a constant step in velocity. 
We used the python function \texttt{scipy.signal.correlate} to perform the correlation between two spectra.
An example of the process in two different spaxels is shown in Fig. \ref{fig:method_complete}, one at the position of the planet in red, and the other one at an arbitrary position, in a noise-limited region, in pink.
The cross-correlation function shows a peak of correlation at a radial velocity $\delta V=0$. Looking at the spaxel away from the position of the planet, no peak of correlation is observed. 
We note that the MRS does not have a high enough spectral resolution to resolve the Doppler shift of the known imaged planets. Therefore, no Doppler shift is included in our simulation, and we focus only on the value at $\delta V=0$ of the correlation function.
The method is repeated independently on each spaxel to derive a correlation coefficient map at $\delta V=0$ in which a planet would correspond to the highest correlation in the FoV (Fig. \ref{fig:method_complete}).
The value of the correlation map in each position $i, j$ is given by the equation \ref{equation_cc}, with M the model spectra and S the spectra from the data. 
\begin{equation}
    C_{i,j} = \frac{\sum_\lambda S(\lambda)_{i,j}\times M(\lambda)_{i,j}}{\sqrt{\sum_\lambda S(\lambda)_{i,j}^2\times \sum_\lambda M(\lambda)_{i,j}^2}}
    \label{equation_cc}
\end{equation}

\subsection{Signal to noise ratio calculations}
\label{sec:S/N}
\begin{figure}[h]
     \centering
     \includegraphics[width=90mm]{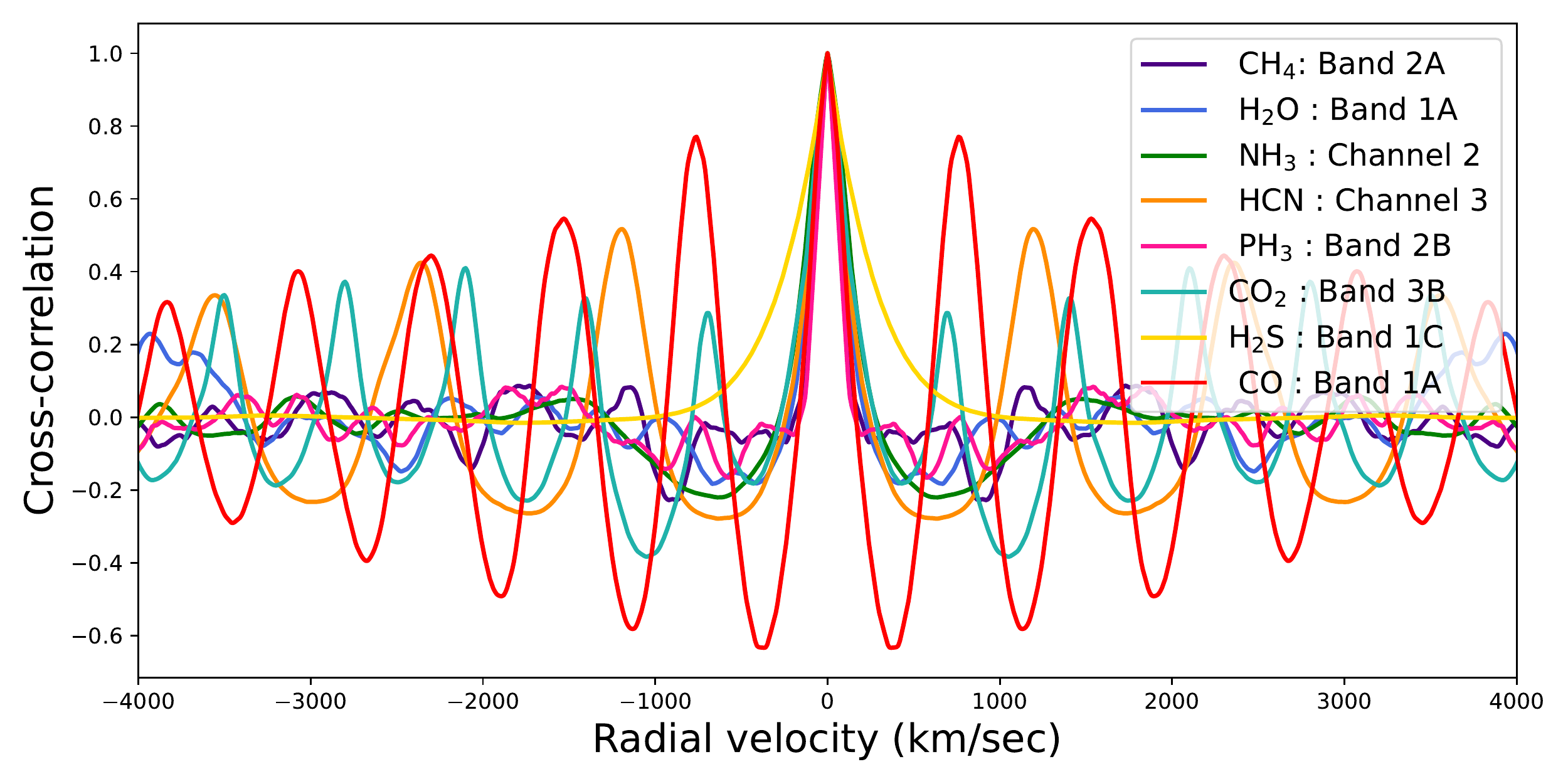}
     \caption{Autocorrelation of each molecule spectrum (at T = 550 K) in the band of the spectral features.}
     \label{fig:ccf_autocorr}
\end{figure}

\begin{figure}[h]
     \centering
     \includegraphics[width=85mm]{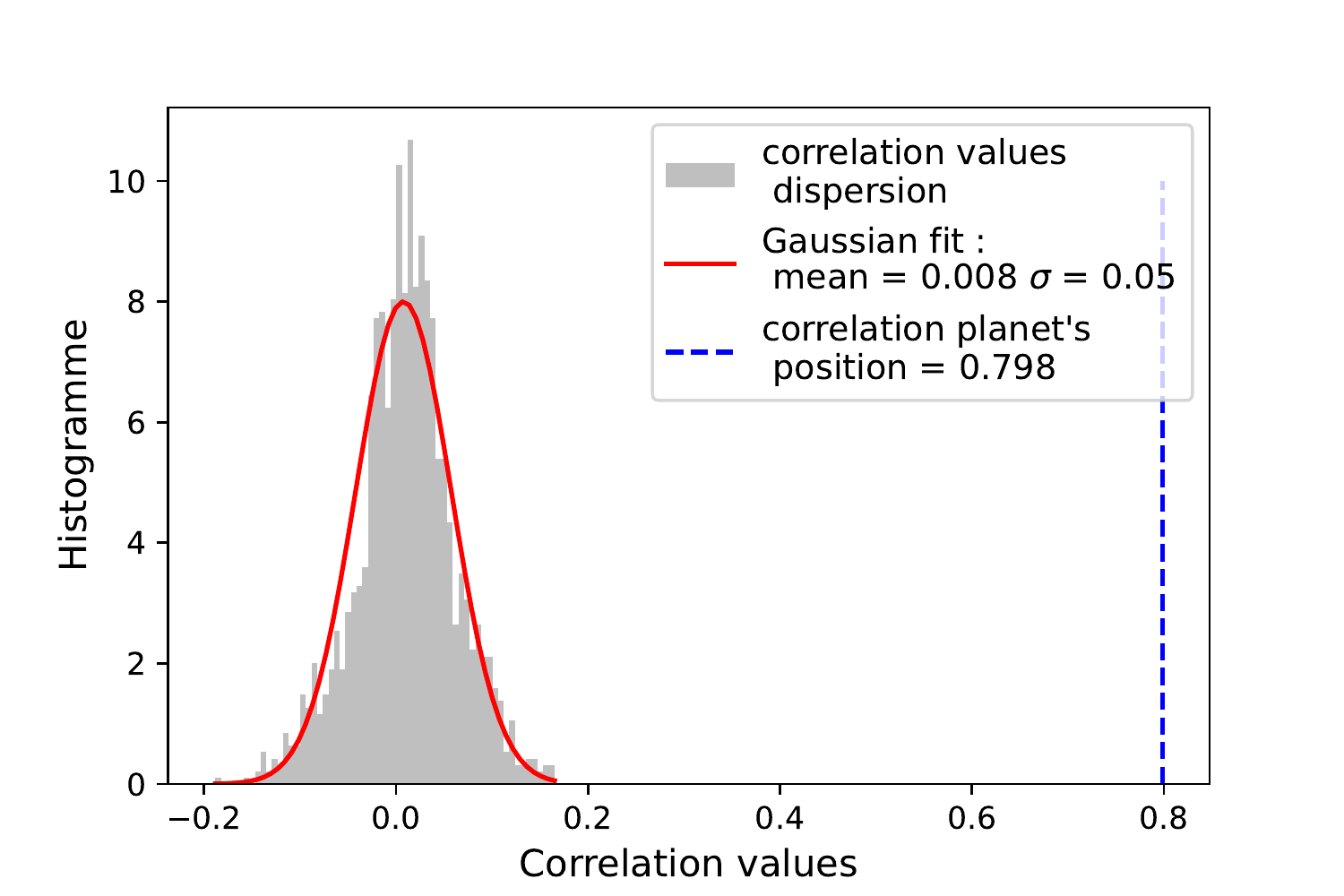}
     \caption{Histogram of the pixels in the correlation map (grey area and red model), compared to the correlation coefficient at the planet's location, as proposed in \citet{petrus_medium-resolution_2021} and used in \cite{patapis_direct_2022}.}
     \label{fig:SNR_method}
\end{figure}

To evaluate the signal-to-noise ($S/N$) ratio, first \cite{hoeijmakers_medium-resolution_2018} measured the average standard deviation of the CCF in an annulus away from the peak of correlation, and away from the planet's position, to avoid systematic variations in the CCF at the location of the planet due to the autocorrelation. 
The autocorrelation function arises from all the harmonics in a molecule's spectrum which produce a non-zero correlation signal away from $\delta V=0$. For instance, the CO generates secondary correlation peaks which can be almost as strong as the main correlation peak (Fig. \ref{fig:ccf_autocorr}).
To account for the autocorrelation, \cite{cugno_molecular_2021} added a correction. The autocorrelation function of the model spectrum is calculated and subtracted away from the peak of the CCF.
However, the impact of the autocorrelation signal justifies taking into account the spatial dimension in estimating the noise.
\citet{petrus_medium-resolution_2021} measured the noise as the standard deviation of a Gaussian distribution derived from all the spaxels, excluding those containing the planet's signal (Fig. \ref{fig:SNR_method}). The correlation signal of the planet is averaged in the velocity space around the correlation peak and spatially in a region centered on the planet.
This method is also used by \citet{patapis_direct_2022}, who measured the signal as the mean value of the CCF in an aperture centered at the position of the planet. This measurement assumes that the noise follows a Gaussian distribution which is not always the case depending on the instrument and on the residuals left after stellar subtraction.

These methods are also conservative since the planet's signal can be integrated in principle over several pixels. 
Nevertheless, measuring the $S/N$ of a planet in the footprint of its correlation pattern is not straightforward. 
In Appendix \ref{sec:appendix_correlation}, we derive how the size of the correlation pattern varies at first order.
We find that it is a function of the wavelength-dependent PSF size, the template used for the correlation, and the noise level. 
However, the derived formula is too approximate to be used to measure the size of the correlation pattern in the data.

\begin{figure}[h]
    \centering
    \includegraphics[width=90mm]{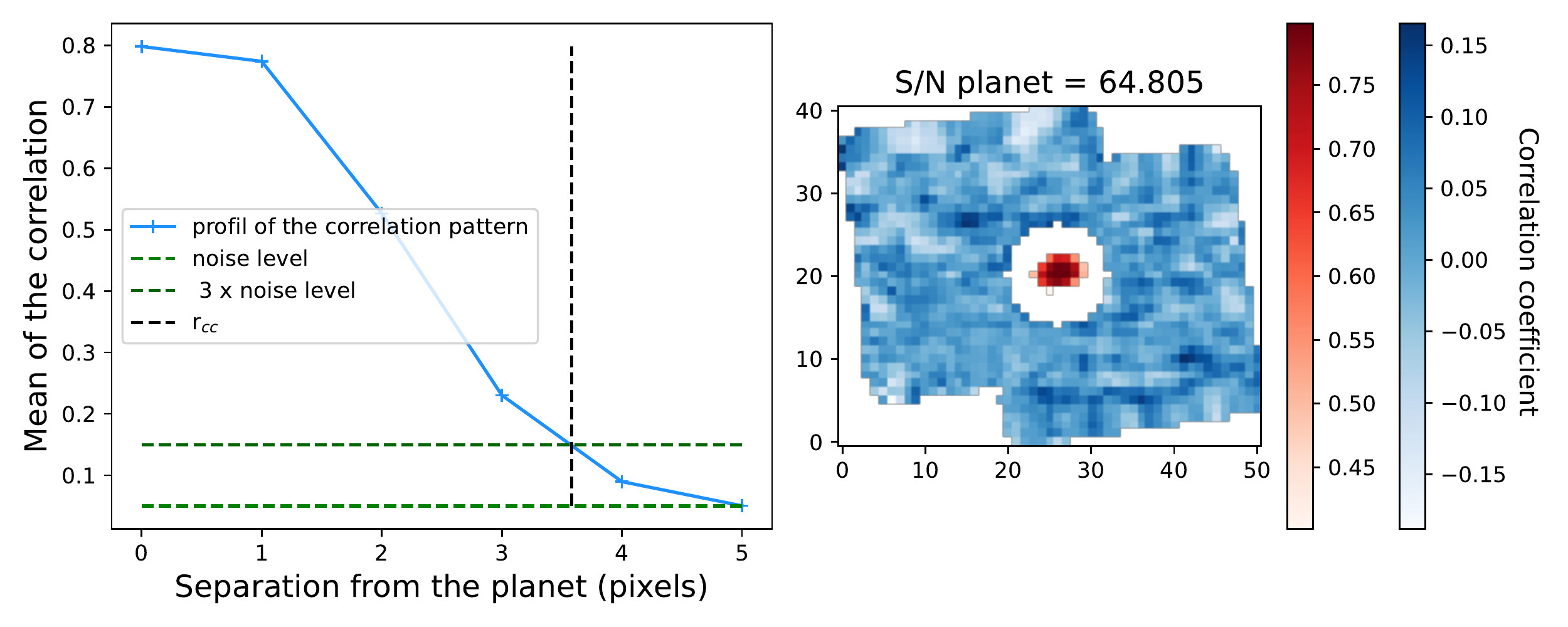}
     \includegraphics[width=90mm]{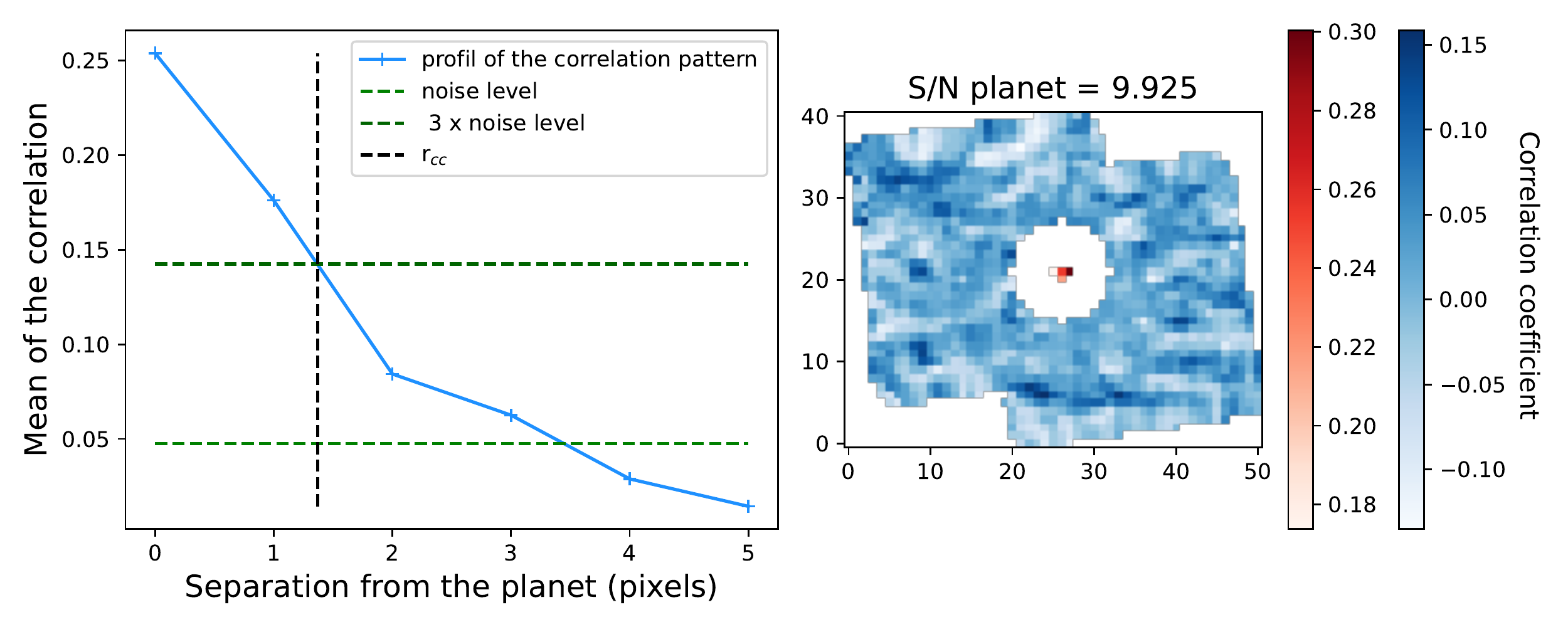}
     \caption{Mean azimuthal profile of the correlation pattern (left) and typical correlation map to illustrate the $S/N$ measurement (right), the same configuration as in Fig. \ref{fig:method_complete}.
    In blue the pixels used to evaluate the noise as the standard deviation of the distribution. In red, the pixels that are considered for the signal of the planet, as defined in Sect. \ref{sec:S/N}.
    Top: Simulation with a low noise level (example with a star at 1.8" from the planet)
    Bottom: Simulation with a higher noise level ( example with a star at 0.6" from the planet).}
     \label{fig:S/N_high_noise}
\end{figure}

To obtain a robust measurement of the $S/N$, we defined a new method compliant with both a low level and a high level of detection, and also able to deal with the residual correlations of the stellar spectrum itself, particularly important if the planet's and star's temperatures are close, like for late-type stars.
Our method also accounts for the spatial variation of the noise to avoid being limited by the autocorrelation signal as noticed in \cite{petrus_medium-resolution_2021}.
Hiding the planets with a radius of 6 spaxels (maximum size of the correlation pattern estimated experimentally for a single planet in the image), we measured the noise as the standard deviation of all the other spaxels, ie. in the correlation map at $\delta$V = 0.
Based on the parametric study (Sect. \ref{sec:parametric}), we note that the strength of the correlation depends on the separation between the star and the planet. In addition, if the data are noisier, the correlation pattern is smaller. Finally, the width of the correlation pattern also scales with the wavelength as does the PSF. To complement the formalism in Appendix \ref{sec:appendix_correlation}, we present more figures in Appendix \ref{sec:appendix_fig_corr} to demonstrate these effects on simulated data.
Concerning the astrometry, the maximum of the correlation pattern does not necessarily correspond to the real position of the planet, and this effect is more important at small angular separations.
Indeed, the stellar flux can contaminate the spaxels located at the planet's position, so the net effect is a higher correlation value further out in the planet's signature. Therefore, we stress that the astrometry of a companion based on the correlation map is unreliable at high noise levels and/or short angular separations.
Finally, in all of our MRS simulations, we notice that the correlation pattern decreases at increasing wavelengths, especially because of a loss of sensitivity and higher background level impacting the longer wavelengths. We also note that molecular features tend to become shallower at longer wavelengths.
Given these observed behaviors, we chose to measure the $S/N$ in correlation maps only spatially, and we define the size of the planet's correlation pattern (containing $N_S$ spaxels) experimentally based on its radial profile. 
We imposed a maximum size of 6 spaxels for all channels. As a first criterion, we selected the spaxels (red area Fig. \ref{fig:S/N_high_noise} to display a case with higher noise level) whose correlation value is higher than 3 times the noise (measured in the blue area of Fig. \ref{fig:S/N_high_noise}).
To ensure that we do not integrate noise in the signal since the correlation pattern is not circular, we imposed a second criterion by selecting the spaxels in which the correlation is larger than 50\% of the maximum correlation.
In addition, to account for the particular situation where the whole profile is above 50\% of the maximum correlation, for instance, if the correlation with the star itself dominates the pattern (mainly the case of CO, or a hot planet around a cold star), we only use the central spaxel to measure the correlation peak.
Finally, the $S/N$ is calculated with Eq. \ref{eq:equation_SNR}, where $\sigma$ is the standard deviation of the noise, and $C_{i}$ the correlation values for the $N_S$ spaxels. 
\begin{equation}
    S/N =\frac{\sum_{i}C_{i}}{\sqrt{N_S}\times \sigma}
    \label{eq:equation_SNR}
\end{equation}

\section{Parametric analysis of MIRI/MRS detection capacity}

To evaluate the detection limit of the MRS with molecular mapping, we run two sets of simulations, and we study the impact of the spectral type and the angular separation on a planet's detection, and on the detection of each molecule included in Exo-REM. The first set of simulations (Sect. \ref{subsec:sp_type}) allows us to restrain the parameters space to pursue this parametric study.

\label{sec:parametric}
\subsection{Impact of the  spectral type}
\label{subsec:sp_type}
As the molecular mapping method relies on the fact that planetary and stellar spectra are different, we investigated the impact of the spectral type for both the star and the planet. We defined a set of 21 simulations (with a star and a planet in each simulation) by varying the planet temperature from 500\,K to 2000\,K by steps of 250\,K, and we assumed 3 stellar temperatures of  3000\,K, 6000\,K and 9000\,K, typically corresponding to M, G, and A type stars. 
In order to study only the impact of the spectral features, we considered a non-realistic situation in which the stellar flux, as well as the planet-to-star contrast, are kept constant for all 21 cases. 
The data simulation is done with $N_{group}=26$ and $N_{int}=13$ for a total exposure of 1 hour determined to achieve sufficient S/N on the planet (located at an angular separation of $1.4''$), in a reasonable computing time.
The contrast at 5\,$\muup$m is set to $10^3$. The model spectrum to calculate the correlation is identical to the input spectrum.

\begin{figure*}[h]
     \centering
     \includegraphics[width=175mm]{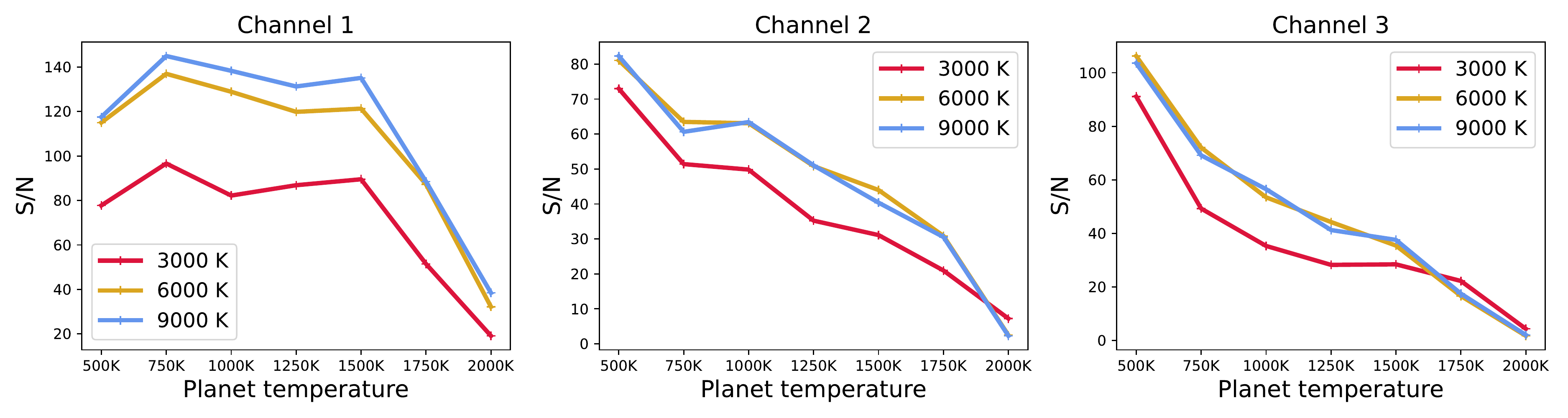}
     \caption{$S/N$ as simulated for a range of planet and star temperatures in the first 3 MRS channels (built over the three sub-bands).}
     \label{fig:impact_sp_type_stars}
\end{figure*}

The S/N measured for the 3 first MRS channels are displayed in Fig. \ref{fig:impact_sp_type_stars} (similar results are observed if we look only at a single band). Globally, we find that colder planets are easier to detect with molecular mapping, as a result of molecular lines being more pronounced in the planetary spectrum. On the contrary, the hottest planet in our sample ($T_{p}=2000$\,K) features a much higher correlation with the stellar spectrum and may become almost undetectable, a feature that is even more remarkable in channels 2 and 3.
These results stand regardless of the stellar spectrum, but we note that, as expected, the detection is globally worse for a colder star, which has more spectral features ($T_{s}=3000$\,K).
We also observe a general trend of a lower correlation signal with increasing wavelength, with channel 1 providing the highest detection.

\subsection{Planets spectral type and angular separation}
Guided by the former analysis, we chose a single stellar temperature ($T_{s}=6000$\,K) and considered more realistic simulations in which the system is located at 30\,pc. Planet fluxes were calculated for the same temperature range as in Sect. \ref{subsec:sp_type} and for a radius of 1\,$R_{Jup}$. We tested the dependency of the molecular mapping efficiency to the planet's temperature and to its angular separation from the star. For convenience, the planet was positioned at the center of the FoV, while the position of the star was offset from 0.2$''$ to 3.2$''$. 
Compared to the simulations in Sect. \ref{subsec:sp_type}, the planets here have potentially lower fluxes, so we generated 2 hours of observation with $N_{group}=33$, and $N_{int}=19$ (again with the goal to minimize computing time). 

\begin{figure*}[h]
     \centering
     \includegraphics[width=175mm]{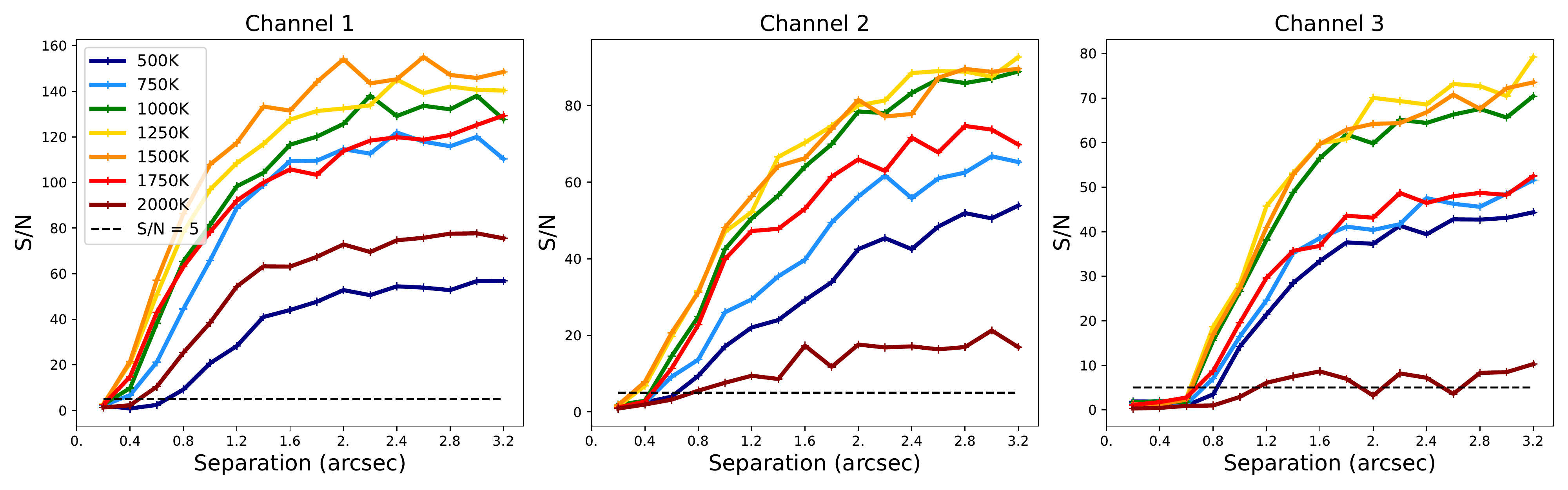}
     \caption{$S/N$ obtained as a function of angular separation from the star for a range of planet temperatures in the first 3 MRS channels.}
     \label{fig:var_param_temp_planet}
\end{figure*}

Fig. \ref{fig:var_param_temp_planet} displays the $S/N$ in channels 1, 2, and 3, for each planet's temperature as a function of the angular separation from the star. 
In general, the $S/N$ does not depend only on the planet's flux, since the continuum is filtered out while suppressing the star's contribution so the high frequency is the dominant factor in the correlation.
The highest $S/N$ values are observed for temperatures ranging between 750\,K and 1750\,K, while lower performances are obtained for the coldest ($T_{p}$ = 500\,K) and the warmest ($T_{p}$ = 2000\,K) planets, because of a lower absolute flux, and respectively, a lower level of correlation due to fewer absorption lines.
The $S/N$ increases rapidly with the angular separation up to about $\sim 1.5''$, and then becomes asymptotic (at least in channels 1 and 3). Channel 2 shows a more gradual increase of the $S/N$ with the separation.\\

\begin{figure*}[h]
     \centering
     \includegraphics[width=175mm]{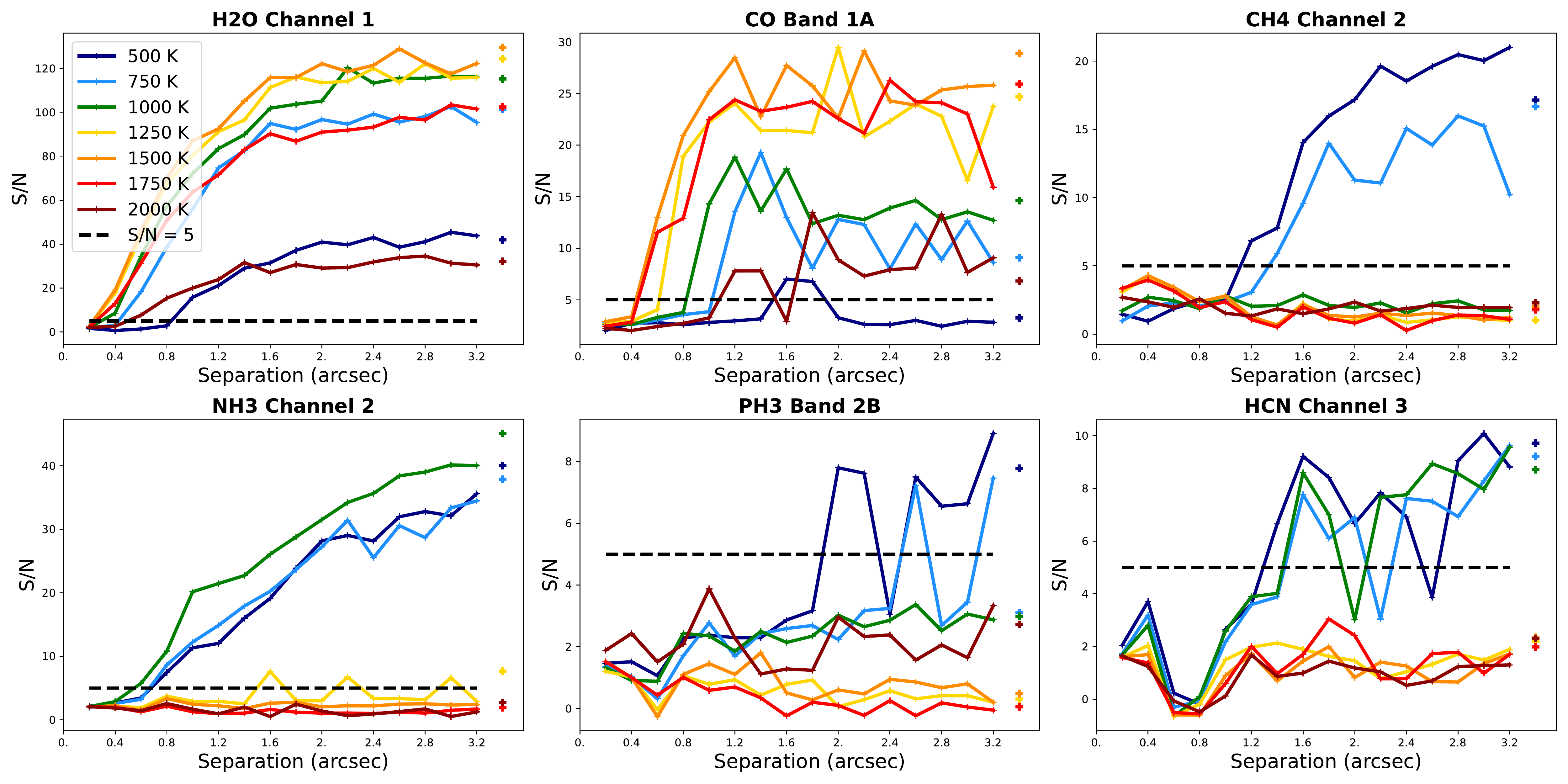}
     \caption{Signal-to-noise ratio obtained at each separation from the star for a range of planet's temperature for the following molecules H$_2$O, CO, CH$_4$, HCN, NH$_3$, PH$_3$ (in the band or channel where the spectral features are the strongest), the cross represents simulation without any star.}
     \label{fig:var_param_temp_planet_molec}
\end{figure*}

The same parametric study is performed for each individual molecule, focusing on the channel or the band in which the detection is optimal (Fig. \ref{fig:var_param_temp_planet_molec}).
The molecules are detected in the bands for which the absorption is the largest, and obviously departing the most from the stellar spectrum, as long as this absorption is not hidden by another molecule's absorption.  
Molecules with spectral features spanning a wider range of wavelengths will benefit from calculating the $S/N$ in the cube built over the three sub-bands of one channel.

- H$_2$O is the prominent molecule in a planet's spectrum for any temperature. It is detected in every channel but mostly in the first one, except for cold planets where CH$_4$ will dominate in channel 2 and hide H$_2$O features. We observe the same trend (rapid and then asymptotic increase versus angular separation) as in the case of the full atmospheric model, although with a slightly lower $S/N$ (120 at maximum). 

- CO is well detected ($S/N= 10 \sim 30$) for the warmest planets, from 1250\,K to 1750\,K (but not in the hottest one at 2000\,K) and as close as $0.6-0.8''$. For colder planets at 750\,K and  1000\,K, CO is detectable for separations larger than 0.8$''$. Because the molecule's spectrum is featureless at wavelengths longer than 6\,$\muup$m we present the result for the band 1A only, which globally presents the highest $S/N$. 
We found that the star itself produces a non-negligible correlation with CO, yielding some spatial residuals in the correlation map responsible for strong variations in the $S/N$ curves. 

- CH$_4$ is only detectable in cold planets (500\,K and 750\,K) in channel 2, and for separations larger than 1.4$''$ as a result of fewer spectral features as compared for instance with H$_2$O.

- NH$_3$ is detected in channel 2 for planets with T$< 1000$\,K, farther than 0.8$''$. The detection of NH$_3$ will be a good tracer to discriminate between several assumptions of a planet's temperature such as 2M\,1207\,b (see Sect. \ref{sec:planets}).

\begin{figure}[h]
     \centering
     \includegraphics[width=80mm]{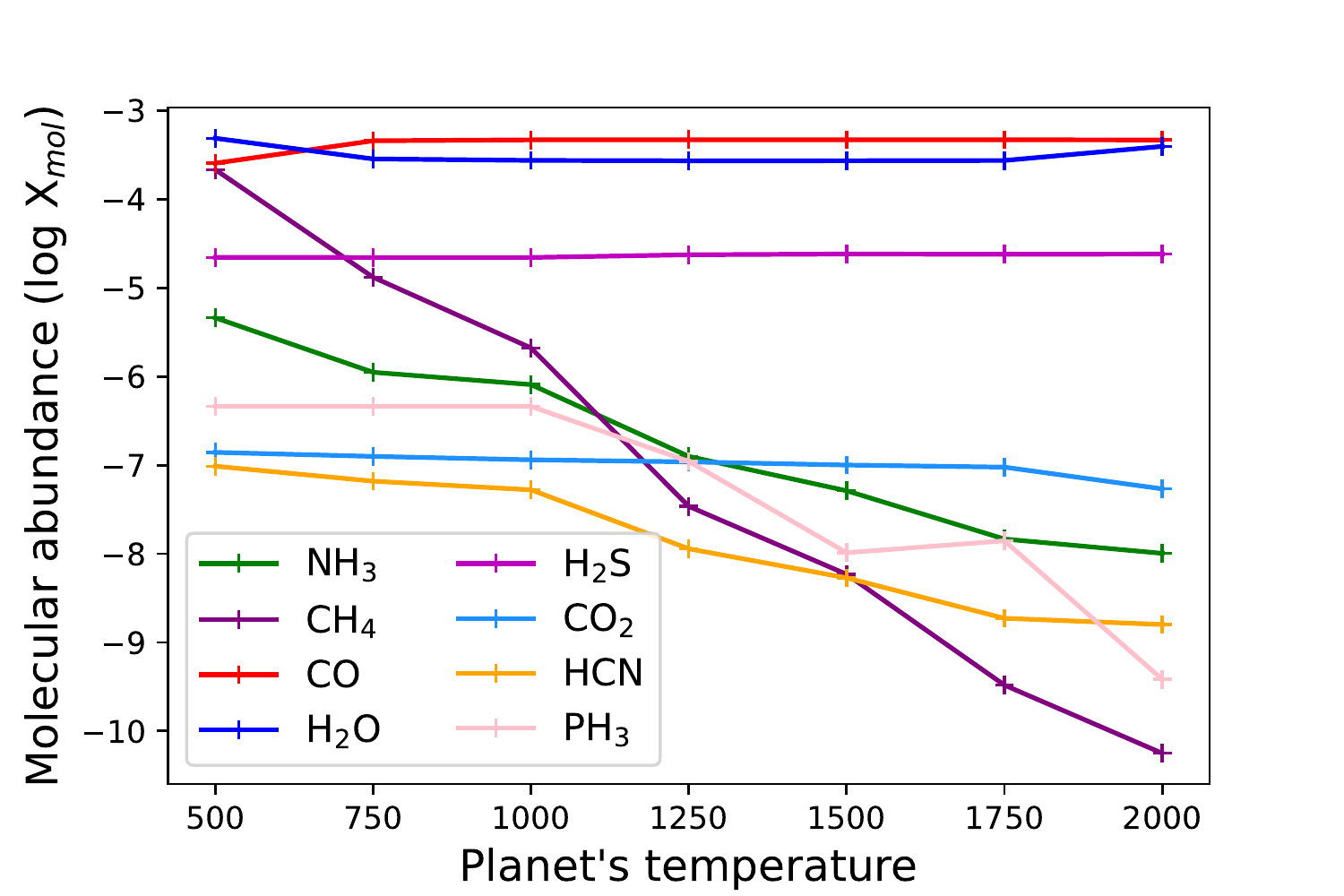}
     \caption{Molecular abundance in Exo-REM models of each molecule as a function of the planetary temperature.}
     \label{fig:molec_abundance_temp}
\end{figure}

- PH$_3$ and HCN have fewer features than the previous molecules, they are by nature more difficult to detect. According to the $S/N$ analysis, we expect potential detection for the coldest objects (T$< 1000$\,K) at rather large separations ($>2''$). For PH$_3$, we restrict the analysis to the individual band 2B and HCN in channel 3.\\
The abundances of each molecule as a function of the planet's temperature are indicated in Fig. \ref{fig:molec_abundance_temp}, justifying the detection of NH$_3$, CH$_4$, HCN, and PH$_3$ only in cold planets. For hot planets, clouds are masking the absorption lines, explaining that fewer molecular features are detected.

To provide a reference $S/N$, we performed the very same simulations without any star, but just a planet in the center of the FoV, indicated as a cross in Fig. \ref{fig:var_param_temp_planet_molec}.
This confirms that PH$_3$ and HCN should be detected for cold planets, in the case where the planet is not contaminated by stellar speckles. 
It also confirms that H$_2$S and CO$_2$, even if quite abundant, are not accessible to the MRS for this range of planet temperature and brightness.
As for H$_2$S, most spectral features are localized between $\sim$5 and 8\,$\muup$m, where the signature of H$_2$O is dominating the spectrum hence masking H$_2$S features. This molecule might be detectable in the case of very bright targets.
The detection of CO$_2$ is limited by shallow spectral features at 15\,$\muup$m, low sensitivity of the instrument at such wavelengths, and stellar contamination. 
Therefore, CO$_2$ could be detected only in very bright objects, if we can manage to strongly attenuate the star's contamination. 
For instance, an optimistic simulation, with a bright system at 22\,pc, in which the star/planet contrast is favorable ($R_{star} = 0.8\,R_{sun}$, $R_{planet} = 1.2\,R_{jup}$, separation=1.5$''$) confirms this assumption: in this case CO$_2$, is clearly detected.
Other molecules present in the models do not show any detection. TiO has spectral features in channel 2 for the hot planets (T > 1750 K), however, it is too faint to be detected. FeH, K, VO, and Na have either no spectral features or signatures too faint to be detected.\\
We notice that for CO, CH$_4$, PH$_3$, at some temperatures the $S/N$ of the simulation without the star is smaller than the $S/N$ with the star.
This is explained by the non-zeros correlation of the star's spectrum with these molecular features which results in a broader correlation pattern and tends to increase the $S/N$. This means that the $S/N$ calculation method isn't perfect yet and could still be improved. More efficient attenuation of the star prior to applying molecular mapping could also help to reduce these effects.

\section{Molecules detection based on known systems}

\label{sec:atm_charac_mm}
\subsection{Choice of targets based on observational limits}
\begin{figure}[h]
     \centering
     \includegraphics[width=90mm]{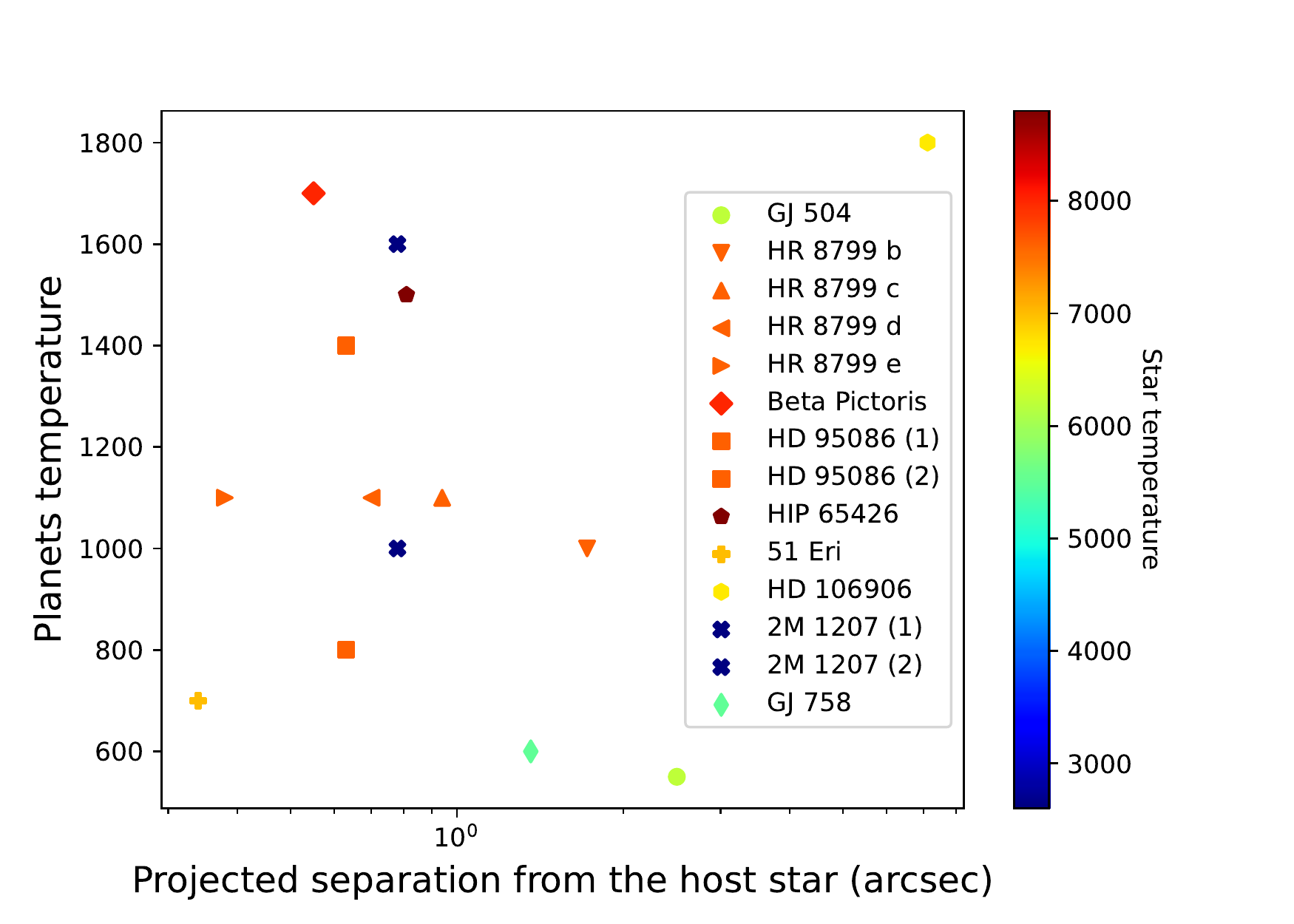}
     \caption{Temperature of the planet as a function of its projected separation with its host star, for all of the planets from the sample. Stellar temperature is color-coded.}
     \label{fig:planet_sample}
\end{figure}

To complement the parametric analysis in Sect. \ref{sec:parametric}, we explored the performance with known directly imaged planets in order to estimate the relevance of future programs with JWST/MIRI. The sample
was defined to fulfill observational requirements: first, the angular resolution that JWST can achieve in the mid-IR imposes angular separations larger than $\sim0.3''$ (which is about the angular resolution for the mean MRS wavelength), and second, the sensitivity of the MRS allows us to observe targets with flux larger than $30\,\muup$Jy \citep[10 $\sigma$ signal in 10000 sec,][]{glasse_mid-infrared_2015}.
Therefore, we considered the following systems: GJ\,504, HR\,8799, $\beta$\,Pic, HD\,95086, HIP\,65426, 51\,Eri HD\,106906, 2M\,1207 and the brown dwarf companion GJ\,758, the characteristics of which are provided in Tab. \ref{table_param_stars} for the stars, and Tab. \ref{table_param_planets} for the planets. These systems cover a broad range of temperatures, angular separations, and stellar types (See Fig. \ref{fig:planet_sample}), hence are meaningful to test the ability to characterize atmospheric parameters in the mid-IR, as compared with previous analysis in the near IR. Furthermore, all of these planets will be observed in the GTO programs with coronagraphs, either with MIRI or NIRCam.\\

\begin{table*}[t]
\begin{tabular}{c c c c c c c c c c}
\hline
\hline
Parameters & GJ\,504 & HR\,8799 & $\beta$ Pictoris & HD\,95086 & HIP\,65426 & 51\, Eri & HD\,106906 & 2M\,1207 & GJ\,758\\
\hline
Spectral type & G0V &  A5V & A6V & A8III & A2V & F0IV & F5V & M8 & G9V\\
Temperature (K) & 6200 & 7600 & 8000 & 7600 & 8800 & 7000 & 6700 & 2600 & 5500\\
R (R$_{\sun}$) & 1.35 & 1.5 & 1.8 & 1.6 & 1.77 & 1.5 & 1.4 & 0.25 & 0.88\\
Distance (pc) & 17.56 & 39.4 & 19.45 & 86.2 & 111.4 & 29.4 & 102.8 & 52.4 & 15.5\\
\hline
\end{tabular}
\caption{Stars parameters chosen for simulation, all have [M/H] = 0 and $\mathrm{log} g = 4.0$.\\
References for stellar distances \cite{van_leeuwen_validation_2007} and \cite{gaia_collaboration_gaia_2016}
}
\label{table_param_stars}
\end{table*}

\begin{table*}[t]
\begin{tabular}{c c c c c c c}
\hline
\hline
Planet's parameters & GJ\,504 b $^{1}$ & HR\,8799 b $^{2}$ & HR\,8799 cde $^{2}$ & $\beta$ Pictoris b $^{3}$ & HD\,95086 b $^{4}$ & HD\,95086 b $^{4}$\\
\hline
T(K)  & 550 & 1000  & 1100  & 1700 & 800 & 1400  \\
$\mathrm{log} g$ & 4.0 & 3.5 & 4.0  &  4.0 & 4.0 & 4.0 \\
Separation (au) & 43 & 68 & 43 - 27 - 16  & 9 & 62 & 62\\
Angular separation (arcsec) & 2.5 & 1.72 & 0.94 - 0.7 - 0.38 &  0.55 & 0.63 & 0.63\\
\hline
Molecules & & & & & & \\
H$_2$O & -3.27 & -3.62 & -3.62 &  -3.62 & -3.53 & -3.62 \\
CO  & -3.68 & -3.3  & -3.3 &  -3.3 & -3.35 & -3.3 \\
CO$_2$ & -6.95 & -6.96 & -7.0 &  -7.01 & -6.91 & -7.0 \\
CH$_4$ & -3.53 & -5.76 & -5.76 &  -7.52 & -4.23 & -7.17\\
HCN & -6.92 & -7.3 & -7.3 &  -7.78 & -6.79 & -7.67\\
NH$_3$ &-5.17 & -6.27 & -6.27 &  -6.94 & -5.61 & -6.75\\
H$_2$S & -4.66 & -4.66 & -4.66 & -4.63 & -4.66 & -4.64\\
PH$_3$ & -6.34 & -6.34 & -6.34 & -7.50 & -6.34 & -6.76\\
\hline
\end{tabular}
\end{table*}
\begin{table*}[t]
\begin{tabular}{c c c c c c c}
\hline
\hline
Planet's parameters & HIP\,65426 b $^{5}$ & 51 Eri b $^{6}$ & HD\,106906 $^{7}$ & 2M\,1207 $^{8}$ & 2M\,1207 $^{9}$ & GJ\,758 $^{10}$\\
\hline
T(K)  & 1500 & 700 & 1800 & 1000 & 1600 & 600\\
$\mathrm{log} g $ & 4.0 & 4.0 & 4.0 & 4.0 & 4.0 & 3.5\\
Separation (au) & 110 & 11 & 850 & 125 & 125 & 226\\
Angular separation (arcsec) & 0.81 & 0.34 & 7.11 & 0.78 & 0.78 & 1.36\\
\hline
Molecules & & & & & & \\
H$_2$O & -3.62 & -3.44 & -3.62 & - 3.62& -3.62 & -3.32\\
CO  & -3.3 & -3.42 & -3.3 & -3.3 & -3.3 & -3.58\\
CO$_2$ & -6.96 & -6.89 & -7.06 & -6.98 &-7.02 & -6.91\\
CH$_4$ & -7.06 & -3.9 & -7.83 & -5.30 &-7.61& -3.61\\
HCN & -7.6 & -6.78 & -7.88 & -7.06 &-7.81& -6.84\\
NH$_3$ & -6.71 & -5.46 & -7.05 & -5.97 &-6.93 &-5.22\\
H$_2$S & -4.65 & -4.66 & -4.63 & -4.66 &-4.63 &-4.66\\
PH$_3$ & -6.47 & -6.34 & -7.83 & -6.34 &-7.36&-6.34\\
\hline
\end{tabular}
\caption{Parameters of the simulated planets. All of them are simulated with [M/H] = 1 and C/O = 0.5. The volume mixing ratio (in log scale) of molecules present in the Exo-REM models are given, for each of the simulated planets, at the top of the atmosphere ($\log X_{mol}$ at 10$^{-2}$ bar). Other molecules used in the model have a volume mixing ratio less than $10^{-20}$. \\
References : $^{1}$\cite{bonnefoy_gj_2018}, $^{2}$\cite{konopacky_detection_2013}, $^{3}$\cite{bonnefoy_near-infrared_2013}, $^{4}$\cite{desgrange_-depth_2022}, $^{5}$\cite{petrus_medium-resolution_2021} and \cite{chauvin_discovery_2017}, $^{6}$\cite{samland_spectral_2017} , $^{7}$\cite{daemgen_high_2017}, $^{8}$\cite{barman_young_2011}, $^{9}$\cite{patience_highest_2010}, $^{10}$\cite{vigan_first_2016}}
\label{table_param_planets} 
\end{table*}

Stellar spectra are defined with the parameters from Tab. \ref{table_param_stars} and normalized to the mean flux density values at 5.03\,$\muup$m\footnote{we choose the shortest MRS wavelengths to be representative of the actual saturation level of the targets in the sample.} as measured in the M band of the Johnson photometric band and tabulated at the SIMBAD astronomical database \citep{wenger_simbad_2000}.
Since we are studying young systems, we can expect unresolved inner dust rings to contribute to the mid-IR flux, which needs to be taken into account in the global stellar flux.
Planetary spectra are generated with the Exo-REM model using a set of temperature, $\mathrm{log} g$, [M/H], C/O ratio, as listed in Tab. \ref{table_param_planets}, and their flux density is scaled to the distance of the system and the planet's radius. For some systems, the flux level is adapted so that the models globally match the near IR data.
These data are then converted to MIRISim input requirements: $\muup$Jy for the flux density, and $\muup$m for the wavelengths.
We used \url{whereistheplanet.com} \citep{wang_whereistheplanet_2021} to infer the astrometry of the planet for an arbitrary date of June 2023 (likely the start of JWST cycle 2). For long-period planets, their projected positions do not vary significantly with the date (except for $\beta$ Pictoris b).

The position and the spectrum of each object are used in the Exposure Time Calculator (ETC) to calculate the observational parameters ($N_{group}$, $N_{int}$).
The number of groups per integration $N_{group}$ was determined to avoid saturation while maximizing counts. Then, we chose the number of integrations to reach a $S/N$ larger than 3 (and ideally above 5) on the detector for the planet's flux in each spectral band for the complete observation. The $S/N$ is extracted on an aperture of 0.4$''$ centered on the planet. These parameters are indicated for each simulation in Tab \ref{tab:param_simu}. 
The ETC is also convenient to check, based on the astrometry, that the planet is contained within the FoV. 
When needed, we adapted the telescope pointing, to position either the planet at a suitable location on the detector (especially if the angular separation is of the order of the size of the FoV) or to move away from a bright star which may cause saturation and latency.
The $S/N$ of the detection for each molecule of each system is summed up in Tab. \ref{tab:s_n_molec_planet}.

\subsection{Simulations and molecular mapping analysis of this planet sample}
\label{sec:planets}
\begin{figure*}[h]
     \centering
     \includegraphics[width=190mm]{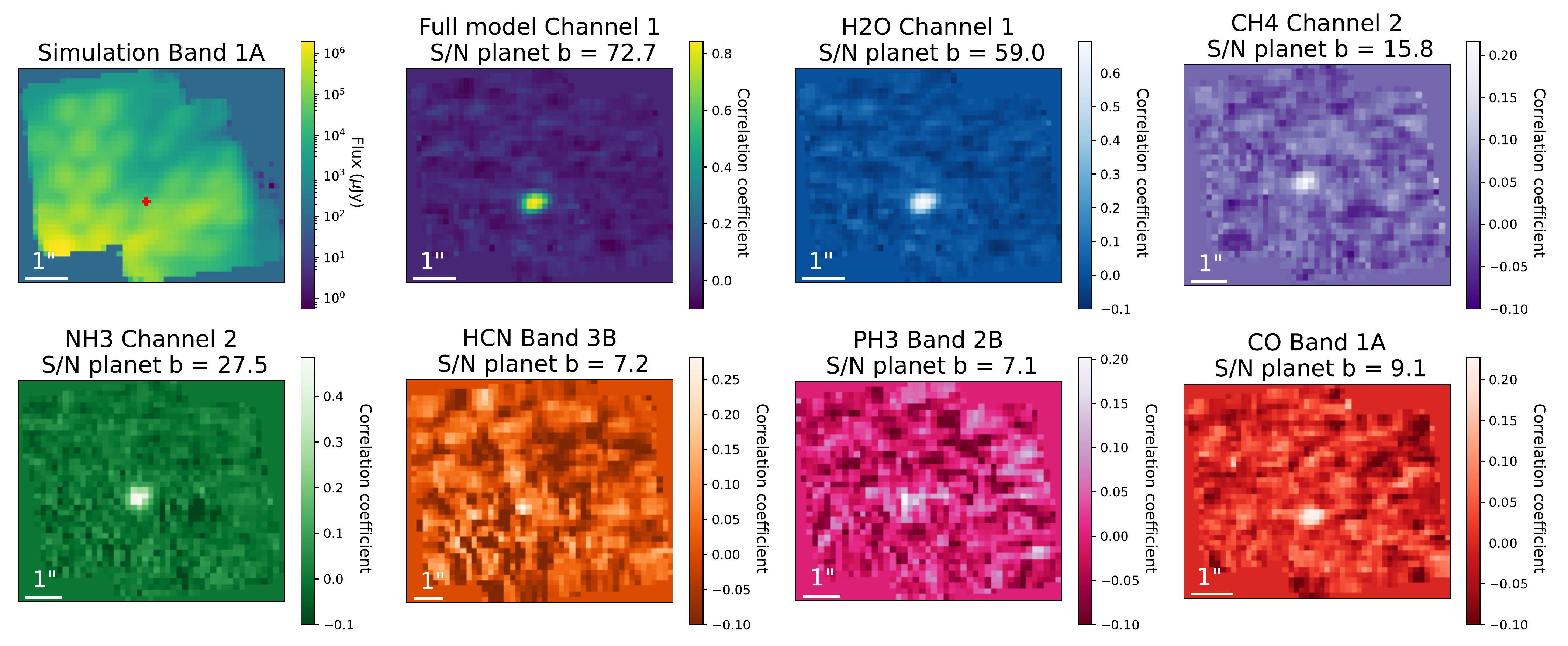}
     \caption{Example of correlation maps for the simulated system GJ\,504 with Exo-REM full atmospheric template and molecular template. Each molecule is shown in the channel/ band where the S/N is the highest.}
     \label{fig:GJ504_cc_maps}
\end{figure*}

\textbf{GJ\,504\,b} is a T8-T9.5 object, discovered by \citep{kuzuhara_direct_2013}. \cite{bonnefoy_gj_2018} has analyzed the system in detail, aiming at constraining atmospheric parameters with near IR data (from 1 to 2.5 $\mu$m). The uncertainty on the age (21 Myr to 4 Gyr) of this system gives two mass regimes (1 $M_{jup}$ or 23 $M_{jup}$), making this object either a young exoplanet or an older brown dwarf. More measurements on the molecular abundances and metallicity are needed to put more robust constraints on the planet and thus on its formation. 
Methane has been detected in the atmosphere of this planet by \cite{janson_direct_2013}, but no other molecular feature has been detected yet.\\
The system is simulated by offsetting the star outside of the FoV at coordinates (2.0, -2.5)$''$. As it is a nearby system, the star is too bright for the MRS and the detector would saturate in a few groups. Offsetting the star allows longer integrations. 
Processing the simulated data with the molecular mapping method, we obtained the correlation maps shown in Fig. \ref{fig:GJ504_cc_maps}, which displays the $S/N$ for each detection. 
\begin{figure}[h]
     \centering
     \includegraphics[width=80mm]{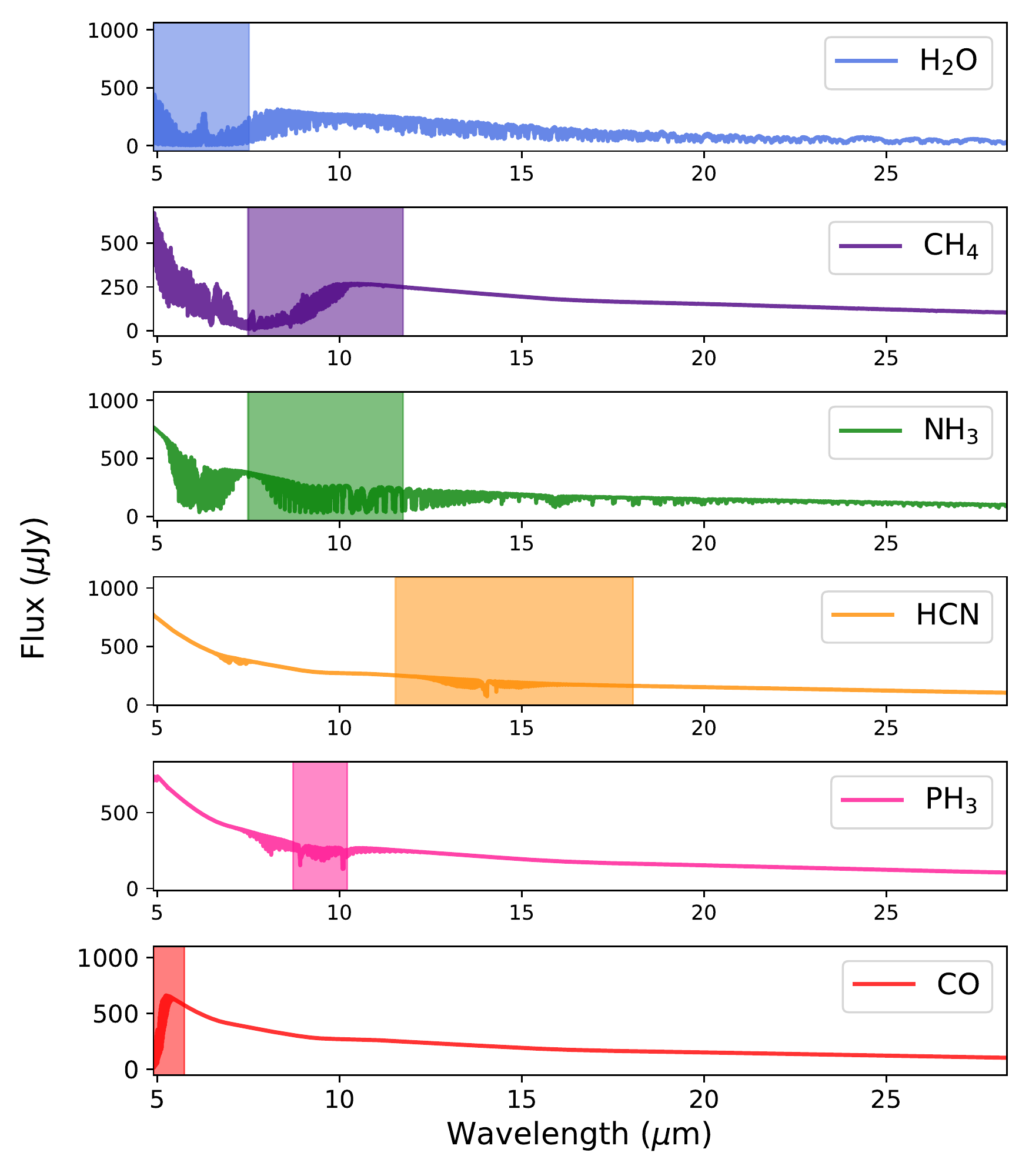}
     \caption{Exo-REM molecular template spectra of GJ\,504 b The parts of the spectra corresponding to the channels/ bands shown in Fig.\ \ref{fig:GJ504_cc_maps} are highlighted.}
     \label{fig:GJ504_spectra}
\end{figure}
We were able to detect: H$_2$O, CO, CH$_4$, NH$_3$, HCN, and PH$_3$. 
Moreover, the correlation with the full model allows for the detection of wavelengths up to 18 $\mu$m 
(Appendix Fig. \ref{fig:cc_maps_GJ504_all}).\\
As a test case, we ran the simulation without the star to assess the impact of the stellar contribution. We found an improvement of the $S/N$ by a factor of 3, respectively 2, for NH$_3$ and HCN, respectively, CH$_4$. Other molecules are also easier to detect, and in particular, a $S/N$ = 6.8 is achieved for CO$_2$. These results argue for better stellar removal to improve the detections and possibly detect CO$_2$ in this system and other similar systems.\\

\textbf{HR\,8799} system harbors 4 young giant planets with similar characteristics in terms of temperature and luminosity. They have been discovered by \citet{marois_direct_2008,marois_images_2010}.
The presence of a planetesimal belt has been inferred from sub-millimetric observations with ALMA \citep{booth_resolving_2016}.
Water and carbon monoxide have been clearly identified at high $S/N$ in HR\,8799\,b,\,c, and d \citep{petit_dit_de_la_roche_molecule_2018, barman_simultaneous_2015,ruffio_deep_2021}. However, the presence of methane is still debated. In the case of HR\,8799\,b, it was claimed by \cite{barman_simultaneous_2015} using cross-correlation with a model spectrum on Osiris Keck data, but not confirmed using molecular mapping on the same data \citep{petit_dit_de_la_roche_molecule_2018} neither from complementary data \citep{ruffio_deep_2021}. 
Broadband photometry of planets b,\,c, and d has provided evidence of significant atmospheric cloud coverage, while spectroscopy of planets b and c shows evidence for non-equilibrium CO/CH$_4$ chemistry \citep{janson_spatially_2010,hinz_thermal_2010}.\\
In the simulation, the star is offset to the coordinates (-0.4, 0.0)$''$ from the center of the FoV to be sure that all four planets are contained in a single dither position. 
The planet e cannot be detected, as it is too close to the star to be resolved with the MRS.  
The correlation maps of these simulated data are displayed in Fig. \ref{fig:cc_maps_HR8799}. 
We detect H$_2$O and CO (for planets b, c, and d). There is also a faint detection of NH$_3$ and CH$_4$ for planet b. No other molecule is detected.\\

\textbf{$\beta$ Pictoris} system has two discovered planets \citet{lagrange_probable_2009, lagrange_giant_2010, lagrange_post-conjunction_2019, nowak_direct_2020}. Planet c is too close to the star and cannot be resolved with the MRS, therefore we focus on planet b. This planet has a dusty atmosphere \citep{bonnefoy_near-infrared_2013}. Previously, water and carbon monoxide have been detected in its atmosphere with SINFONI data using molecular mapping \citep{hoeijmakers_medium-resolution_2018}.\\
Using the observation parameters of the GTO 1294 (PI: C. Chen) we did not manage to detect the planet, therefore we chose to increase the number of integrations. 
From the angular separation and temperature of the planet, we do not expect a strong detection.
However, the brightness of the target still allows us to detect the planet with the full atmospheric model, while H$_2$O is the only detected molecule.
The correlation maps corresponding to this system are presented in Fig. \ref{fig:cc_maps_beta_pictoris}.\\

\textbf{HD\,95086\,b}'s detection has been presented in \cite{rameau_confirmation_2013}, they showed that it has a cool and dusty atmosphere, where the effects of possible non-equilibrium chemistry, reduced surface gravity, and methane bands in the near-infrared might be explored in the future. \cite{chauvin_investigating_2018} found that its near-infrared spectral energy distribution is well-fitted by spectral models of dusty and/or young L7-L9 dwarfs.
Here, we aim at testing the two scenarios highlighted in \cite{desgrange_-depth_2022} using SPHERE observations combined with archival observations from VLT/NaCo and Gemini/GPI. These scenarios indicate that the color of the planet can be explained by the presence of a circumplanetary disk around planet b, with a range of high-temperature solutions (1400–1600\,K) and significant extinction, or by a super-solar metallicity atmosphere but lower temperatures (800–1300\,K), and small to medium amount of extinction.\\
We performed two simulations, one with a planet at the temperature of 800\,K and a second with a planet of 1400\,K.
With the full spectrum, the planet is only detected in the coldest scenario.
In terms of molecules, we secured the detection of H$_2$O together with a suspicion of NH$_3$. 
These results are presented in Fig. \ref{fig:cc_maps_HD95086_800} for the cold planet scenario and in Fig. \ref{fig:cc_maps_HD95086_1400} for the warm planet scenario.
The planet is close to its star (one of the closest in our sample) and this system is located at a large distance, which explains the globally faint detection of the planet and non-detection of other molecules.\\

\textbf{HIP\,65426\,b} has been discovered by \cite{chauvin_discovery_2017} with VLT/SPHERE, it is a young giant exoplanet. The Y- to H- band photometry and low-resolution spectrum indicate a $L6 \pm 1$ spectral type and a warm, dusty atmosphere.
\cite{petrus_medium-resolution_2021} studied this system with different methods including molecular mapping using VLT/SINFONI data and detected carbon monoxide and water vapor.\\
The planet is detected in channel 1, and only the molecule H$_2$O is detected.
At this temperature, the detection of CO was expected from our parametric study. However, in the case of this system, which is almost 4 times more distant than the systems simulated in the parametric studies, the fainter flux of the planet explains why we do not have better detection of the planet and no detection of CO.\\

\textbf{51\,Eri\,b} has been discovered with GPI in the near IR, \cite{macintosh_discovery_2015} indicates that it has a T4.5-T6 spectral type and the J-band spectroscopy shows methane absorption. From VLT/SPHERE data in the near IR, \cite{samland_spectral_2017} derived the presence of a vertically extended, optically thick cloud cover with small particles.\\
51\,Eri is a bright star, the number of groups is small to avoid saturation so that the integration time is larger than for other sources to achieve the $S/N$ required to detect the planet. The correlation maps are shown in Fig. \ref{fig:cc_map_51_Eri}.
Even though the planet is bright, it is not detected as it is too close to the star to be observable with this method, as expected from our parametric study in Sect. \ref{sec:parametric}.\\

\textbf{HD\,106906\,b} is a young low-mass companion near the deuterium burning limit \citep{daemgen_high_2017}. It has been characterized spectrally in the near IR with VLT/SINFONI. This planet is the hottest and the most distant from its star in our sample.\\
Therefore, to simulate this system, we chose to have the planet in the center of the FoV and the star's PSF located outside the FoV. 
In principle, molecular mapping is not required to detect the planet as it is far from its host star and sufficiently bright.
Applying molecular mapping, the planet is detected in all 3 channels, we can detect H$_2$O and CO.
At high temperatures, we do not expect other molecules to be detected, as we can see in Fig. \ref{fig:cc_maps_HD106906}.\\

\textbf{2M\,1207\,b} is the first planet ever detected with direct imaging by \cite{chauvin_giant_2004} and will be one of the first exoplanets targeted with the MRS (GTO 1270, PI : Stephan Birkmann). 
The atmospheric properties of 2M\,1207\,b are not well constrained, and the MRS observations have the ability to break the degeneracy between two radically different models.
On the one hand, \cite{barman_young_2011} proposed a temperature of 1000\,K, log(g) = 4 and 1.5\,$R_{Jup}$ which is in agreement with the first estimate of \cite{chauvin_giant_2004}, and on the other hand, \cite{patience_highest_2010} found a best fit model at about 1600\,K and log(g) = 4.5 with a smaller radius of 0.5\,$R_{Jup}$.\\
The correlation maps are shown in the appendix Fig. \ref{fig:cc_maps_2M1207_1000K} and Fig \ref{fig:cc_maps_2M1207_1600K}.
Considering the model at 1000\,K, we obtained a detection of H$_2$O, CO, and NH$_3$. The planet is detected at a very high S/N with the full model.
For the model at 1600\,K, the detection of H$_2$O is much weaker 
and CO is undetected simply because the planet is smaller than in the former case. NH$_3$ is also undetected, as expected for such a high temperature based on the parametric study. 
The star being an M8 brown dwarf, the warmer planet scenario represents an extreme case for the molecular mapping method, but based on our simulations the MRS has the ability to provide a definitive answer about the planet's temperature. The detection of NH$_3$ can be a good indicator of the temperature of the planet.\\

\textbf{GJ\,758\,B} is a brown dwarf companion to a solar-type star. It has been discovered with Subaru/HiCIAO \citep{thalmann_discovery_2009} and characterized with VLT/SPHERE in \citet{vigan_first_2016}. 
No atmospheric model perfectly reproduces the measured fluxes of GJ\,758 B in the near IR. As one of the coldest companions that have been directly imaged, it also appears to be an interesting target to apply molecular mapping.\\
The star is bright, therefore we offset it outside the FoV at coordinates (2.3, 1.3)$''$, and we used $N_{group}=59$, and $N_{int}=5$ for a total exposure time of 3374.45\,s for one observation. The dithering pattern is modified from the default pattern to be optimized for the system.
Correlation with the full spectrum yields a detection in the three channels, while both H$_2$O and NH$_3$ are clearly detected, and CH$_4$ can be suspected. 
These results are displayed in Fig. \ref{fig:cc_maps_GJ758}.

\begin{table}[t]
\centering
\begin{tabular}{c c c c}
\hline
\hline
System & $N_{group}$ & $N_{int}$ & Exposure time (s)\\
\hline
GJ\,504 & 52 & 9 & 5649.98\\
HR\,8799 & 21 & 44 & 10733.85\\
$\beta$ Pictoris & 5 & 100 & 6649.00\\
HD\,95086 & 76 & 20 & 7083.15\\
HIP\,65426  & 79 & 10 & 8869.0\\
51 Eri  & 10 & 200 & 24409.25\\
HD\,106906 & 100 & 4 & 473.36\\
2M\,1207 & 76 & 1 & 843.61\\
GJ\,758 & 59 & 5 & 3374.45\\
\hline
\end{tabular}
\caption{Simulation parameters for each target. Exposure times are indicated for one observation (three observations are required to obtain the full wavelength range). The settings of the simulation for 2M\,1207 are the ones that will be used for the GTO program.}
\label{tab:param_simu} 
\end{table}

\begin{table*}[t]
\centering
\begin{tabular}{c c c c c c c c }
\hline
\hline
Planets & Full model & H$_2$O & CO & NH$_3$ & CH$_4$ & HCN &PH$_3$\\
 & Channel 1 & Channel 1 & Band 1A & Channel 2 & Channel 2 & Channel 3 & Band 2B \\
\hline
GJ\,504 b & 72.9 & 59.0 & 9.1 & 27.4 & 15.8 & 7.9 & 7.1 \\
HR\,8799 b & 66.0 & 64.6 & 10.4 & 7.7 & 4.1 & -- & -- \\
HR\,8799 c & 34.8 & 34.4 & 9.5 &  -- & -- & -- & -- \\
HR\,8799 d & 34.7 & 36.6 & 7.5 & -- & -- & -- & -- \\
$\beta$ Pictoris b & 6.6 & 7.2 & -- & -- & -- & -- & -- \\
HD\,95086 b (800\,K) & 5.9 & 6.8 & -- & -- & 3.6 & -- & -- \\
HD\,95086 b (1400\,K) & -- & 3.5 & -- & -- & -- & -- & -- \\
HIP\,65426 b & 13.6 & 13.7 & -- & -- & -- & -- & -- \\
51 Eri b & -- & -- & -- & -- & -- & -- & -- \\
HD\,106906 b & 43.5 & 39.0 & 18.2 & -- & -- & -- & -- \\
2M\,1207 b (1000\,K) & 57.4 & 52.9 & 9.7 & 10.8 & -- & -- & -- \\
2M\,1207 b (1600\,K) & 7.9 & 6.7 & -- & -- & -- & -- & -- \\
GJ\,758 B & 17.5 & 17.0 & -- & 9.7 & 3.5 & -- & -- \\

\hline
\end{tabular}
\caption{$S/N$ detection measured in correlation maps with the full atmospheric model and molecules templates for each planet of the sample. Only $S/N$ > 3 are indicated.}
\label{tab:s_n_molec_planet} 
\end{table*}

\section{Atmospheric characterization using grids of Exo-REM models for GJ\,504 b}
\label{sec:atm_charac_GJ504}

The former section suggests that GJ\,504 b is the main interesting target of our sample for molecular mapping with the MRS. 
Here, we explore the potential of characterization on this specific system using two methods, one using the correlation maps and the other one with $\chi^{2}$ minimization. 

\subsection{Correlation maps with grid of models}
After subtracting the low frequencies on data and models (same method as in Sect. \ref{sec:method_sub_stellar}), the data are correlated with a grid of Exo-REM models, varying the temperature, the metallicity, the surface gravity, and the C/O ratio of the models. Models are high-pass filtered in the same way as the data.
For each model, we calculate the correlation map with the correlation coefficients using Eq. \ref{eq:C}, where $\sigma_S(\lambda)$ is the uncertainty on the flux at each wavelength, as extracted from the ERR extension of the cubes (see Sect. \ref{sec:jwst_pipeline}). This value takes into account the photon noise and detector noise in each spaxel.
\begin{equation}
C= \frac{\sum_\lambda S(\lambda) \times M(\lambda) /\sigma_S(\lambda)^2}{\sqrt{\sum_\lambda S(\lambda)^2 /\sigma_S(\lambda)^2 \times \sum_\lambda M(\lambda)^2/\sigma_S(\lambda)^2}}
\label{eq:C}
\end{equation}

\begin{figure}[h]
     \centering
     \includegraphics[width=90mm]{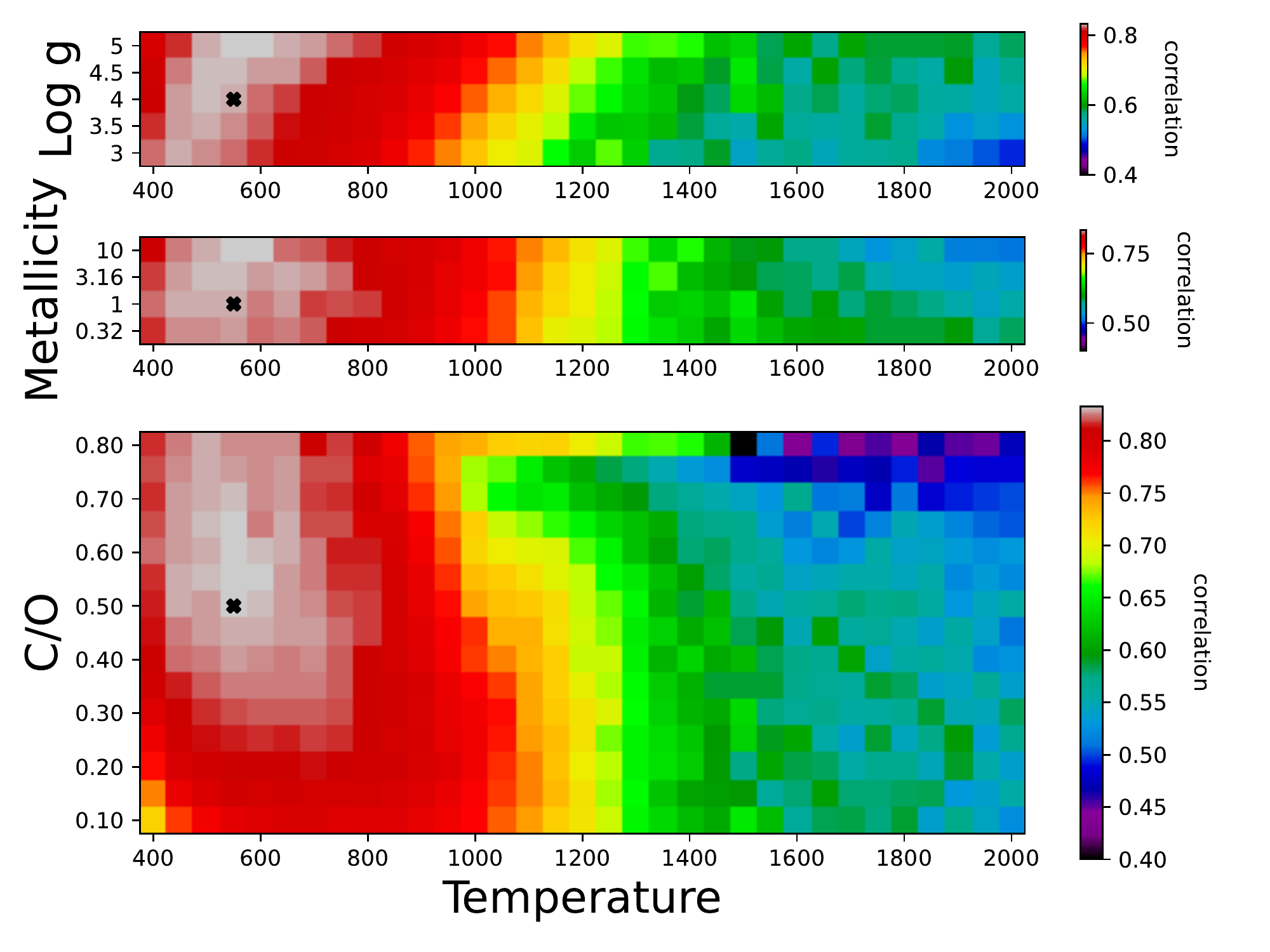}
     \caption{Grids of the correlation coefficient between the spectrum at the position of the planet and the Exo-REM models.
     Case of GJ\,504 in channel 1. Similar results are observed on the other bands or reconstructed channels.}
     \label{fig:GJ504_cc_grid}
\end{figure}

Fig. \ref{fig:GJ504_cc_grid} shows the grids of correlation with the correlation value at the position of the planet, that we obtained when exploring two parameters at once. 
In practice, for each coordinate in any of the grids, corresponding to a couple of parameter values, we took the maximum correlation coefficient obtained when varying the two other parameters.
The real parameters of the planet in input to the simulation are indicated with black crosses.

In general, a significant range of models is producing high correlation values, resulting in a broad peak around the input parameter values, which implies a relatively low accuracy on the retrieved parameters. 
Still, the C/O vs. temperature correlation grid matches reasonably well the input parameters (bottom subplot in Fig. \ref{fig:GJ504_cc_grid}). On the contrary, we observed a tendency for higher metallicity and higher surface gravity in the two upper subplots in Fig. \ref{fig:GJ504_cc_grid}. This apparent mismatch will be discussed in the next section.

\subsection{$\chi^{2}$ minimization with grid of models}
The former method based on the correlation with models is not sufficient to evaluate the reliability of the best model (maximum of correlation) with respect to the data. In addition, the noise estimation does not take into account the spatial noise induced by speckles.
We now investigated the characterization capabilities using $\chi^2$ minimization
Starting from the high-frequency spectrum extracted at the planet's position, we compared it to the same grid of Exo-REM models high-pass filtered. 
We use the correlation map to define where the planet is located, and then extract a high-frequency spectrum in the cubes after the high-pass spectral filtering on the spaxels. 
We extracted that spectrum for the planet's signal by co-adding the flux in the spaxels that are defined by the $S/N$ analysis (section \ref{sec:S/N}). 

We note that this high-frequency spectrum does not contain any information on the total flux of the planet, due to the subtraction of low frequencies. 
We introduced a factor R, chosen to minimize the $\chi^2$ for a given model, as given in Eq. \ref{eq:equation_R}. 
It corresponds to a global scaling parameter that does not influence the shape of the synthetic spectra, but allows us to take into account the planet's radius \citep[as done in][]{baudino_interpreting_2015}, and also captures some photometric calibration issues between the MIRISim simulated model and the actual model.

\begin{figure}[h]
     \centering
     \includegraphics[width=90mm]{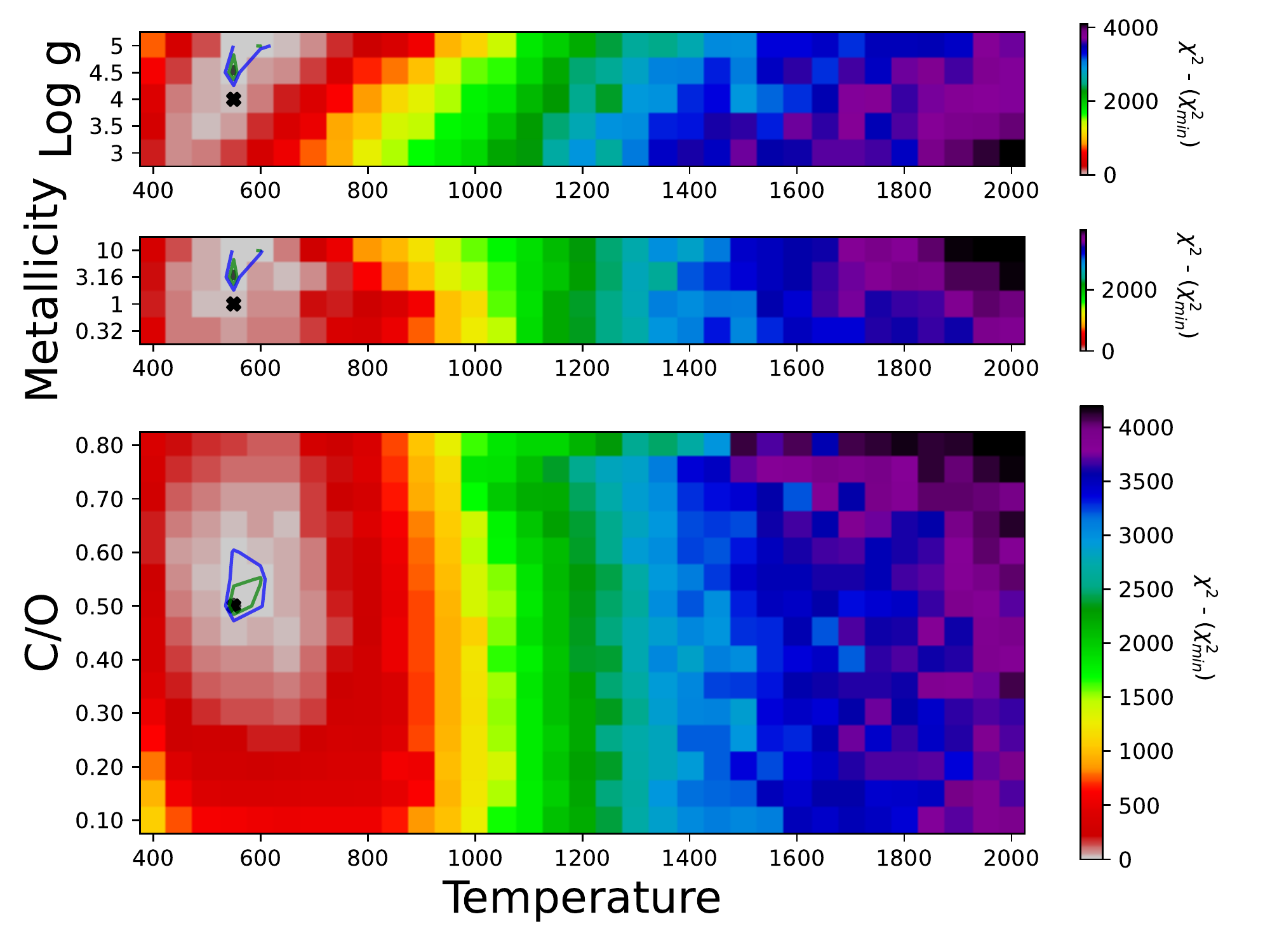}
     \caption{Values $\chi^{2}$ - $\chi^2_{min}$ using the high-frequency spectrum and a grid of Exo-REM models. Green, blue and black lines are respectively 1$\sigma$, 2$\sigma$, and 3$\sigma$ confidence regions (2.3, 6.18, and 11.8).}
     \label{fig:GJ504_chi2}
\end{figure}

\begin{figure}[h]
     \centering
     \includegraphics[width=90mm]{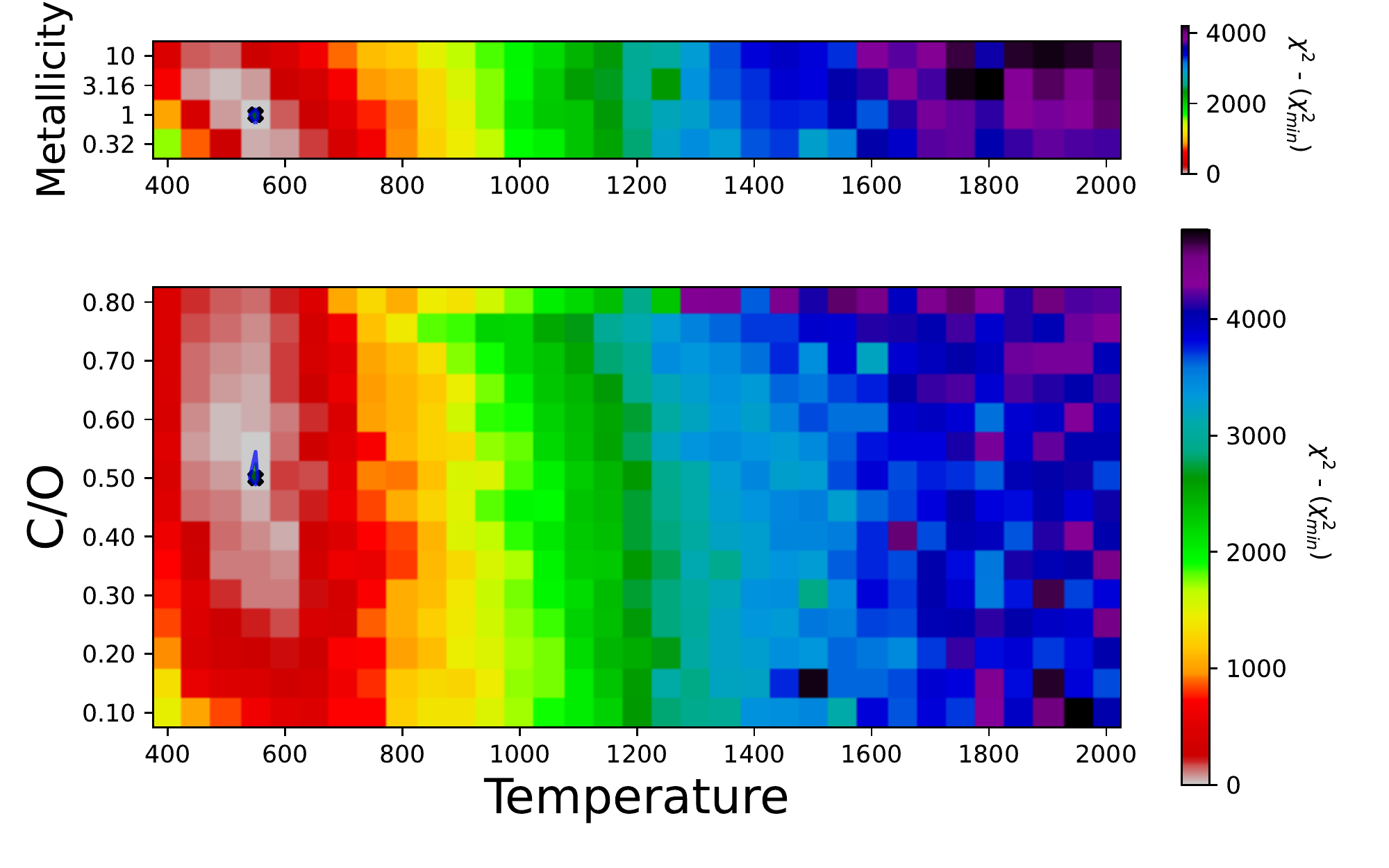}
     \caption{Values $\chi^{2}$ - $\chi^2_{min}$ using the high-frequency spectrum and a grid of Exo-REM models. The log of the surface gravity is fixed at 4.0 to correspond to the input value in each subplot. Green, blue and black lines are respectively 1$\sigma$,2$\sigma$, and 3$\sigma$ confidence regions (2.3, 6.18, and 11.8).}
     \label{fig:GJ504_chi2_fixed}
\end{figure}

For each model, we determined the $\chi^2$ using the equations \ref{eq:equation_R} and \ref{eq:equation_chi2}, in which $\sigma_F(\lambda)$ is the uncertainty on the flux measured at each wavelength in an annulus at the same planet's separation from the star in the high frequencies cubes. 
The noise $\sigma_S$ extracted from the JWST pipeline is now negligible, and this noise $\sigma_F$ takes into account the spatial variation in the high-frequency cubes.
Similarly to Fig. \ref{fig:GJ504_cc_grid}, we display the $\chi^2-\chi_{min}^2$ values of the grid of models in Fig. \ref{fig:GJ504_chi2}.
\begin{equation}
    \chi^2 = \sum_\lambda \bigg( \frac{S(\lambda) - R\times M(\lambda)}{\sigma_F(\lambda)}\bigg)^2
    \label{eq:equation_chi2}
\end{equation}
\begin{equation}
    R = \frac{\sum_\lambda \big(S(\lambda)\times M(\lambda)\big) /\sigma_F(\lambda)^2}{\sum_\lambda M(\lambda)^2/\sigma_F(\lambda)^2}   
    \label{eq:equation_R}
\end{equation}
As a sanity check, we obtained $\chi^2$ values that are in agreement with the number of independent points in the spectrum. 
The $\chi^2$ minimization gives similar results compared to the grid of correlation values.
Again, several models yield $\chi^2$ values that are close to the one of the input model, although the $1\sigma$, $2\sigma$, and $3\sigma$ contours point to a more restrained region in the C/O vs. temperature grid, when compared to Fig. \ref{fig:GJ504_chi2}. However, this is not the case for the metallicity and the surface gravity that are not well constrained, likely due to degeneracies between these atmospheric parameters. 
Indeed, the metallicity and the surface gravity information are mostly contained in the relative depth of the lines, which are affected by the filtering stage. Hence, it is more difficult to disentangle between two models that have different metallicity or surface gravity values.
On the contrary, the abundance of molecules, such as H$_2$O, CO, and CH$_4$ that define the C/O ratio, depends on the temperature of the planet. Even though the continuum of the planetary spectrum is lost in the filtering, they still have a net effect in the high-frequency spectrum of the planet, explaining the rather good match obtained for the temperature and the C/O ratio.
In the event that one of these parameters is well determined by other methods or other observations, the remaining parameters can be well constrained with high confidence as illustrated in Fig. \ref{fig:GJ504_chi2_fixed} in which the surface gravity is fixed at the input values ($\mathrm{log} g = 4.0$) in each subplot.

\section{Discussion}
\label{sec:discussion}
\subsection{Pros and cons of the mid IR for the molecular mapping}

The Mid-IR spectral range offers advantages when combined with molecular mapping with respect to the near IR. For instance a planet at 700\,K has its emission peak at 5.2 $\muup$m while the stellar contribution is getting less dominant. According to the parametric study (systems at 30\,pc, planets of 1 R$_{jup}$ around a star at 6000\,K, 2 hours per observation), the MRS should allow us to detect planets at separations larger than 0.5$''$, (except the hottest planet around the coldest star in our sample), as well as to allow the characterization of those that are further away than 1$''$ from their star. We found that planets with temperatures above 500\,K and below 1500\,K will be easier to detect especially if they orbit G-type stars or younger. In addition, the mid-IR gives access to molecules that are not easily accessible at shorter wavelengths because of fainter absorption features (such as NH$_3$, HCN).

On the contrary, the angular resolution is worse than in the near-IR, so stellar contamination becomes a more important issue. For this reason, we have chosen GJ\,504 b to perform a detailed characterization study as it features favorable configurations in terms of angular separations and temperature. The comparison to a grid of models with a $\chi^2$ approach provides reasonable results (section \ref{sec:atm_charac_GJ504}) but for other targets which are too close to their star, the confidence regions in the $\chi^2$ maps are too large to reduce the accuracy on the atmospheric parameters. Only planets with  $S/N$ larger than 30 in our sample (HR\,8799 planets, 2M\,1207 b at 1000\,K) are amenable to atmospheric characterization with the MRS. We note that the extracted spectra of the others planets (with $S/N < 30 $) are too contaminated by stellar residuals.
Although well detached, HD\,106906 b is a special case because Exo-REM is not well adapted to model a planet at this temperature, therefore other atmospheric models would be needed to characterize this planet.

Moreover, having priors on the atmospheric parameters to correlate our data with models is necessary to reduce the explored parameter range.
As a first approximation, one can use the high-frequency spectrum before applying the molecular mapping, to restrain this range using the $\chi^2$ minimization.
We will benefit from the use of both analyses: molecular mapping and $\chi^2$ minimization, other methods of atmospheric retrievals would be relevant to further study of these planets. 

Finally, Molecular mapping relies on the fact that the spectrum has lots of spectral features due to the molecular absorption and might be less efficient for young dusty planets since dust and clouds tend to flatten the spectrum, as noted for the PDS 70 planets in \cite{cugno_molecular_2021}, (extinction due to dust has not been taken into account in the present work).
Further works should be continued to test the impact of clouds and extinction on molecular mapping.

\subsection{Possible improvements in the data reduction and molecular mapping}

As a source of problems for the characterization, stellar subtraction should be tackled with more efficient methods, especially for planets located at separations below 1$''$ which are more affected by the starlight.
Subtracting a scaled template from each spectrum in the data cube as did \cite{hoeijmakers_medium-resolution_2018} gives a slightly higher $S/N$ for the detection of the planets closer than 1$''$ (such as $\beta$ Pictoris b) but shows no improvement for more distant targets.
As we noticed from the ideal simulations with no star in the scene and providing that the photon noise and the speckle noise are below the background noise, a significant improvement of the S/N is possible for the most impacted molecules (NH$_3$, CH$_4$, HCN, CO$_2$), which argues in favor of developing more efficient PSF subtraction algorithms.
Consequently, more targets would be accessible for characterization with the MRS and molecular mapping.
Other interesting targets might come from future ground-based surveys, such as The Young Suns Exoplanet Survey \citep{bohn_young_2020}, and SHINE with VLT/SPHERE \citep{vigan_sphere_2021}.

A second avenue for improving the performance of molecular mapping with MIRI is to take advantage of the full spectral range. 
Hitherto, because of current limitations in the pipeline, we have restrained the method to the MRS channels, but exploiting the full spectral range of MIRI should increase the detection of molecules that exhibit features at multiple wavelengths in different channels (such as NH$_3$). Two solutions are envisioned, either a full concatenation of the data cubes from each channel or the extraction of the planet spectrum channel per channel, prior to cross-correlation with the model. The first method is limited by interpolation artifacts while the second one requires the planet to be detected in each of the three channels.

\subsection{Interpretation of the detection of the molecules in the targets sample}

All planets in our sample, except 51\,Eri b for which it is too challenging to disentangle the planet from the star's signal, are detected using the correlation with a full atmospheric model.
Depending on the distance of the system, long integrations (more groups) are possible, effectively improving the ramp fitting. However, the planets will be fainter if distant, thus requiring longer exposure time (more than 4 hours per observation for HD\,95086). In contrast, bright stars like $\beta$ Pic and 51\,Eri impose fewer groups per integration, which has a definitive impact on the planet detection, while at the same time the angular separation is smaller than 0.5$''$, so that the star cannot be moved out of the FoV.

The planet GJ\,504 b, although being the target with the largest contrast with its star, is the target with the largest $S/N$.
On the contrary, 2M\,1207 b is the one with the lowest contrast (in both temperature hypotheses) but its detection is much weaker. $S/N$ is 8 times larger for GJ\,504 b than for 2M\,1207 b if the planet is at 1600\,K.

We note that H$_2$O, the most abundant molecule in the atmosphere of giant planets and also the one with the most numerous spectral features, is always detected, even in planets that are close to their star and thus heavily contaminated by the starlight ($\beta$\,Pic\,b, HD\,95086\,b, HIP\,65426\,b). It is the only molecule unambiguously detected in these systems.
This result is consistent with the parametric study. 
It is even detected with a slightly higher $S/N$ than the correlation with the full atmospheric model, which can be explained by the fact that H$_2$O does not correlate with stellar residuals contrary to the full atmospheric model.
CO is the second most abundant molecule in most planets of the sample. However, it is detected only for planets with separations larger than $0.8''$. Indeed, as confirmed by the parametric study, it is the molecule that suffers the most from stellar contamination, which also explains non-detection when the planet has a low flux (such as HIP\,65426, although it is at 0.8$''$ and 2M\,1207 b at 1600\,K small radius hypothesis). 

On the contrary, PH$_3$ and HCN are among the less abundant molecules, but their spectral features are such that they can be detected in cold planets, providing they are bright (nearby and well-separated planet like the case of GJ\,504 b), in agreement with the parametric study.

The access to these molecules is a unique feature of JWST observations as it allows us to derive new constraints on planetary atmospheres as well as on planetary formation.
In complement to NIRSpec data that can access the bright features of PH$_3$ around 4.3 $\muup$m, we show that MIRI/MRS can also detect or confirm the presence of this molecule.
\cite{mukherjee_probing_2022} showed the importance of PH$_3$ in the emission spectrum of an atmosphere in thermochemical disequilibrium (based on MIRI-LRS simulations).
Furthermore, measuring the phosphine abundance, which is the dominant phosphorus molecule, can provide an estimation of the P/O ratio. These abundances ratios are of interest to determine if one element is depleted compared to oxygen which provides valuable information on the planetary formation processes. 
In addition, \cite{zahnle_methane_2014} details the importance and implication of NH$_3$, CH$_4$, and HCN in the atmospheres of young giant planets and brown dwarfs.
These molecules are keys to accurately deriving the C/O and N/O ratios, all being accessible to the MRS. They could also be indicators of chemical disequilibrium and provide constraints on the deep atmospheric temperatures and strength of vertical mixing that characterize the so-called "quench level".

NH$_3$ is also a powerful indicator of both planetary effective temperature and atmospheric chemical equilibrium state.
Differences between two model spectra calculated at equilibrium and disequilibrium chemistry arise between 5–9 $\muup$m. These differences mainly arise from different CH$_4$, H$_2$O, and NH$_3$ abundances due to differences in the quench level \citep{mukherjee_probing_2022}.

\subsection{Influence of the atmospheric models}
The molecules that are detected with molecular mapping are obviously strongly related to the atmospheric model that is used to generate the templates.
This explains some differences in the case of HR\,8799 and GJ\,504 with \cite{patapis_direct_2021}, who used \texttt{petitRADTRANS} model.
In their analysis, CO is not detected in GJ\,504 b.
Indeed, their molecular abundances computed with \texttt{petitRADTRANS} have significant differences with values computed with Exo-REM (for example lower CO abundance for a planet at 550\,K)
Similarly, in the case of the HR\,8799 system, they obtained a higher detection of CO for planet c than for planet b, the former having a higher temperature and hence more CO according to \texttt{petitRADTRANS} as opposed to Exo-REM.
In addition, CH$_4$ and NH$_3$ are less abundant in Exo-REM-based models with respect to models from \texttt{petitRADTRANS}, which explains why the detection of NH$_3$ and no CH$_4$ is lower in our analysis of the HR\,8799 planets. 
These differences in the molecular compositions mostly come from a different treatment of the chemistry. In Exo-REM, chemical abundances are computed assuming chemical disequilibrium with quenching levels (generally between 1 and 10 bars) derived from a parametrization of the vertical mixing. In contrast, in \cite{patapis_direct_2022}, the quenching level is fixed at 10$^{-2}$ bar.

Also, we can expect that a family of models that best fits the ground-based near-IR data might have large differences in the mid-IR regime. Depending on model assumptions, there is a large range of atmospheric parameter values that can reproduce the data. Therefore, the simulations presented in this paper should not be considered a perfect reproduction of each target, as long as we currently have little or no constraint in the mid-IR range for the known directly imaged planets.

\subsection{Complementarity with MIRI coronagraphy}
Although molecular mapping is powerful to identify the presence of molecules it comes with difficulties to put meaningful constraints on atmospheric parameters, as long as the planet continuum is significantly affected by the first step in which the stellar contribution is subtracted.
In that respect, complementary observations would be useful, in particular, those with MIRI's coronagraphs, which are observing in the same spectral range as the MRS but with specific wavelengths and broader bandwidths. The 4QPM coronagraphs \citep{boccaletti_mid-infrared_2015} can provide photometric measurements of the planet's continuum flux (11.4 $\muup$m and 15.5 $\muup$m), as well as the NH$_3$ feature at 10.65 $\muup$m, which are certainly useful to provide priors for the MRS data analysis. The combination of observations with the MRS and the coronagraphs will be decisive in further constraining the atmospheres of known directly imaged planets as illustrated in \citet{danielski_atmospheric_2018}.

\subsection{Complementarity with future ground-based projects}

MIRI/MRS has the advantage of not being impacted by telluric lines in comparison to the data taken with ground-based instruments.
Still, future ELT instruments are highly complementary to the MRS in providing higher spectral resolution and different wavelength coverage.

In the near IR, HARMONI allows a range of spectral resolution settings (3000, 7000, 18000). \cite{houlle_direct_2021} simulated molecular mapping observations combined with a matched-filter approach showing that planets with contrasts up to 16 mag and separations down to 75 mas can be detected at $>5\sigma$.  

In the mid-IR, METIS provides high-resolution spectro-imaging (R $\sim$ 100 000) at L/M band, including a mode with extended instantaneous wavelength coverage (assisted by a coronagraph).
\cite{snellen_combining_2015} showed that an Earth-like planet orbiting $\alpha$Cen could be detected at a $S/N$ of 5 with an instrument like METIS. Note that the METIS M band overlaps with the shortest wavelength of the MRS. 

\section{Conclusions}

In this paper, we showed that future JWST observations with MIRI-MRS are capable to provide clear detection of molecules in the atmospheres of young giant planets which is key in measuring abundance ratios and ultimately providing constraints on their formation and evolution. In the following, we sum up the main specific conclusions of our study.

\begin{itemize}

    \item 
    As already showcase in \cite{patapis_direct_2022}, we confirmed that the MIRI MRS mode has the potential to enhance planet detection owing to the molecular mapping technique which performs cross-correlation with atmospheric models.\\

    \item To determine the significance of the detection, we propose a data-driven experimental method, which takes into account the size and the shape of the correlation pattern, to avoid the impact of the autocorrelation, which arises for molecules that have harmonics in their spectra.
    While satisfactory for a majority of cases, the $S/N$ may be over-estimated in particular situations: with a CO template, or either with a hot planet ($T = 2000$\,K) around a cold star. \\

    \item The correlation pattern differs from a PSF and scales with the wavelength at a high signal-to-noise ratio but also depends on the strength of the correlation, the atmospheric template, and the noise level.
    Importantly, the maximum of the correlation pattern does not necessarily correspond to the real position of the planet (stronger effect at small angular separations). As a result, astrometry is unreliable with correlation maps.\\
     
    \item From the parametric study, which samples a range of planetary temperatures and angular separations, we concluded that, while planets are detected as close as 0.5$''$, a good level of characterization requires more angularly separated objects, typically larger than 1$''$. 
    The stellar spectral type has little impact on the performance, as opposed to the planetary temperature. 
    The highest $S/N$ values are achieved for planet temperatures ranging between 750\,K and 1750\,K, while lower performances are obtained for the coldest ($T_{p}$ = 500\,K), and the warmest ($T_{p}$ = 2000\,K) planets, respectively because of a lower absolute flux, and a lower level of correlation due to fewer spectral features. \\

    \item 
    For planets typically colder than 1500\,K, the following molecules are detectable: H$_2$O, CO, NH$_3$, CH$_4$, HCN, PH$_3$, CO$_2$. 
    Some of these molecules have never been confirmed or even detected in the atmosphere of an exoplanet.\\

    \item 
    We propose three directions for improving the performance of the molecular mapping method with MIRI. 
    Firstly, exploring further the subtraction of the stellar contribution can allow the detection and characterization of planets closer than 1$"$ as well as the detection of the molecules that have less spectral features or that are more hidden by other molecules. 
    Secondly, a robust estimation of the $S/N$ would require developing a more sophisticated analytical approximation to model the size and shape of the correlation pattern. 
    And thirdly, we plan to perform Bayesian analysis, to better constrain the atmospheric parameters and to better evaluate the uncertainties of each parameter.\\

    \item 
    Complementary data, such as coronagraphic mid-IR photometry with MIRI, 
    for measuring a planet's continuum will provide temperature and surface gravity estimates, which in turn can be taken as constraints on the atmospheric parameters for characterization purposes using molecular mapping.\\
    
    \item The interpretation of the data processing with molecular mapping strongly depends on the assumptions of the models to generate synthetic spectra. A pilot program dedicated to a specific target will represent a benchmark for systematically comparing several models.
    
\end{itemize}

\begin{acknowledgements}
M.M. acknowledges the Centre National d’Études Spatiales (CNES, Toulouse, France) for supporting the PhD fellowship.

B.C. and B.B. acknowledge financial support from the Programme National de Planétologie (PNP) of CNRS/INSU.

This work was granted access to the HPC resources of MesoPSL financed by the Région Ile de France and the project Equip@Meso (reference ANR-10-EQPX-29-01) of the programme Investissements d’Avenir supervised by the Agence Nationale pour la Recherche.

This project has received funding from the European Research Council (ERC) under the European Union’s Horizon 2020 research and innovation programme (COBREX; grant agreement n◦ 885593; EPIC, grant agreement n◦ 819155).

This publication makes use of data products from the Wide-field Infrared Survey Explorer, which is a joint project of the University of California, Los Angeles, and the Jet Propulsion Laboratory/California Institute of Technology, funded by the National Aeronautics and Space Administration.

The results presented in this research paper were obtained using the python packages numpy (Harris et al. 2020), scipy (Virtanen et al. 2020), matplotlib (Hunter 2007) and astropy (Astropy Collaboration et al. 2013, 2018).

\end{acknowledgements}

%
%

\bibliographystyle{aa}
\bibliography{references}

\begin{thebibliography}{72}
\expandafter\ifx\csname natexlab\endcsname\relax\def\natexlab#1{#1}\fi

\bibitem[{Barman {et~al.}(2015)Barman, Konopacky, Macintosh, \&
  Marois}]{barman_simultaneous_2015}
Barman, T.~S., Konopacky, Q.~M., Macintosh, B., \& Marois, C. 2015, The
  Astrophysical Journal, 804, 61

\bibitem[{Barman {et~al.}(2011)Barman, Macintosh, Konopacky, \&
  Marois}]{barman_young_2011}
Barman, T.~S., Macintosh, B., Konopacky, Q.~M., \& Marois, C. 2011, The
  Astrophysical Journal, 735, L39

\bibitem[{Baudino {et~al.}(2015)Baudino, Bézard, Boccaletti, Bonnefoy,
  Lagrange, \& Galicher}]{baudino_interpreting_2015}
Baudino, J.-L., Bézard, B., Boccaletti, A., {et~al.} 2015, Astronomy \&
  Astrophysics, 582, A83

\bibitem[{Birkby {et~al.}(2013)Birkby, de~Kok, Brogi, de~Mooij, Schwarz,
  Albrecht, \& Snellen}]{birkby_detection_2013}
Birkby, J.~L., de~Kok, R.~J., Brogi, M., {et~al.} 2013, Monthly Notices of the
  Royal Astronomical Society: Letters, 436, L35

\bibitem[{Blain {et~al.}(2021)Blain, Charnay, \& Bézard}]{blain_1d_2021}
Blain, D., Charnay, B., \& Bézard, B. 2021, Astronomy \& Astrophysics, 646,
  A15

\bibitem[{Boccaletti {et~al.}(2015)Boccaletti, Lagage, Baudoz, Beichman,
  Bouchet, Cavarroc, Dubreuil, Glasse, Glauser, Hines, Lajoie, Lebreton,
  Perrin, Pueyo, Reess, Rieke, Ronayette, Rouan, Soummer, \&
  Wright}]{boccaletti_mid-infrared_2015}
Boccaletti, A., Lagage, P.-O., Baudoz, P., {et~al.} 2015, Publications of the
  Astronomical Society of the Pacific, 127, 633

\bibitem[{Bohn {et~al.}(2020)Bohn, Kenworthy, Ginski, Manara, Pecaut, de Boer,
  Keller, Mamajek, Meshkat, Reggiani, Todorov, \& Snik}]{bohn_young_2020}
Bohn, A.~J., Kenworthy, M.~A., Ginski, C., {et~al.} 2020, Monthly Notices of
  the Royal Astronomical Society, 492, 431

\bibitem[{Bonnefoy {et~al.}(2013)Bonnefoy, Boccaletti, Lagrange, Allard,
  Mordasini, Beust, Chauvin, Girard, Homeier, Apai, Lacour, \&
  Rouan}]{bonnefoy_near-infrared_2013}
Bonnefoy, M., Boccaletti, A., Lagrange, A.-M., {et~al.} 2013, Astronomy \&
  Astrophysics, 555, A107

\bibitem[{Bonnefoy {et~al.}(2018)Bonnefoy, Perraut, Lagrange, Delorme, Vigan,
  Line, Rodet, Ginski, Mourard, Marleau, Samland, Tremblin, Ligi, Cantalloube,
  Mollière, Charnay, Kuzuhara, Janson, Morley, Homeier, D’Orazi, Klahr,
  Mordasini, Lavie, Baudino, Beust, Peretti, Musso~Bartucci, Mesa, Bézard,
  Boccaletti, Galicher, Hagelberg, Desidera, Biller, Maire, Allard, Borgniet,
  Lannier, Meunier, Desort, Alecian, Chauvin, Langlois, Henning, Mugnier,
  Mouillet, Gratton, Brandt, Mc~Elwain, Beuzit, Tamura, Hori, Brandner,
  Buenzli, Cheetham, Cudel, Feldt, Kasper, Keppler, Kopytova, Meyer, Perrot,
  Rouan, Salter, Schmidt, Sissa, Zurlo, Wildi, Blanchard, De~Caprio,
  Delboulbé, Maurel, Moulin, Pavlov, Rabou, Ramos, Roelfsema, Rousset,
  Stadler, Rigal, \& Weber}]{bonnefoy_gj_2018}
Bonnefoy, M., Perraut, K., Lagrange, A.-M., {et~al.} 2018, Astronomy \&
  Astrophysics, 618, A63

\bibitem[{Booth {et~al.}(2016)Booth, Jordán, Casassus, Hales, Dent, Faramaz,
  Matrà, Barkats, Brahm, \& Cuadra}]{booth_resolving_2016}
Booth, M., Jordán, A., Casassus, S., {et~al.} 2016, Monthly Notices of the
  Royal Astronomical Society: Letters, 460, L10

\bibitem[{Brogi {et~al.}(2012)Brogi, Snellen, de~Kok, Albrecht, Birkby, \&
  de~Mooij}]{brogi_signature_2012}
Brogi, M., Snellen, I. A.~G., de~Kok, R.~J., {et~al.} 2012, Nature, 486, 502

\bibitem[{Carter {et~al.}(2022)Carter, Hinkley, Kammerer, Skemer, Biller,
  Leisenring, Millar-Blanchaer, Petrus, Stone, Ward-Duong, Wang, Girard, Hines,
  Perrin, Pueyo, Balmer, Bonavita, Bonnefoy, Chauvin, Choquet, Christiaens,
  Danielski, Kennedy, Matthews, Miles, Patapis, Ray, Rickman, Sallum,
  Stapelfeldt, Whiteford, Zhou, Absil, Boccaletti, Booth, Bowler, Chen, Currie,
  Fortney, Grady, Greenbaum, Henning, Hoch, Janson, Kalas, Kenworthy, Kervella,
  Kraus, Lagage, Liu, Macintosh, Marino, Marley, Marois, Matthews, Mawet,
  McElwain, Metchev, Meyer, Molliere, Moran, Morley, Mukherjee, Pantin,
  Quirrenbach, Rebollido, Ren, Schneider, Vasist, Worthen, Wyatt,
  Briesemeister, Bryan, Calissendorff, Cantalloube, Cugno, De~Furio, Dupuy,
  Factor, Faherty, Fitzgerald, Franson, Gonzales, Hood, Howe, Kuzuhara,
  Lagrange, Lawson, Lazzoni, Lew, Liu, Lloyd, Martinez, Mazoyer, Quanz, Redai,
  Samland, Schlieder, Tamura, Tan, Uyama, Vigan, Vos, Wagner, Wolff, Ygouf,
  Zhang, Zhang, \& Zhang}]{carter_jwst_2022}
Carter, A.~L., Hinkley, S., Kammerer, J., {et~al.} 2022, The {JWST} {Early}
  {Release} {Science} {Program} for {Direct} {Observations} of {Exoplanetary}
  {Systems} {I}: {High} {Contrast} {Imaging} of the {Exoplanet} {HIP} 65426 b
  from 2-16 \${\textbackslash}mu\$m, arXiv:2208.14990 [astro-ph]

\bibitem[{Charbonneau {et~al.}(2002)Charbonneau, Brown, Noyes, \&
  Gilliland}]{charbonneau_detection_2002}
Charbonneau, D., Brown, T.~M., Noyes, R.~W., \& Gilliland, R.~L. 2002, The
  Astrophysical Journal, 568, 377

\bibitem[{Charnay {et~al.}(2018)Charnay, Bézard, Baudino, Bonnefoy,
  Boccaletti, \& Galicher}]{charnay_self-consistent_2018}
Charnay, B., Bézard, B., Baudino, J.-L., {et~al.} 2018, The Astrophysical
  Journal, 854, 172

\bibitem[{Chauvin {et~al.}(2017)Chauvin, Desidera, Lagrange, Vigan, Gratton,
  Langlois, Bonnefoy, Beuzit, Feldt, Mouillet, Meyer, Cheetham, Biller,
  Boccaletti, D’Orazi, Galicher, Hagelberg, Maire, Mesa, Olofsson, Samland,
  Schmidt, Sissa, Bonavita, Charnay, Cudel, Daemgen, Delorme, Janin-Potiron,
  Janson, Keppler, Le~Coroller, Ligi, Marleau, Messina, Mollière, Mordasini,
  Müller, Peretti, Perrot, Rodet, Rouan, Zurlo, Dominik, Henning, Menard,
  Schmid, Turatto, Udry, Vakili, Abe, Antichi, Baruffolo, Baudoz, Baudrand,
  Blanchard, Bazzon, Buey, Carbillet, Carle, Charton, Cascone, Claudi,
  Costille, Deboulbe, De~Caprio, Dohlen, Fantinel, Feautrier, Fusco, Gigan,
  Giro, Gisler, Gluck, Hubin, Hugot, Jaquet, Kasper, Madec, Magnard, Martinez,
  Maurel, Le~Mignant, Möller-Nilsson, Llored, Moulin, Origné, Pavlov, Perret,
  Petit, Pragt, Puget, Rabou, Ramos, Rigal, Rochat, Roelfsema, Rousset, Roux,
  Salasnich, Sauvage, Sevin, Soenke, Stadler, Suarez, Weber, Wildi, Antoniucci,
  Augereau, Baudino, Brandner, Engler, Girard, Gry, Kral, Kopytova, Lagadec,
  Milli, Moutou, Schlieder, Szulágyi, Thalmann, \&
  Wahhaj}]{chauvin_discovery_2017}
Chauvin, G., Desidera, S., Lagrange, A.-M., {et~al.} 2017, Astronomy \&
  Astrophysics, 605, L9

\bibitem[{Chauvin {et~al.}(2018)Chauvin, Gratton, Bonnefoy, Lagrange, de~Boer,
  Vigan, Beust, Lazzoni, Boccaletti, Galicher, Desidera, Delorme, Keppler,
  Lannier, Maire, Mesa, Meunier, Kral, Henning, Menard, Moor, Avenhaus, Bazzon,
  Janson, Beuzit, Bhowmik, Bonavita, Borgniet, Brandner, Cheetham, Cudel,
  Feldt, Fontanive, Ginski, Hagelberg, Janin-Potiron, Lagadec, Langlois,
  Le~Coroller, Messina, Meyer, Mouillet, Peretti, Perrot, Rodet, Samland,
  Sissa, Olofsson, Salter, Schmidt, Zurlo, Milli, van Boekel, Quanz, Feautrier,
  Le~Mignant, Perret, Ramos, \& Rochat}]{chauvin_investigating_2018}
Chauvin, G., Gratton, R., Bonnefoy, M., {et~al.} 2018, Astronomy \&
  Astrophysics, 617, A76

\bibitem[{Chauvin {et~al.}(2004)Chauvin, Lagrange, Dumas, Zuckerman, Mouillet,
  Song, Beuzit, \& Lowrance}]{chauvin_giant_2004}
Chauvin, G., Lagrange, A.-M., Dumas, C., {et~al.} 2004, Astronomy \&
  Astrophysics, 425, L29

\bibitem[{Cugno {et~al.}(2021)Cugno, Patapis, Stolker, Quanz, Boehle,
  Hoeijmakers, Marleau, Mollière, Nasedkin, \& Snellen}]{cugno_molecular_2021}
Cugno, G., Patapis, P., Stolker, T., {et~al.} 2021, Astronomy \& Astrophysics,
  653, A12

\bibitem[{Daemgen {et~al.}(2017)Daemgen, Todorov, Quanz, Meyer, Mordasini,
  Marleau, \& Fortney}]{daemgen_high_2017}
Daemgen, S., Todorov, K., Quanz, S.~P., {et~al.} 2017, Astronomy \&
  Astrophysics, 608, A71

\bibitem[{Danielski {et~al.}(2018)Danielski, Baudino, Lagage, Boccaletti,
  Gastaud, Coulais, \& Bézard}]{danielski_atmospheric_2018}
Danielski, C., Baudino, J.-L., Lagage, P.-O., {et~al.} 2018, The Astronomical
  Journal, 156, 276

\bibitem[{Delorme {et~al.}(2017)Delorme, Schmidt, Bonnefoy, Desidera, Ginski,
  Charnay, Lazzoni, Christiaens, Messina, D’Orazi, Milli, Schlieder, Gratton,
  Rodet, Lagrange, Absil, Vigan, Galicher, Hagelberg, Bonavita, Lavie, Zurlo,
  Olofsson, Boccaletti, Cantalloube, Mouillet, Chauvin, Hambsch, Langlois,
  Udry, Henning, Beuzit, Mordasini, Lucas, Marocco, Biller, Carson, Cheetham,
  Covino, De~Caprio, Delboulbe, Feldt, Girard, Hubin, Maire, Pavlov, Petit,
  Rouan, Roelfsema, \& Wildi}]{delorme_-depth_2017}
Delorme, P., Schmidt, T., Bonnefoy, M., {et~al.} 2017, Astronomy \&
  Astrophysics, 608, A79

\bibitem[{Deming {et~al.}(2013)Deming, Wilkins, McCullough, Burrows, Fortney,
  Agol, Dobbs-Dixon, Madhusudhan, Crouzet, Desert, Gilliland, Haynes, Knutson,
  Line, Magic, Mandell, Ranjan, Charbonneau, Clampin, Seager, \&
  Showman}]{deming_infrared_2013}
Deming, D., Wilkins, A., McCullough, P., {et~al.} 2013, The Astrophysical
  Journal, 774, 95

\bibitem[{Desgrange {et~al.}(2022)Desgrange, Chauvin, Christiaens, Cantalloube,
  Lefranc, Le~Coroller, Rubini, Otten, Beust, Bonavita, Delorme, Devinat,
  Gratton, Lagrange, Langlois, Mesa, Milli, Szulágyi, Nowak, Rodet, Rojo,
  Petrus, Janson, Henning, Kral, van Holstein, Ménard, Beuzit, Biller,
  Boccaletti, Bonnefoy, Brown, Costille, Delboulbe, Desidera, D’Orazi, Feldt,
  Fusco, Galicher, Hagelberg, Lazzoni, Ligi, Maire, Messina, Meyer, Potier,
  Ramos, Rouan, Schmidt, Vigan, \& Zurlo}]{desgrange_-depth_2022}
Desgrange, C., Chauvin, G., Christiaens, V., {et~al.} 2022, Astronomy \&
  Astrophysics, 664, A139

\bibitem[{{Gaia Collaboration} {et~al.}(2016){Gaia Collaboration}, Brown,
  Vallenari, Prusti, de~Bruijne, Mignard, Drimmel, Babusiaux, Bailer-Jones,
  Bastian, Biermann, Evans, Eyer, Jansen, Jordi, Katz, Klioner, Lammers,
  Lindegren, Luri, O’Mullane, Panem, Pourbaix, Randich, Sartoretti, Siddiqui,
  Soubiran, Valette, van Leeuwen, Walton, Aerts, Arenou, Cropper, Høg,
  Lattanzi, Grebel, Holland, Huc, Passot, Perryman, Bramante, Cacciari,
  Castañeda, Chaoul, Cheek, De~Angeli, Fabricius, Guerra, Hernández,
  Jean-Antoine-Piccolo, Masana, Messineo, Mowlavi, Nienartowicz,
  Ordóñez-Blanco, Panuzzo, Portell, Richards, Riello, Seabroke, Tanga,
  Thévenin, Torra, Els, Gracia-Abril, Comoretto, Garcia-Reinaldos, Lock,
  Mercier, Altmann, Andrae, Astraatmadja, Bellas-Velidis, Benson, Berthier,
  Blomme, Busso, Carry, Cellino, Clementini, Cowell, Creevey, Cuypers,
  Davidson, De~Ridder, de~Torres, Delchambre, Dell’Oro, Ducourant, Frémat,
  García-Torres, Gosset, Halbwachs, Hambly, Harrison, Hauser, Hestroffer,
  Hodgkin, Huckle, Hutton, Jasniewicz, Jordan, Kontizas, Korn, Lanzafame,
  Manteiga, Moitinho, Muinonen, Osinde, Pancino, Pauwels, Petit, Recio-Blanco,
  Robin, Sarro, Siopis, Smith, Smith, Sozzetti, Thuillot, van Reeven, Viala,
  Abbas, Abreu~Aramburu, Accart, Aguado, Allan, Allasia, Altavilla, Álvarez,
  Alves, Anderson, Andrei, Anglada~Varela, Antiche, Antoja, Antón, Arcay,
  Bach, Baker, Balaguer-Núñez, Barache, Barata, Barbier, Barblan, Barrado~y
  Navascués, Barros, Barstow, Becciani, Bellazzini, Bello~García, Belokurov,
  Bendjoya, Berihuete, Bianchi, Bienaymé, Billebaud, Blagorodnova,
  Blanco-Cuaresma, Boch, Bombrun, Borrachero, Bouquillon, Bourda, Bouy,
  Bragaglia, Breddels, Brouillet, Brüsemeister, Bucciarelli, Burgess, Burgon,
  Burlacu, Busonero, Buzzi, Caffau, Cambras, Campbell, Cancelliere,
  Cantat-Gaudin, Carlucci, Carrasco, Castellani, Charlot, Charnas, Chiavassa,
  Clotet, Cocozza, Collins, Costigan, Crifo, Cross, Crosta, Crowley, Dafonte,
  Damerdji, Dapergolas, David, David, De~Cat, de~Felice, de~Laverny, De~Luise,
  De~March, de~Martino, de~Souza, Debosscher, del Pozo, Delbo, Delgado,
  Delgado, Di~Matteo, Diakite, Distefano, Dolding, Dos~Anjos, Drazinos, Duran,
  Dzigan, Edvardsson, Enke, Evans, Eynard~Bontemps, Fabre, Fabrizio, Faigler,
  Falcão, Farràs~Casas, Federici, Fedorets, Fernández-Hernández, Fernique,
  Fienga, Figueras, Filippi, Findeisen, Fonti, Fouesneau, Fraile, Fraser,
  Fuchs, Gai, Galleti, Galluccio, Garabato, García-Sedano, Garofalo, Garralda,
  Gavras, Gerssen, Geyer, Gilmore, Girona, Giuffrida, Gomes, González-Marcos,
  González-Núñez, González-Vidal, Granvik, Guerrier, Guillout, Guiraud,
  Gúrpide, Gutiérrez-Sánchez, Guy, Haigron, Hatzidimitriou, Haywood, Heiter,
  Helmi, Hobbs, Hofmann, Holl, Holland, Hunt, Hypki, Icardi, Irwin, Jevardat~de
  Fombelle, Jofré, Jonker, Jorissen, Julbe, Karampelas, Kochoska, Kohley,
  Kolenberg, Kontizas, Koposov, Kordopatis, Koubsky, Krone-Martins,
  Kudryashova, Kull, Bachchan, Lacoste-Seris, Lanza, Lavigne,
  Le~Poncin-Lafitte, Lebreton, Lebzelter, Leccia, Leclerc, Lecoeur-Taibi,
  Lemaitre, Lenhardt, Leroux, Liao, Licata, Lindstrøm, Lister, Livanou, Lobel,
  Löffler, López, Lorenz, MacDonald, Magalhães~Fernandes, Managau, Mann,
  Mantelet, Marchal, Marchant, Marconi, Marinoni, Marrese, Marschalkó,
  Marshall, Martín-Fleitas, Martino, Mary, Matijevič, Mazeh, McMillan,
  Messina, Michalik, Millar, Miranda, Molina, Molinaro, Molinaro, Molnár,
  Moniez, Montegriffo, Mor, Mora, Morbidelli, Morel, Morgenthaler, Morris,
  Mulone, Muraveva, Musella, Narbonne, Nelemans, Nicastro, Noval, Ordénovic,
  Ordieres-Meré, Osborne, Pagani, Pagano, Pailler, Palacin, Palaversa,
  Parsons, Pecoraro, Pedrosa, Pentikäinen, Pichon, Piersimoni, Pineau, Plachy,
  Plum, Poujoulet, Prša, Pulone, Ragaini, Rago, Rambaux, Ramos-Lerate,
  Ranalli, Rauw, Read, Regibo, Reylé, Ribeiro, Rimoldini, Ripepi, Riva, Rixon,
  Roelens, Romero-Gómez, Rowell, Royer, Ruiz-Dern, Sadowski,
  Sagristà~Sellés, Sahlmann, Salgado, Salguero, Sarasso, Savietto,
  Schultheis, Sciacca, Segol, Segovia, Segransan, Shih, Smareglia, Smart,
  Solano, Solitro, Sordo, Soria~Nieto, Souchay, Spagna, Spoto, Stampa, Steele,
  Steidelmüller, Stephenson, Stoev, Suess, Süveges, Surdej, Szabados,
  Szegedi-Elek, Tapiador, Taris, Tauran, Taylor, Teixeira, Terrett, Tingley,
  Trager, Turon, Ulla, Utrilla, Valentini, van Elteren, Van~Hemelryck, van
  Leeuwen, Varadi, Vecchiato, Veljanoski, Via, Vicente, Vogt, Voss, Votruba,
  Voutsinas, Walmsley, Weiler, Weingrill, Wevers, Wyrzykowski, Yoldas, Žerjal,
  Zucker, Zurbach, Zwitter, Alecu, Allen, Allende~Prieto, Amorim,
  Anglada-Escudé, Arsenijevic, Azaz, Balm, Beck, Bernstein, Bigot, Bijaoui,
  Blasco, Bonfigli, Bono, Boudreault, Bressan, Brown, Brunet, Bunclark,
  Buonanno, Butkevich, Carret, Carrion, Chemin, Chéreau, Corcione, Darmigny,
  de~Boer, de~Teodoro, de~Zeeuw, Delle~Luche, Domingues, Dubath, Fodor,
  Frézouls, Fries, Fustes, Fyfe, Gallardo, Gallegos, Gardiol, Gebran, Gomboc,
  Gómez, Grux, Gueguen, Heyrovsky, Hoar, Iannicola, Isasi~Parache, Janotto,
  Joliet, Jonckheere, Keil, Kim, Klagyivik, Klar, Knude, Kochukhov, Kolka, Kos,
  Kutka, Lainey, LeBouquin, Liu, Loreggia, Makarov, Marseille, Martayan,
  Martinez-Rubi, Massart, Meynadier, Mignot, Munari, Nguyen, Nordlander,
  Ocvirk, O’Flaherty, Olias~Sanz, Ortiz, Osorio, Oszkiewicz, Ouzounis,
  Palmer, Park, Pasquato, Peltzer, Peralta, Péturaud, Pieniluoma, Pigozzi,
  Poels, Prat, Prod’homme, Raison, Rebordao, Risquez, Rocca-Volmerange,
  Rosen, Ruiz-Fuertes, Russo, Sembay, Serraller~Vizcaino, Short, Siebert,
  Silva, Sinachopoulos, Slezak, Soffel, Sosnowska, Straižys, ter Linden,
  Terrell, Theil, Tiede, Troisi, Tsalmantza, Tur, Vaccari, Vachier, Valles,
  Van~Hamme, Veltz, Virtanen, Wallut, Wichmann, Wilkinson, Ziaeepour, \&
  Zschocke}]{gaia_collaboration_gaia_2016}
{Gaia Collaboration}, Brown, A. G.~A., Vallenari, A., {et~al.} 2016, Astronomy
  \& Astrophysics, 595, A2

\bibitem[{Glasse {et~al.}(2015)Glasse, Rieke, Bauwens, García-Marín, Ressler,
  Rost, Tikkanen, Vandenbussche, \& Wright}]{glasse_mid-infrared_2015}
Glasse, A., Rieke, G.~H., Bauwens, E., {et~al.} 2015, Publications of the
  Astronomical Society of the Pacific, 127, 686

\bibitem[{Hinkley {et~al.}(2022)Hinkley, Carter, Ray, Skemer, Biller, Choquet,
  Millar-Blanchaer, Sallum, Miles, Whiteford, Patapis, Perrin, Pueyo,
  Schneider, Stapelfeldt, Wang, Ward-Duong, Bowler, Boccaletti, H.~Girard,
  Hines, Kalas, Kammerer, Kervella, Leisenring, Pantin, Zhou, Meyer, Liu,
  Bonnefoy, Currie, McElwain, Metchev, Wyatt, Absil, Adams, Barman, Baraffe,
  Bonavita, Booth, Bryan, Chauvin, Chen, Danielski, Furio, Factor, Fitzgerald,
  Fortney, Grady, Greenbaum, Henning, Hoch, Janson, Kennedy, Kenworthy, Kraus,
  Kuzuhara, Lagage, Lagrange, Launhardt, Lazzoni, Lloyd, Marino, Marley,
  Martinez, Marois, Matthews, Matthews, Mawet, Mazoyer, Phillips, Petrus,
  Quanz, Quirrenbach, Rameau, Rebollido, Rickman, Samland, Sargent, Schlieder,
  Sivaramakrishnan, Stone, Tamura, Tremblin, Uyama, Vasist, Vigan, Wagner, \&
  Ygouf}]{hinkley_jwst_2022}
Hinkley, S., Carter, A.~L., Ray, S., {et~al.} 2022, Publications of the
  Astronomical Society of the Pacific, 134, 095003

\bibitem[{Hinz {et~al.}(2010)Hinz, Rodigas, Kenworthy, Sivanandam, Heinze,
  Mamajek, \& Meyer}]{hinz_thermal_2010}
Hinz, P.~M., Rodigas, T.~J., Kenworthy, M.~A., {et~al.} 2010, The Astrophysical
  Journal, 716, 417

\bibitem[{Hoeijmakers {et~al.}(2018)Hoeijmakers, Schwarz, Snellen, de~Kok,
  Bonnefoy, Chauvin, Lagrange, \& Girard}]{hoeijmakers_medium-resolution_2018}
Hoeijmakers, H.~J., Schwarz, H., Snellen, I. A.~G., {et~al.} 2018, Astronomy \&
  Astrophysics, 617, A144

\bibitem[{Houllé {et~al.}(2021)Houllé, Vigan, Carlotti, Choquet, Cantalloube,
  Phillips, Sauvage, Schwartz, Otten, Baraffe, Emsenhuber, \&
  Mordasini}]{houlle_direct_2021}
Houllé, M., Vigan, A., Carlotti, A., {et~al.} 2021, Astronomy \& Astrophysics,
  652, A67

\bibitem[{Janson {et~al.}(2010)Janson, Bergfors, Goto, Brandner, \&
  Lafrenière}]{janson_spatially_2010}
Janson, M., Bergfors, C., Goto, M., Brandner, W., \& Lafrenière, D. 2010, The
  Astrophysical Journal, 710, L35

\bibitem[{Janson {et~al.}(2013)Janson, Brandt, Kuzuhara, Spiegel, Thalmann,
  Currie, Bonnefoy, Zimmerman, Sorahana, Kotani, Schlieder, Hashimoto, Kudo,
  Kusakabe, Abe, Brandner, Carson, Egner, Feldt, Goto, Grady, Guyon, Hayano,
  Hayashi, Hayashi, Henning, Hodapp, Ishii, Iye, Kandori, Knapp, Kwon, Matsuo,
  McElwain, Mede, Miyama, Morino, Moro-Martín, Nakagawa, Nishimura, Pyo,
  Serabyn, Suenaga, Suto, Suzuki, Takahashi, Takami, Takato, Terada, Tomono,
  Turner, Watanabe, Wisniewski, Yamada, Takami, Usuda, \&
  Tamura}]{janson_direct_2013}
Janson, M., Brandt, T.~D., Kuzuhara, M., {et~al.} 2013, The Astrophysical
  Journal, 778, L4

\bibitem[{Klaassen {et~al.}(2020)Klaassen, Geers, Beard, O’Brien, Cossou,
  Gastaud, Coulais, Schreiber, Kavanagh, Topinka, Azzollini, De Meester,
  Bouwman, Glasse, Glauser, Law, Cracraft, Murray, Sargent, Jones, \&
  Wright}]{klaassen_span_2020}
Klaassen, P.~D., Geers, V.~C., Beard, S.~M., {et~al.} 2020, Monthly Notices of
  the Royal Astronomical Society, 500, 2813

\bibitem[{Konopacky {et~al.}(2013)Konopacky, Barman, Macintosh, \&
  Marois}]{konopacky_detection_2013}
Konopacky, Q.~M., Barman, T.~S., Macintosh, B., \& Marois, C. 2013, Science,
  339, 1398

\bibitem[{Kuzuhara {et~al.}(2013)Kuzuhara, Tamura, Kudo, Janson, Kandori,
  Brandt, Thalmann, Spiegel, Biller, Carson, Hori, Suzuki, Burrows, Henning,
  Turner, McElwain, Moro-Martín, Suenaga, Takahashi, Kwon, Lucas, Abe,
  Brandner, Egner, Feldt, Fujiwara, Goto, Grady, Guyon, Hashimoto, Hayano,
  Hayashi, Hayashi, Hodapp, Ishii, Iye, Knapp, Matsuo, Mayama, Miyama, Morino,
  Nishikawa, Nishimura, Kotani, Kusakabe, Pyo, Serabyn, Suto, Takami, Takato,
  Terada, Tomono, Watanabe, Wisniewski, Yamada, Takami, \&
  Usuda}]{kuzuhara_direct_2013}
Kuzuhara, M., Tamura, M., Kudo, T., {et~al.} 2013, The Astrophysical Journal,
  774, 11

\bibitem[{Labiano {et~al.}(2021)Labiano, Argyriou, Álvarez Márquez, Glasse,
  Glauser, Patapis, Law, Brandl, Justtanont, Lahuis, Martínez-Galarza,
  Mueller, Noriega-Crespo, Royer, Shaughnessy, \&
  Vandenbussche}]{labiano_wavelength_2021}
Labiano, A., Argyriou, I., Álvarez Márquez, J., {et~al.} 2021, Astronomy \&
  Astrophysics, 656, A57

\bibitem[{Labiano-Ortega {et~al.}(2016)Labiano-Ortega, Dicken, Vandenbussche,
  Lahuis, Muller, Beard, Justtanont, Azzollini, Law, Gordon, Glasse, Wright,
  Rieke, Klaassen, Glauser, Morrison, Geers, Bailey, \&
  Garcia-Marin}]{labiano-ortega_miri_2016}
Labiano-Ortega, A., Dicken, D., Vandenbussche, B., {et~al.} 2016, in
  Observatory {Operations}: {Strategies}, {Processes}, and {Systems} {VI}, ed.
  A.~B. Peck, C.~R. Benn, \& R.~L. Seaman (Edinburgh, United Kingdom: SPIE),
  117

\bibitem[{Lagrange {et~al.}(2019)Lagrange, Boccaletti, Langlois, Chauvin,
  Gratton, Beust, Desidera, Milli, Bonnefoy, Cheetham, Feldt, Meyer, Vigan,
  Biller, Bonavita, Baudino, Cantalloube, Cudel, Daemgen, Delorme, D’Orazi,
  Girard, Fontanive, Hagelberg, Janson, Keppler, Koypitova, Galicher, Lannier,
  Le~Coroller, Ligi, Maire, Mesa, Messina, Müeller, Peretti, Perrot, Rouan,
  Salter, Samland, Schmidt, Sissa, Zurlo, Beuzit, Mouillet, Dominik, Henning,
  Lagadec, Ménard, Schmid, Turatto, Udry, Bohn, Charnay, Gomez~Gonzales, Gry,
  Kenworthy, Kral, Mordasini, Moutou, van~der Plas, Schlieder, Abe, Antichi,
  Baruffolo, Baudoz, Baudrand, Blanchard, Bazzon, Buey, Carbillet, Carle,
  Charton, Cascone, Claudi, Costille, Deboulbe, De~Caprio, Dohlen, Fantinel,
  Feautrier, Fusco, Gigan, Giro, Gisler, Gluck, Hubin, Hugot, Jaquet, Kasper,
  Madec, Magnard, Martinez, Maurel, Le~Mignant, Möller-Nilsson, Llored,
  Moulin, Origné, Pavlov, Perret, Petit, Pragt, Szulagyi, \&
  Wildi}]{lagrange_post-conjunction_2019}
Lagrange, A.-M., Boccaletti, A., Langlois, M., {et~al.} 2019, Astronomy \&
  Astrophysics, 621, L8

\bibitem[{Lagrange {et~al.}(2010)Lagrange, Bonnefoy, Chauvin, Apai, Ehrenreich,
  Boccaletti, Gratadour, Rouan, Mouillet, Lacour, \&
  Kasper}]{lagrange_giant_2010}
Lagrange, A.-M., Bonnefoy, M., Chauvin, G., {et~al.} 2010, Science, 329, 57

\bibitem[{Lagrange {et~al.}(2009)Lagrange, Gratadour, Chauvin, Fusco,
  Ehrenreich, Mouillet, Rousset, Rouan, Allard, Gendron, Charton, Mugnier,
  Rabou, Montri, \& Lacombe}]{lagrange_probable_2009}
Lagrange, A.-M., Gratadour, D., Chauvin, G., {et~al.} 2009, Astronomy \&
  Astrophysics, 493, L21

\bibitem[{Macintosh {et~al.}(2015)Macintosh, Graham, Barman, De~Rosa,
  Konopacky, Marley, Marois, Nielsen, Pueyo, Rajan, Rameau, Saumon, Wang,
  Patience, Ammons, Arriaga, Artigau, Beckwith, Brewster, Bruzzone, Bulger,
  Burningham, Burrows, Chen, Chiang, Chilcote, Dawson, Dong, Doyon, Draper,
  Duchêne, Esposito, Fabrycky, Fitzgerald, Follette, Fortney, Gerard,
  Goodsell, Greenbaum, Hibon, Hinkley, Cotten, Hung, Ingraham, Johnson-Groh,
  Kalas, Lafreniere, Larkin, Lee, Line, Long, Maire, Marchis, Matthews, Max,
  Metchev, Millar-Blanchaer, Mittal, Morley, Morzinski, Murray-Clay,
  Oppenheimer, Palmer, Patel, Perrin, Poyneer, Rafikov, Rantakyrö, Rice, Rojo,
  Rudy, Ruffio, Ruiz, Sadakuni, Saddlemyer, Salama, Savransky, Schneider,
  Sivaramakrishnan, Song, Soummer, Thomas, Vasisht, Wallace, Ward-Duong,
  Wiktorowicz, Wolff, \& Zuckerman}]{macintosh_discovery_2015}
Macintosh, B., Graham, J.~R., Barman, T., {et~al.} 2015, Science, 350, 64

\bibitem[{Madhusudhan {et~al.}(2014)Madhusudhan, Amin, \&
  Kennedy}]{madhusudhan_toward_2014}
Madhusudhan, N., Amin, M.~A., \& Kennedy, G.~M. 2014, The Astrophysical
  Journal, 794, L12

\bibitem[{Marois {et~al.}(2006)Marois, Lafreniere, Doyon, Macintosh, \&
  Nadeau}]{marois_angular_2006}
Marois, C., Lafreniere, D., Doyon, R., Macintosh, B., \& Nadeau, D. 2006, The
  Astrophysical Journal, 641, 556

\bibitem[{Marois {et~al.}(2008)Marois, Macintosh, Barman, Zuckerman, Song,
  Patience, Lafreniere, \& Doyon}]{marois_direct_2008}
Marois, C., Macintosh, B., Barman, T., {et~al.} 2008, Science, 322, 1348,
  arXiv: 0811.2606

\bibitem[{Marois {et~al.}(2010)Marois, Zuckerman, Konopacky, Macintosh, \&
  Barman}]{marois_images_2010}
Marois, C., Zuckerman, B., Konopacky, Q.~M., Macintosh, B., \& Barman, T. 2010,
  Nature, 468, 1080

\bibitem[{Miles {et~al.}(2022)Miles, Biller, Patapis, Worthen, Rickman, Hoch,
  Skemer, Perrin, Chen, Mukherjee, Morley, Moran, Bonnefoy, Petrus, Carter,
  Choquet, Hinkley, Ward-Duong, Leisenring, Millar-Blanchaer, Pueyo, Ray,
  Stapelfeldt, Stone, Wang, Absil, Balmer, Boccaletti, Bonavita, Booth, Bowler,
  Chauvin, Christiaens, Currie, Danielski, Fortney, Girard, Greenbaum, Henning,
  Hines, Janson, Kalas, Kammerer, Kenworthy, Kervella, Lagage, Lew, Liu,
  Macintosh, Marino, Marley, Marois, Matthews, Matthews, Mawet, McElwain,
  Metchev, Meyer, Molliere, Pantin, Rebollido, Ren, Vasist, Wyatt, Zhou,
  Briesemeister, Bryan, Calissendorff, Catalloube, Cugno, De~Furio, Dupuy,
  Factor, Faherty, Fitzgerald, Franson, Gonzales, Hood, Howe, Kraus, Kuzuhara,
  Lawson, Lazzoni, Liu, Llop-Sayson, Lloyd, Martinez, Mazoyer, Quanz, Redai,
  Samland, Schlieder, Tamura, Tan, Uyama, Vigan, Vos, Wagner, Wolff, Ygouf,
  Zhang, \& Zhang}]{miles_jwst_2022}
Miles, B.~E., Biller, B.~A., Patapis, P., {et~al.} 2022, The {JWST} {Early}
  {Release} {Science} {Program} for {Direct} {Observations} of {Exoplanetary}
  {Systems} {II}: {A} 1 to 20 {Micron} {Spectrum} of the {Planetary}-{Mass}
  {Companion} {VHS} 1256-1257 b, arXiv:2209.00620 [astro-ph]

\bibitem[{Mordasini {et~al.}(2016)Mordasini, van Boekel, Mollière, Henning, \&
  Benneke}]{mordasini_imprint_2016}
Mordasini, C., van Boekel, R., Mollière, P., Henning, T., \& Benneke, B. 2016,
  The Astrophysical Journal, 832, 41

\bibitem[{Mukherjee {et~al.}(2022)Mukherjee, Fortney, Batalha, Karalidi,
  Marley, Visscher, Miles, \& Skemer}]{mukherjee_probing_2022}
Mukherjee, S., Fortney, J.~J., Batalha, N.~E., {et~al.} 2022, The Astrophysical
  Journal, 938, 107

\bibitem[{Nowak {et~al.}(2020{\natexlab{a}})Nowak, Lacour, Lagrange, Rubini,
  Wang, Stolker, Abuter, Amorim, Asensio-Torres, Bauböck, Benisty, Berger,
  Beust, Blunt, Boccaletti, Bonnefoy, Bonnet, Brandner, Cantalloube, Charnay,
  Choquet, Christiaens, Clénet, Coudé~du Foresto, Cridland, de~Zeeuw, Dembet,
  Dexter, Drescher, Duvert, Eckart, Eisenhauer, Gao, Garcia, Garcia~Lopez,
  Gardner, Gendron, Genzel, Gillessen, Girard, Grandjean, Haubois, Heißel,
  Henning, Hinkley, Hippler, Horrobin, Houllé, Hubert, Jiménez-Rosales,
  Jocou, Kammerer, Kervella, Keppler, Kreidberg, Kulikauskas, Lapeyrère,
  Le~Bouquin, Léna, Mérand, Maire, Mollière, Monnier, Mouillet, Müller,
  Nasedkin, Ott, Otten, Paumard, Paladini, Perraut, Perrin, Pueyo, Pfuhl,
  Rameau, Rodet, Rodríguez-Coira, Rousset, Scheithauer, Shangguan, Stadler,
  Straub, Straubmeier, Sturm, Tacconi, van Dishoeck, Vigan, Vincent, von
  Fellenberg, Ward-Duong, Widmann, Wieprecht, Wiezorrek, Woillez, \& {the
  GRAVITY Collaboration}}]{nowak_direct_2020}
Nowak, M., Lacour, S., Lagrange, A.-M., {et~al.} 2020{\natexlab{a}}, Astronomy
  \& Astrophysics, 642, L2

\bibitem[{Nowak {et~al.}(2020{\natexlab{b}})Nowak, Lacour, Mollière, Wang,
  Charnay, van Dishoeck, Abuter, Amorim, Berger, Beust, Bonnefoy, Bonnet,
  Brandner, Buron, Cantalloube, Collin, Chapron, Clénet, Coudé~du Foresto,
  de~Zeeuw, Dembet, Dexter, Duvert, Eckart, Eisenhauer, Förster~Schreiber,
  Fédou, Garcia~Lopez, Gao, Gendron, Genzel, Gillessen, Haußmann, Henning,
  Hippler, Hubert, Jocou, Kervella, Lagrange, Lapeyrère, Le~Bouquin, Léna,
  Maire, Ott, Paumard, Paladini, Perraut, Perrin, Pueyo, Pfuhl, Rabien, Rau,
  Rodríguez-Coira, Rousset, Scheithauer, Shangguan, Straub, Straubmeier,
  Sturm, Tacconi, Vincent, Widmann, Wieprecht, Wiezorrek, Woillez, Yazici, \&
  Ziegler}]{nowak_peering_2020}
Nowak, M., Lacour, S., Mollière, P., {et~al.} 2020{\natexlab{b}}, Astronomy \&
  Astrophysics, 633, A110

\bibitem[{Oberg \& Bergin(2021)}]{oberg_astrochemistry_2021}
Oberg, K.~I. \& Bergin, E.~A. 2021, Physics Reports, 893, 1, arXiv: 2010.03529

\bibitem[{Patapis {et~al.}(2022)Patapis, Nasedkin, Cugno, Glauser, Argyriou,
  Whiteford, Mollière, Glasse, \& Quanz}]{patapis_direct_2022}
Patapis, P., Nasedkin, E., Cugno, G., {et~al.} 2022, Astronomy \& Astrophysics,
  658, A72

\bibitem[{Patience {et~al.}(2010)Patience, King, De~Rosa, \&
  Marois}]{patience_highest_2010}
Patience, J., King, R.~R., De~Rosa, R.~J., \& Marois, C. 2010, Astronomy and
  Astrophysics, 517, A76

\bibitem[{Petit dit de~la Roche {et~al.}(2018)Petit dit de~la Roche,
  Hoeijmakers, \& Snellen}]{petit_dit_de_la_roche_molecule_2018}
Petit dit de~la Roche, D. J.~M., Hoeijmakers, H.~J., \& Snellen, I. A.~G. 2018,
  Astronomy \& Astrophysics, 616, A146

\bibitem[{Petrus {et~al.}(2021)Petrus, Bonnefoy, Chauvin, Charnay, Marleau,
  Gratton, Lagrange, Rameau, Mordasini, Nowak, Delorme, Boccaletti, Carlotti,
  Houllé, Vigan, Allard, Desidera, D’Orazi, Hoeijmakers, Wyttenbach, \&
  Lavie}]{petrus_medium-resolution_2021}
Petrus, S., Bonnefoy, M., Chauvin, G., {et~al.} 2021, Astronomy \&
  Astrophysics, 648, A59

\bibitem[{Racine {et~al.}(1999)Racine, Walker, Nadeau, Doyon, \&
  Marois}]{racine_speckle_1999}
Racine, R., Walker, G., Nadeau, D., Doyon, R., \& Marois, C. 1999, Publications
  of the Astronomical Society of the Pacific, 111, 587

\bibitem[{Rameau {et~al.}(2013)Rameau, Chauvin, Lagrange, Meshkat, Boccaletti,
  Quanz, Currie, Mawet, Girard, Bonnefoy, \&
  Kenworthy}]{rameau_confirmation_2013}
Rameau, J., Chauvin, G., Lagrange, A.-M., {et~al.} 2013, The Astrophysical
  Journal, 779, L26

\bibitem[{Ruffio {et~al.}(2021)Ruffio, Konopacky, Barman, Macintosh, Hoch,
  De~Rosa, Wang, Czekala, \& Marois}]{ruffio_deep_2021}
Ruffio, J.-B., Konopacky, Q.~M., Barman, T., {et~al.} 2021, The Astronomical
  Journal, 162, 290

\bibitem[{Ruffio {et~al.}(2019)Ruffio, Macintosh, Konopacky, Barman, De~Rosa,
  Wang, Wilcomb, Czekala, \& Marois}]{ruffio_radial_2019}
Ruffio, J.-B., Macintosh, B., Konopacky, Q.~M., {et~al.} 2019, The Astronomical
  Journal, 158, 200

\bibitem[{Samland {et~al.}(2017)Samland, Mollière, Bonnefoy, Maire,
  Cantalloube, Cheetham, Mesa, Gratton, Biller, Wahhaj, Bouwman, Brandner,
  Melnick, Carson, Janson, Henning, Homeier, Mordasini, Langlois, Quanz, van
  Boekel, Zurlo, Schlieder, Avenhaus, Beuzit, Boccaletti, Bonavita, Chauvin,
  Claudi, Cudel, Desidera, Feldt, Fusco, Galicher, Kopytova, Lagrange,
  Le~Coroller, Martinez, Moeller-Nilsson, Mouillet, Mugnier, Perrot, Sevin,
  Sissa, Vigan, \& Weber}]{samland_spectral_2017}
Samland, M., Mollière, P., Bonnefoy, M., {et~al.} 2017, Astronomy \&
  Astrophysics, 603, A57

\bibitem[{Snellen {et~al.}(2015)Snellen, de~Kok, Birkby, Brandl, Brogi, Keller,
  Kenworthy, Schwarz, \& Stuik}]{snellen_combining_2015}
Snellen, I., de~Kok, R., Birkby, J.~L., {et~al.} 2015, Astronomy \&
  Astrophysics, 576, A59

\bibitem[{Snellen {et~al.}(2010)Snellen, de~Kok, de~Mooij, \&
  Albrecht}]{snellen_orbital_2010}
Snellen, I. A.~G., de~Kok, R.~J., de~Mooij, E. J.~W., \& Albrecht, S. 2010,
  Nature, 465, 1049

\bibitem[{Sparks \& Ford(2002)}]{sparks_imaging_2002}
Sparks, W.~B. \& Ford, H.~C. 2002, The Astrophysical Journal, 578, 543

\bibitem[{Thalmann {et~al.}(2009)Thalmann, Carson, Janson, Goto, McElwain,
  Egner, Feldt, Hashimoto, Hayano, Henning, Hodapp, Kandori, Klahr, Kudo,
  Kusakabe, Mordasini, Morino, Suto, Suzuki, \&
  Tamura}]{thalmann_discovery_2009}
Thalmann, C., Carson, J., Janson, M., {et~al.} 2009, The Astrophysical Journal,
  707, L123

\bibitem[{van Leeuwen(2007)}]{van_leeuwen_validation_2007}
van Leeuwen, F. 2007, Astronomy \& Astrophysics, 474, 653

\bibitem[{Vigan {et~al.}(2016)Vigan, Bonnefoy, Ginski, Beust, Galicher, Janson,
  Baudino, Buenzli, Hagelberg, D’Orazi, Desidera, Maire, Gratton, Sauvage,
  Chauvin, Thalmann, Malo, Salter, Zurlo, Antichi, Baruffolo, Baudoz,
  Blanchard, Boccaletti, Beuzit, Carle, Claudi, Costille, Delboulbé, Dohlen,
  Dominik, Feldt, Fusco, Gluck, Girard, Giro, Gry, Henning, Hubin, Hugot,
  Jaquet, Kasper, Lagrange, Langlois, Le~Mignant, Llored, Madec, Martinez,
  Mawet, Mesa, Milli, Mouillet, Moulin, Moutou, Origné, Pavlov, Perret, Petit,
  Pragt, Puget, Rabou, Rochat, Roelfsema, Salasnich, Schmid, Sevin,
  Siebenmorgen, Smette, Stadler, Suarez, Turatto, Udry, Vakili, Wahhaj, Weber,
  \& Wildi}]{vigan_first_2016}
Vigan, A., Bonnefoy, M., Ginski, C., {et~al.} 2016, Astronomy \& Astrophysics,
  587, A55

\bibitem[{Vigan {et~al.}(2021)Vigan, Fontanive, Meyer, Biller, Bonavita, Feldt,
  Desidera, Marleau, Emsenhuber, Galicher, Rice, Forgan, Mordasini, Gratton,
  Le~Coroller, Maire, Cantalloube, Chauvin, Cheetham, Hagelberg, Lagrange,
  Langlois, Bonnefoy, Beuzit, Boccaletti, D’Orazi, Delorme, Dominik, Henning,
  Janson, Lagadec, Lazzoni, Ligi, Menard, Mesa, Messina, Moutou, Müller,
  Perrot, Samland, Schmid, Schmidt, Sissa, Turatto, Udry, Zurlo, Abe, Antichi,
  Asensio-Torres, Baruffolo, Baudoz, Baudrand, Bazzon, Blanchard, Bohn,
  Brown~Sevilla, Carbillet, Carle, Cascone, Charton, Claudi, Costille,
  De~Caprio, Delboulbé, Dohlen, Engler, Fantinel, Feautrier, Fusco, Gigan,
  Girard, Giro, Gisler, Gluck, Gry, Hubin, Hugot, Jaquet, Kasper, Le~Mignant,
  Llored, Madec, Magnard, Martinez, Maurel, Möller-Nilsson, Mouillet, Moulin,
  Origné, Pavlov, Perret, Petit, Pragt, Puget, Rabou, Ramos, Rickman, Rigal,
  Rochat, Roelfsema, Rousset, Roux, Salasnich, Sauvage, Sevin, Soenke, Stadler,
  Suarez, Wahhaj, Weber, \& Wildi}]{vigan_sphere_2021}
Vigan, A., Fontanive, C., Meyer, M., {et~al.} 2021, Astronomy \& Astrophysics,
  651, A72

\bibitem[{Visscher {et~al.}(2010)Visscher, Lodders, \&
  Fegley}]{visscher_atmospheric_2010}
Visscher, C., Lodders, K., \& Fegley, B. 2010, The Astrophysical Journal, 716,
  1060

\bibitem[{Wang {et~al.}(2021)Wang, Kulikauskas, \&
  Blunt}]{wang_whereistheplanet_2021}
Wang, J., Kulikauskas, M., \& Blunt, S. 2021, whereistheplanet: {Predicting}
  positions of directly imaged companions,

\bibitem[{Wells {et~al.}(2015)Wells, Pel, Glasse, Wright, Aitink-Kroes,
  Azzollini, Beard, Brandl, Gallie, Geers, Glauser, Hastings, Henning, Jager,
  Justtanont, Kruizinga, Lahuis, Lee, Martinez-Delgado, Martínez-Galarza,
  Meijers, Morrison, Müller, Nakos, O’Sullivan, Oudenhuysen, Parr-Burman,
  Pauwels, Rohloff, Schmalzl, Sykes, Thelen, van Dishoeck, Vandenbussche,
  Venema, Visser, Waters, \& Wright}]{wells_mid-infrared_2015}
Wells, M., Pel, J.-W., Glasse, A., {et~al.} 2015, Publications of the
  Astronomical Society of the Pacific, 127, 646

\bibitem[{Wenger {et~al.}(2000)Wenger, Ochsenbein, Egret, Dubois, Bonnarel,
  Borde, Genova, Jasniewicz, Laloë, Lesteven, \& Monier}]{wenger_simbad_2000}
Wenger, M., Ochsenbein, F., Egret, D., {et~al.} 2000, Astronomy and
  Astrophysics Supplement Series, 143, 9

\bibitem[{Wright {et~al.}(2015)Wright, Wright, Goodson, Rieke, Aitink-Kroes,
  Amiaux, Aricha-Yanguas, Azzollini, Banks, Barrado-Navascues,
  Belenguer-Davila, Bloemmart, Bouchet, Brandl, Colina, Detre, Diaz-Catala,
  Eccleston, Friedman, García-Marín, Güdel, Glasse, Glauser, Greene,
  Groezinger, Grundy, Hastings, Henning, Hofferbert, Hunter, Jessen,
  Justtanont, Karnik, Khorrami, Krause, Labiano, Lagage, Langer, Lemke, Lim,
  Lorenzo-Alvarez, Mazy, McGowan, Meixner, Morris, Morrison, Müller,
  Nø~rgaard Nielson, Olofsson, O’Sullivan, Pel, Penanen, Petach, Pye, Ray,
  Renotte, Renouf, Ressler, Samara-Ratna, Scheithauer, Schneider, Shaughnessy,
  Stevenson, Sukhatme, Swinyard, Sykes, Thatcher, Tikkanen, van Dishoeck,
  Waelkens, Walker, Wells, \& Zhender}]{wright_mid-infrared_2015}
Wright, G.~S., Wright, D., Goodson, G.~B., {et~al.} 2015, Publications of the
  Astronomical Society of the Pacific, 127, 595

\bibitem[{Zahnle \& Marley(2014)}]{zahnle_methane_2014}
Zahnle, K.~J. \& Marley, M.~S. 2014, The Astrophysical Journal, 797, 41

\end{thebibliography}

\begin{appendix}
    
\section{JWST Pipeline}
\label{sec:appendix_pipeline}
\textbf{Stage 1.} The first stage of the pipeline corrects for the detector effects. The input raw data are in the form of one or more ramps (integration) containing accumulating counts from the non-destructive detector readouts.
The output is a corrected count-rate (slope) image. Corrections are applied group by group: first, the pipeline corrects the quality of the pixels, flagging those that will not be used. The step \textit{dq$\_$init} initializes the data quality using a reference file where known bad pixels are indicated. Saturated pixels are flagged with the step \textit{saturation}. 
The first and the last groups of each integration are suppressed as they are the most affected by detector effects \textit{(firstframe, lastframe)}. 
The \textit{linearity} step applies for each pixel a correction for the non-linear detector response, using the "classic" polynomial method where the coefficients of the polynomial are stored in a reference file.
Dark current is corrected by subtracting dark current reference data from the input science data model \textit{(dark)}. 
On each integration ramp within an exposure, we perform cosmic rays/jumps detection \textit{(jump)} by looking for outliers in the up-the-ramp signal in each pixel.
Finally, ramp fitting determines the mean count rate, in units of counts per second, for each pixel by performing a linear fit to the data in the input file \textit{(ramp$\_$fitting)}. 
This stage 1 takes 4D data in the shape of $(N_{int}, N_{group}, N_{pixel,x}, N_{pixel, y})$ to 2D images for the detector $(N_{pixel,x}, N_{pixel, y})$. Other steps of corrections are available but they will not be useful for simulated data, they will be considered for the real on-orbit data.

\textbf{Stage 2.} The second stage of the pipeline corresponds to the calibration. It includes additional corrections depending on the instrument and the observation mode to produce fully calibrated exposures.
First, the pipeline associates a WCS object with each science exposure, which transforms the positions in the detector frame to positions using the International Celestial Reference System (ICRS) frame and wavelength \textit{(assign$\_$wcs)}. The Source Type \textit{(srctype)} step attempts to determine whether a spectroscopic source should be considered a point or extended object. 
Fringes in spectra are corrected \textit{(fringe)}. 
Finally, photometric calibrations allow converting count rates to surface brightness (in MJy/str) \textit{(photom)}. The outputs are 2D calibrated data of the detector. 

\textbf{Stage 3.} The last stage is intended for combining all calibrated exposures. 
We can also subtract or equalize the sky background in science image \textit{(mrs$\_$imatch)}. 
Outliers, bad pixels, and cosmic rays that would remain are flagged using the \textit{outlier$\_$detection} step. This stage takes into account the different dither positions to build a data cube with \textit{cube$\_$build} and extract a spatially averaged spectrum over the full field of view with \textit{extract$\_$1d}.
We can choose to construct cubes for each band or combine bands and channels with a larger wavelength range. 
The outputs are 3D cubes with two extensions, the SCI image contains the surface brightness of the spaxels, and the ERR image is the uncertainty of the SCI values.
\section{Size and shape of the correlation pattern in a simple case} 
\subsection{Analytical first approximation}
\label{sec:appendix_correlation}

To have an idea of the parameters involved in the correlation that should help to understand the size of the correlation pattern, we derive a formula for the correlation depending on its parameters. 
We defined S as the observed spectrum and M as the model spectrum. The noise in the observed spectrum is $\xi$ $\sim \mathcal{N}(0, \xi(\lambda))$, P is the PSF.
\begin{equation}
    S =  P M + \xi 
\end{equation}
The correlation can be written as :
\begin{equation}
    C = \frac{M\otimes (PM+ \xi )}{\lVert M \rVert \times \lVert (PM+ \xi ) \rVert }
\end{equation}
The numerator can be written as : 
\begin{align}
    M\otimes (PM+ \xi ) =& P M\otimes M + M \otimes \xi\\
    =& \sum_\lambda  P(\lambda) M^2 + \sum_\lambda M \xi\\
    \approx& \sum_\lambda P(\lambda)  M^2
\end{align}
In detail, the norm of the observed spectrum is: 
\begin{align}
    \lVert (PM+ \xi ) \rVert =& \sqrt{\sum_\lambda(PM)^2 + 2\sum_\lambda PM\xi + \sum_\lambda \xi^2}\\
    \approx& \sqrt{\sum_\lambda P(\lambda)^2 M(\lambda)^2 + \sum_\lambda \xi(\lambda)^2}
\end{align}

Finally :\\
\begin{align}
    \lVert M \rVert =& \sqrt{\sum_\lambda M^2 }
\end{align}

If the PSF function $P_x(\lambda)$ is a Gaussian $G_\sigma(x)$ with $x$ the spaxel distance from the centroid and with a wavelength-dependent width $\sigma(\lambda)$=$\sigma_0 \lambda/\lambda_0$, then we may develop $P$ at first order in $(\lambda-\lambda_0)/\lambda_0$ as: 

\begin{align}
P &= G_\sigma(x) \\ 
&= \exp\left(-\frac{{\bf x}^2}{2\,\sigma_0^2 \lambda^2/\lambda_0^2}\right) \\
&\approx \exp\left(-\frac{{\bf x}^2}{2\,\sigma_0^2}\right) \left( 1 + \frac{x^2}{\sigma_0^2} \frac{\lambda-\lambda_0}{\lambda_0}\right) \\
\end{align}

In the following we will write $G_0(x)=\exp\left(-\frac{{\bf x}^2}{2\,\sigma_0^2}\right) $. 

The different terms in the numerator and in the denominator involving $P_x(\lambda)$ can be developed according to the above equation at first order in $(\lambda-\lambda_0)/\lambda_0$. Linearising all first order terms leads to the following equation: 

\begin{align}
C \approx \frac{G_0(x)}{\sqrt{G_0(x)^2 + \delta^2}} + \frac{x^2}{\sigma_0^2}\,G_0(x) \, \frac{\delta^2}{ \left(G_0(x)^2 + \delta^2\right)^{3/2} } \,\beta
\end{align}

where we defined: 

\begin{align}
    \beta &= \frac{\sum_\lambda \frac{\lambda - \lambda_0}{\lambda_0} \, M(\lambda)^2}{\sum_\lambda M(\lambda)^2}\\
    \delta^2 &= \frac{\sum_\lambda \xi(\lambda)^2}{\sum_\lambda M(\lambda)^2}\\
\end{align}

The parameter $\delta^2$ can be considered as a positive constant that depends on the signal-to-noise level in the cube, while $\beta$ can be bounded by the wavelength range of the considered channel. Indeed, since $\lambda_\text{min}<\lambda<\lambda_\text{max}$: 

\begin{equation}
    \frac{\lambda_\text{min}-\lambda_0}{\lambda_0} <\beta< \frac{\lambda_\text{max}-\lambda_0}{\lambda_0} 
\end{equation}

Thus since the channels windows are on the order of $\lambda_0$$\pm$$14 - 18$\%, we expect 
$\beta$<0.18.
The zeroth-order term $C=G_0(x)/\sqrt{G_0(x)^2 + \delta^2}$ is thus a good approximation of the correlation pattern.
It indicates that the correlation pattern will depend on the PSF, but also on the model spectrum and on the noise. However, this formula is insufficient to estimate a correlation radius, and it would result in an oversized correlation pattern. Indeed this is correct in the case of data dominated by Gaussian noise, which is not the case at long wavelengths and when we have more stellar residuals. It also assumes that the PSF is a Gaussian.

\subsection{Example of simulations from the parametric study}
\label{sec:appendix_fig_corr}

\begin{figure}[h!]
    \centering
    \includegraphics[width=90mm]{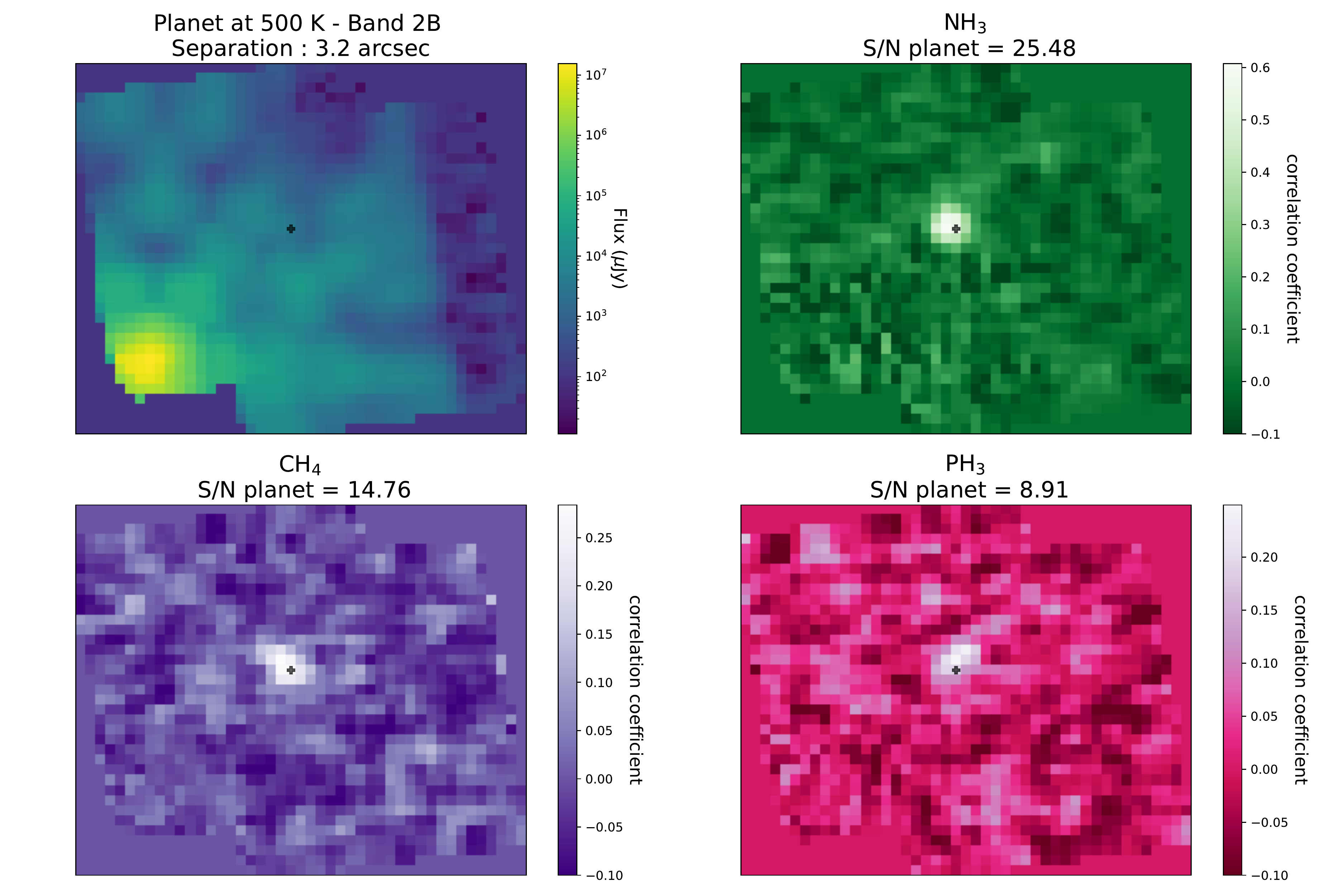}
    \caption{Simulations from the parametric study with the planet at 500\,K and the star separated at 3.2\,". Correlations maps with different molecular templates: NH$_3$, CH$_4$, and PH$_3$ in the band 2B.}
    \label{fig:appendix_molec_500K_cc_maps}
\end{figure}

\begin{figure}[h!]
    \centering
    \includegraphics[width=90mm]{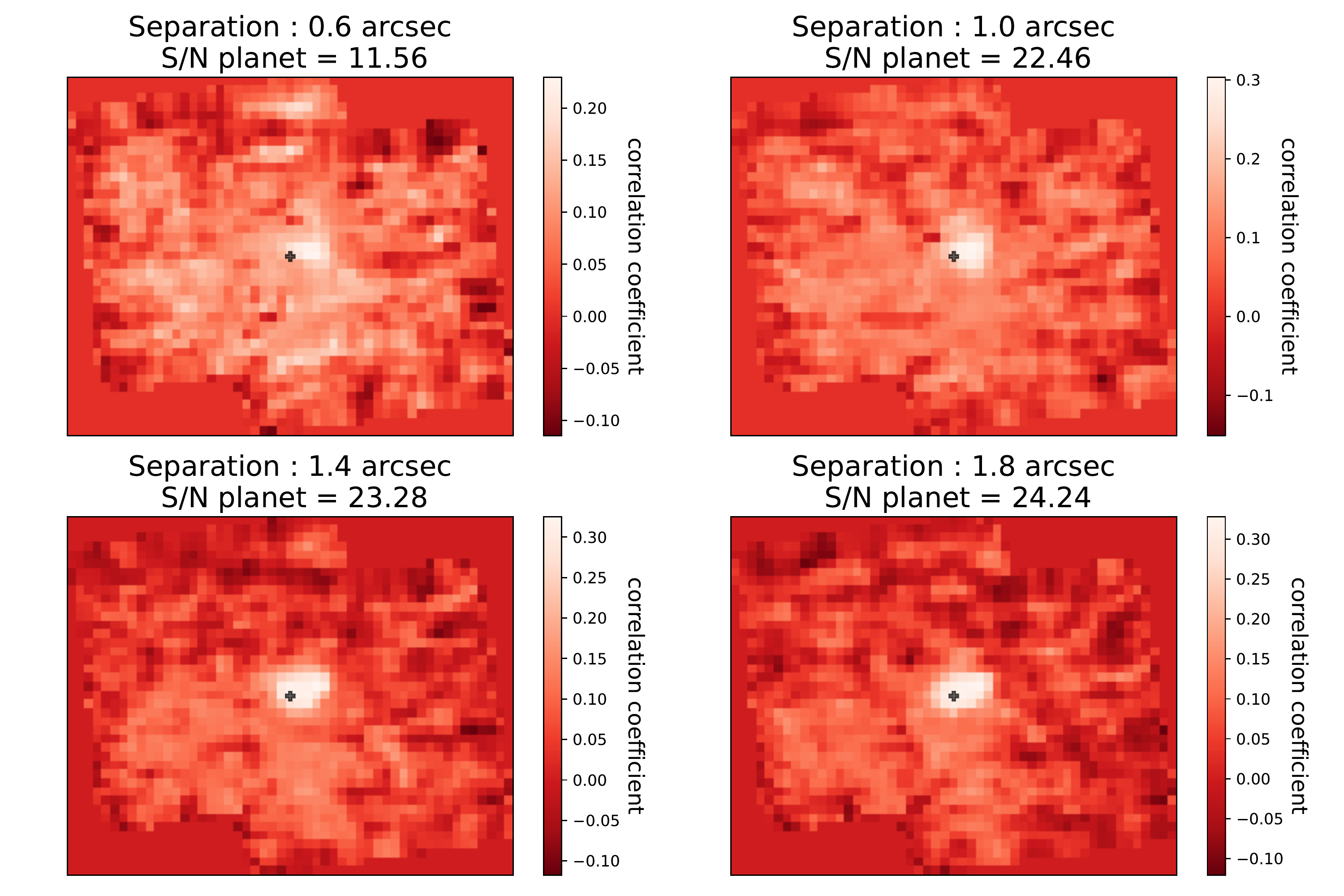}
     \caption{Simulations from the parametric study with the planet at 1750\,K and the star at different separations. Correlation maps with CO in band 1A.}
    \label{fig:appendix_co_1750K}
\end{figure}

\begin{figure}[h!]
    \centering  
    \includegraphics[width=90mm]
    {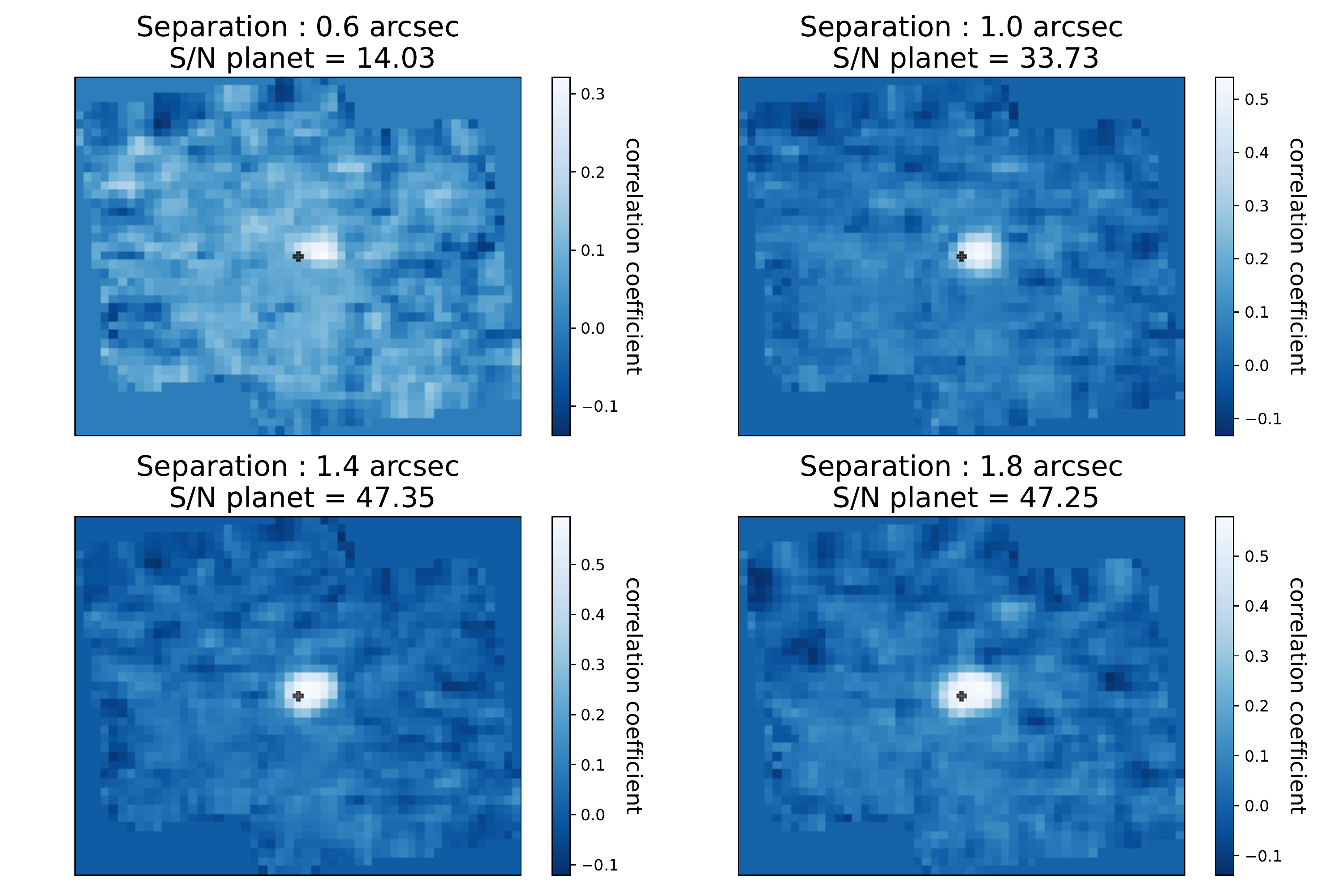}
    \caption{Simulations from the parametric study with the planet at 1750\,K and the star at different separations. Correlation maps with H$_2$O in band 1A.}
    \label{fig:appendix_h2o_1750K}
\end{figure}

Simulations show that we obtain different shapes and sizes of the correlation pattern depending on the noise level and the molecules studied.
Fig. \ref{fig:appendix_molec_500K_cc_maps} presents correlation maps for a planet at 500\,K, it is a comparison between the direct image, and the correlation maps for three detected molecules (NH$_3$, CH$_4$, and PH$_3$) at the same wavelength (band 2B) and same star-planet separation.
Fig. \ref{fig:appendix_co_1750K} and Fig. \ref{fig:appendix_h2o_1750K} is a comparison between CO and H$_2$O for four values of separation between a star and a planet. It corresponds to the simulation with a planet at 1750\,K, in band 1A.
Moreover, it is notable that astrometry is unreliable, mostly for short star-planet separation. The asymmetry of the correlation pattern is clearly visible in Fig. \ref{fig:appendix_co_1750K}.
Concerning the mean azimuthal profile, we note that the profile of the PSF is unchanged whatever the noise level, whereas the profile of the correlation pattern depends on the noise level and on the template with which we correlate the data.

\begin{figure}[h!]
    \centering
    \includegraphics[width=60mm]{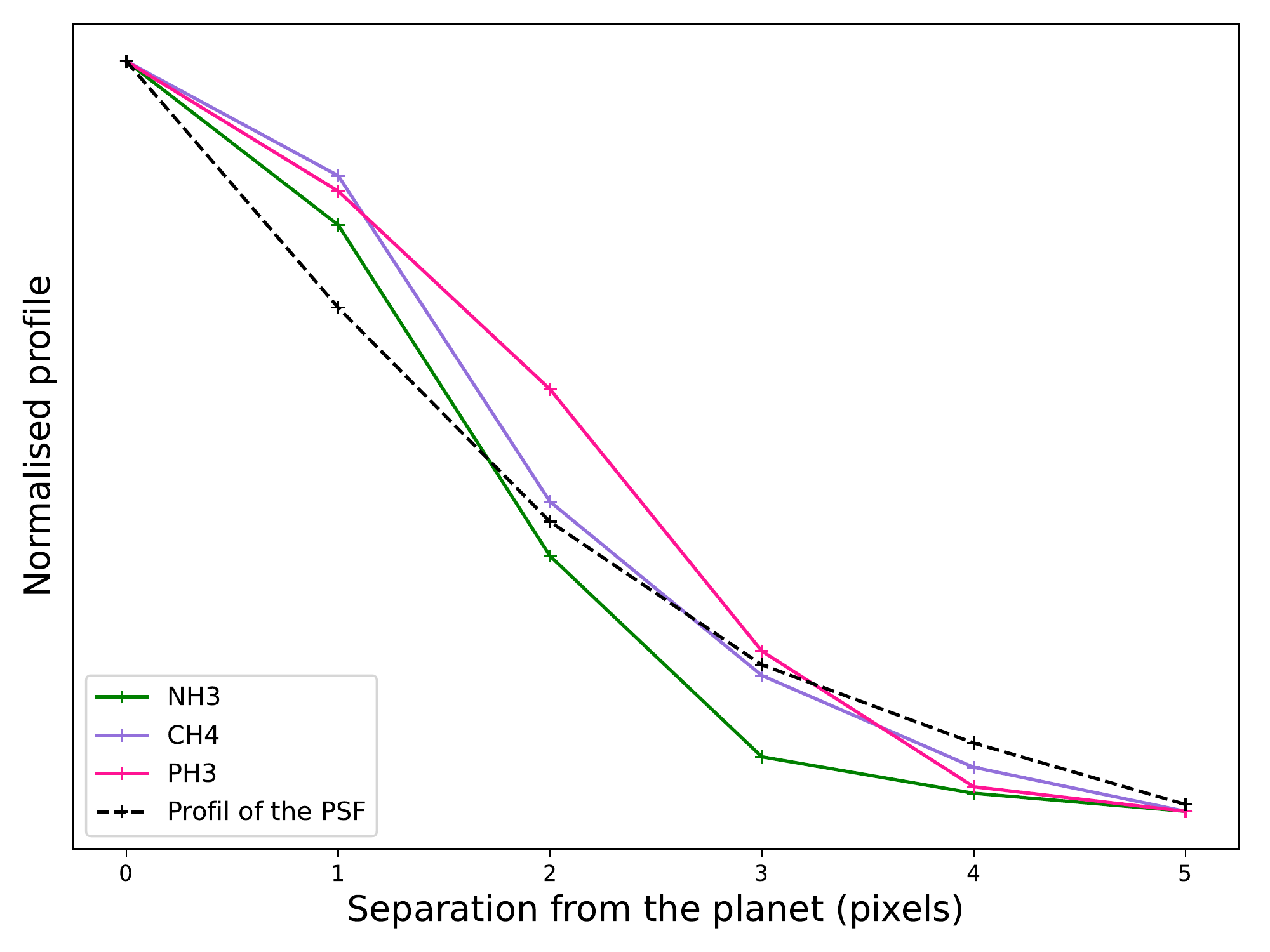}
    \caption{Corresponding mean azimuthal profiles for each correlation maps above Fig. \ref{fig:appendix_molec_500K_cc_maps}.}
    \label{fig:appendix_molec_500K_prodile}
\end{figure}

\begin{figure}[h!]
    \centering
    \includegraphics[width=60mm]{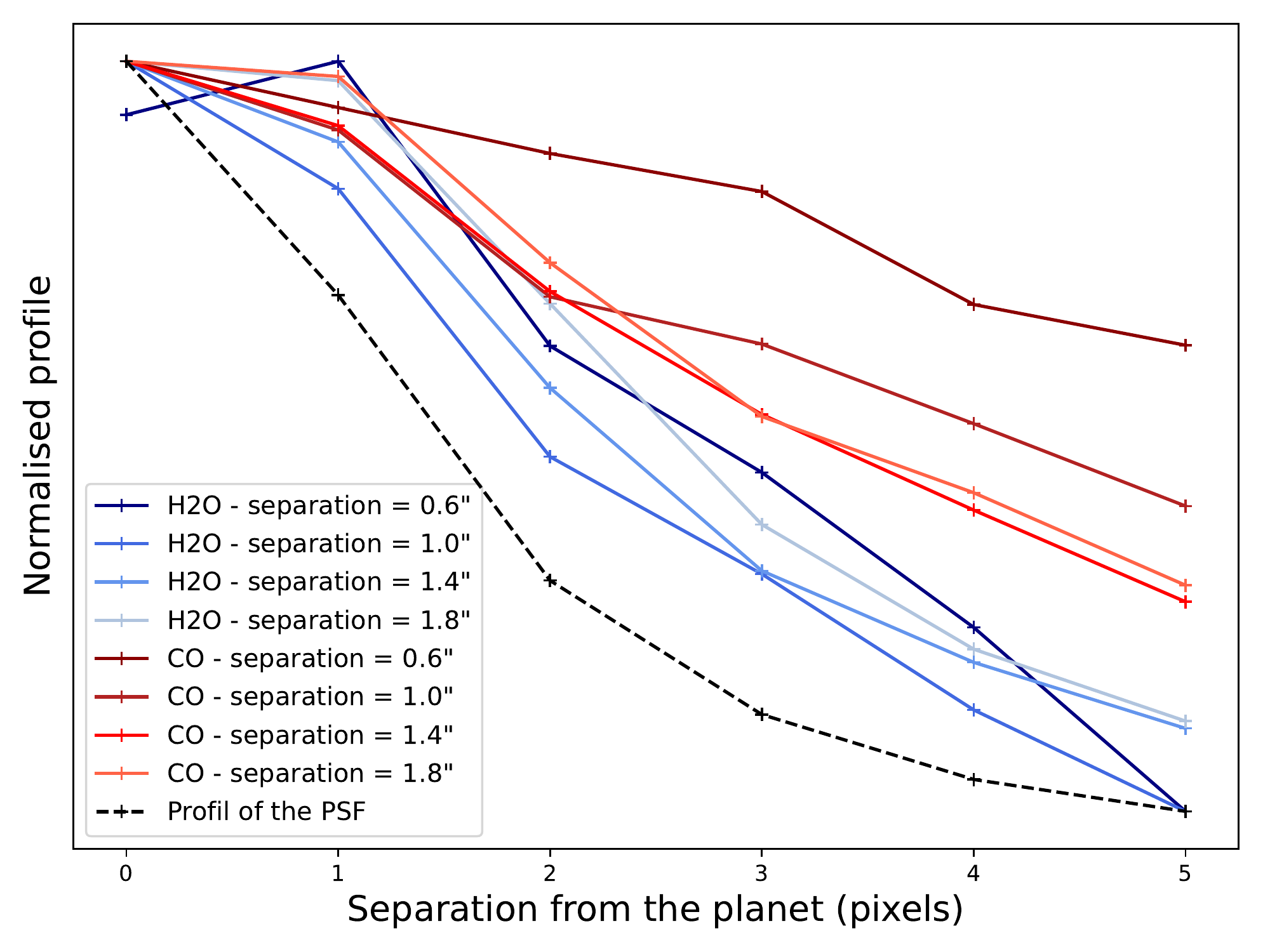}
    \caption{Corresponding radial profiles for each of the correlation maps of Fig. \ref{fig:appendix_co_1750K} and Fig. \ref{fig:appendix_h2o_1750K}.}
\end{figure}

\onecolumn

\section{Simulations and correlation maps for each planet}

\label{sec:appendix_mm}
\begin{figure*}[h!]
     \centering
     \includegraphics[width=190mm]{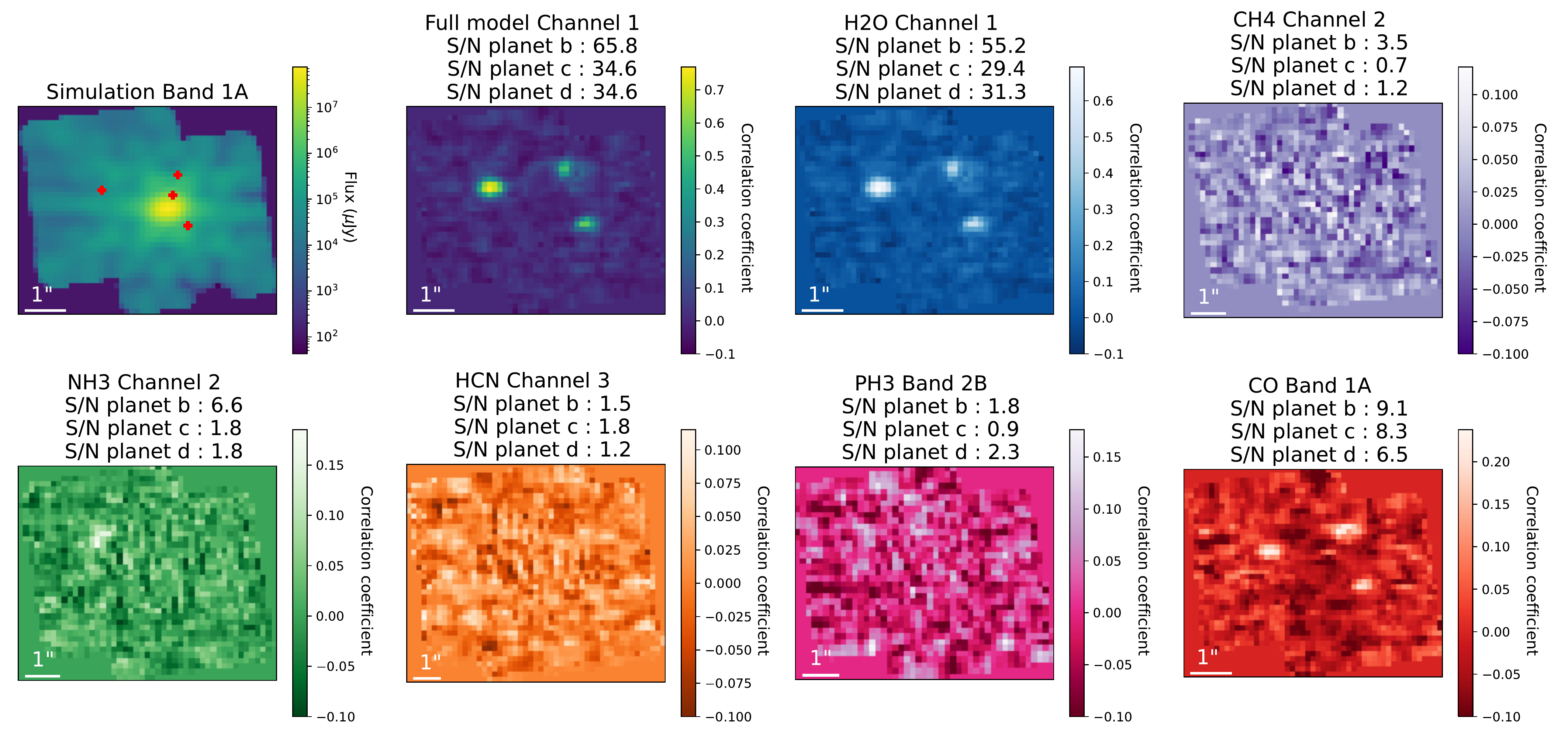}
     \caption{MIRISim simulation and correlation maps of the simulated system HR\,8799.}
     \label{fig:cc_maps_HR8799}
\end{figure*}

\begin{figure*}[h!]
     \centering
     \includegraphics[width=190mm]{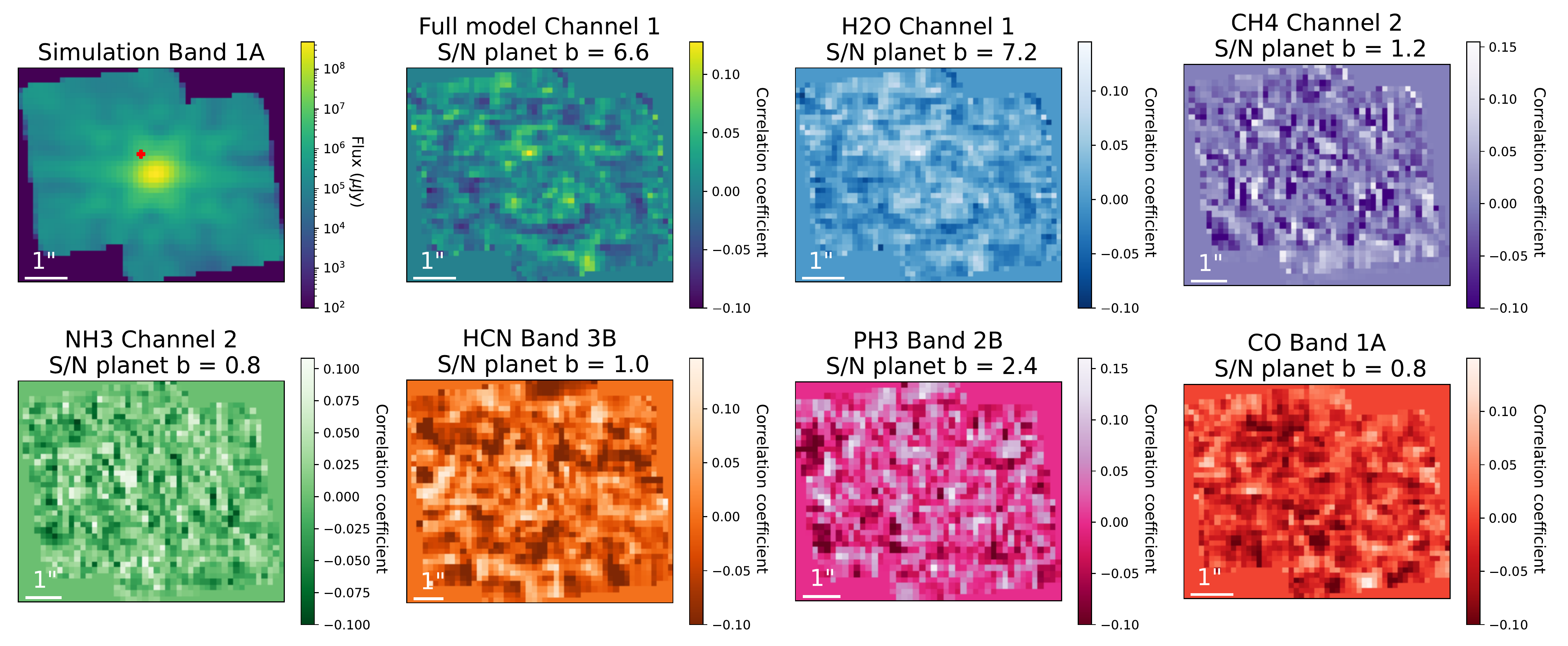}
     \caption{MIRISim simulation and correlation maps of the simulated system $\beta$ Pictoris.}
     \label{fig:cc_maps_beta_pictoris}
\end{figure*}

\begin{figure*}[h!]
     \centering
     \includegraphics[width=190mm]{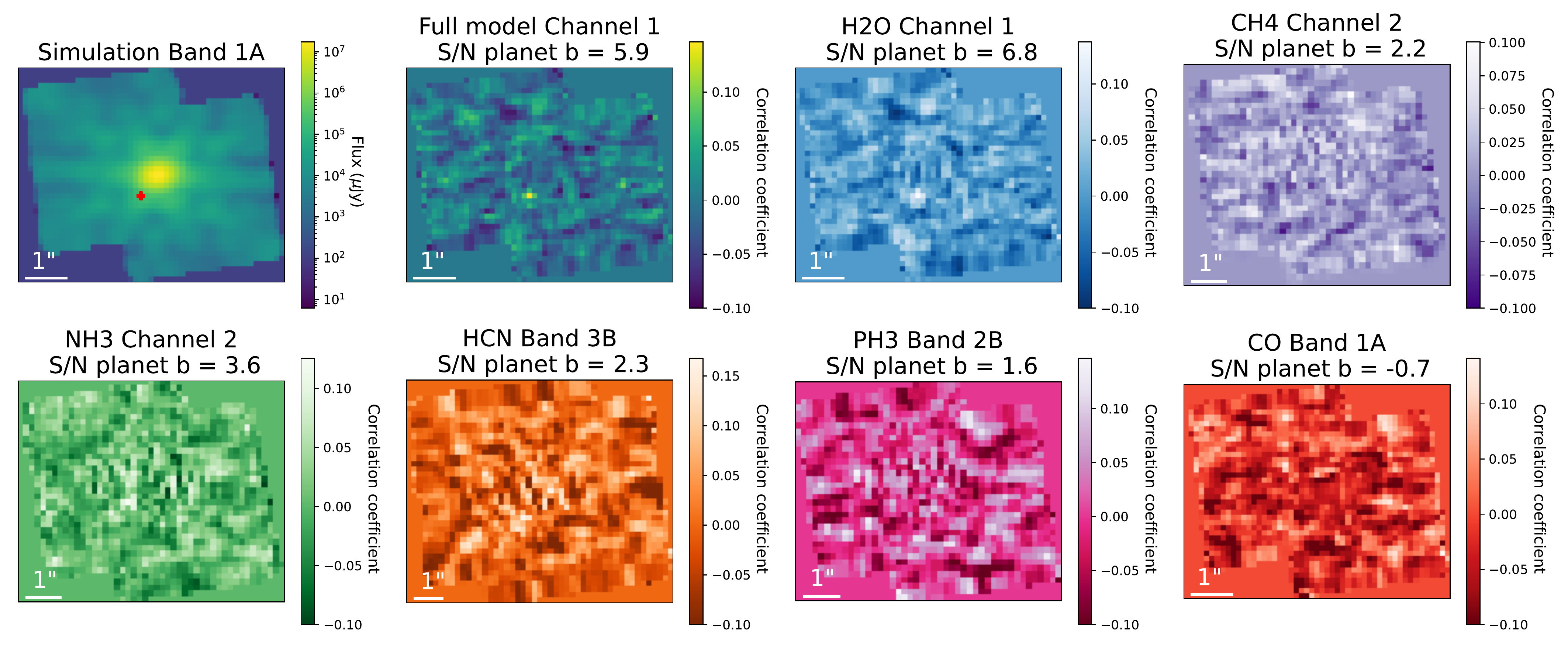}
     \caption{MIRISim simulation and correlation maps of the simulated system HD\,95086 at 800\,K.}
     \label{fig:cc_maps_HD95086_800}
\end{figure*}

\begin{figure*}[h!]
     \centering
     \includegraphics[width=190mm]{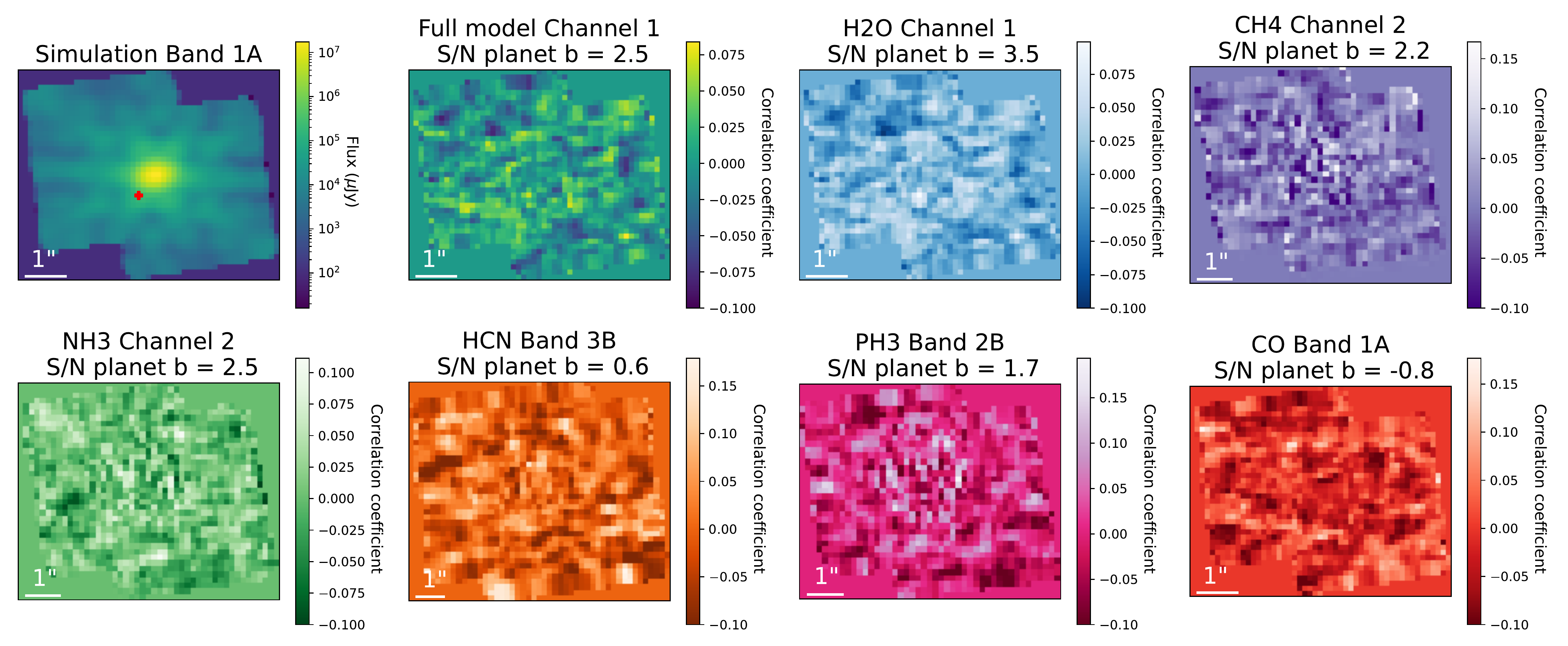}
     \caption{MIRISim simulation and correlation maps of the simulated system HD\,95086 at 1400\,K.}
     \label{fig:cc_maps_HD95086_1400}
\end{figure*}

\begin{figure*}[h!]
     \centering
     \includegraphics[width=190mm]{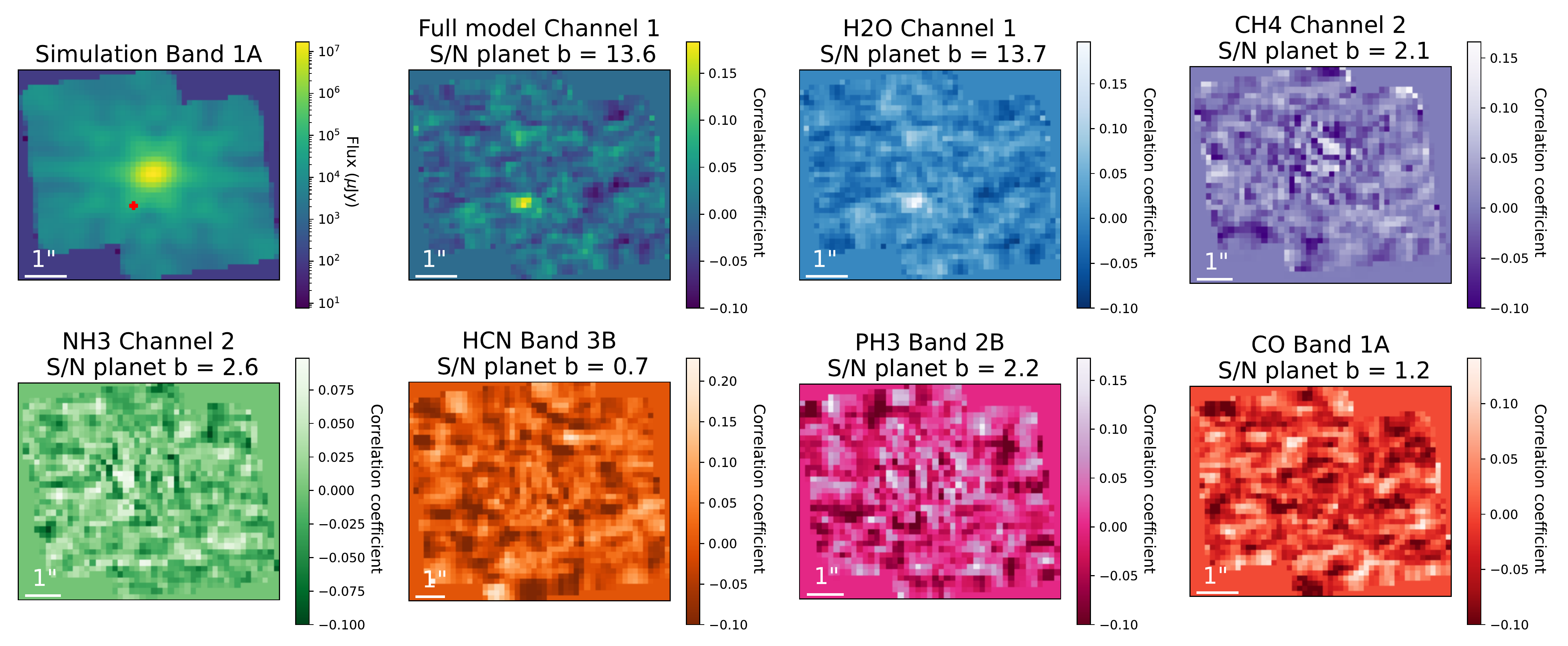}
     \caption{MIRISim simulation and correlation maps of the simulated system HIP\,65426.}
     \label{fig:cc_map_HIP65426}
\end{figure*}

\begin{figure*}[h]
     \centering
     \includegraphics[width=190mm]{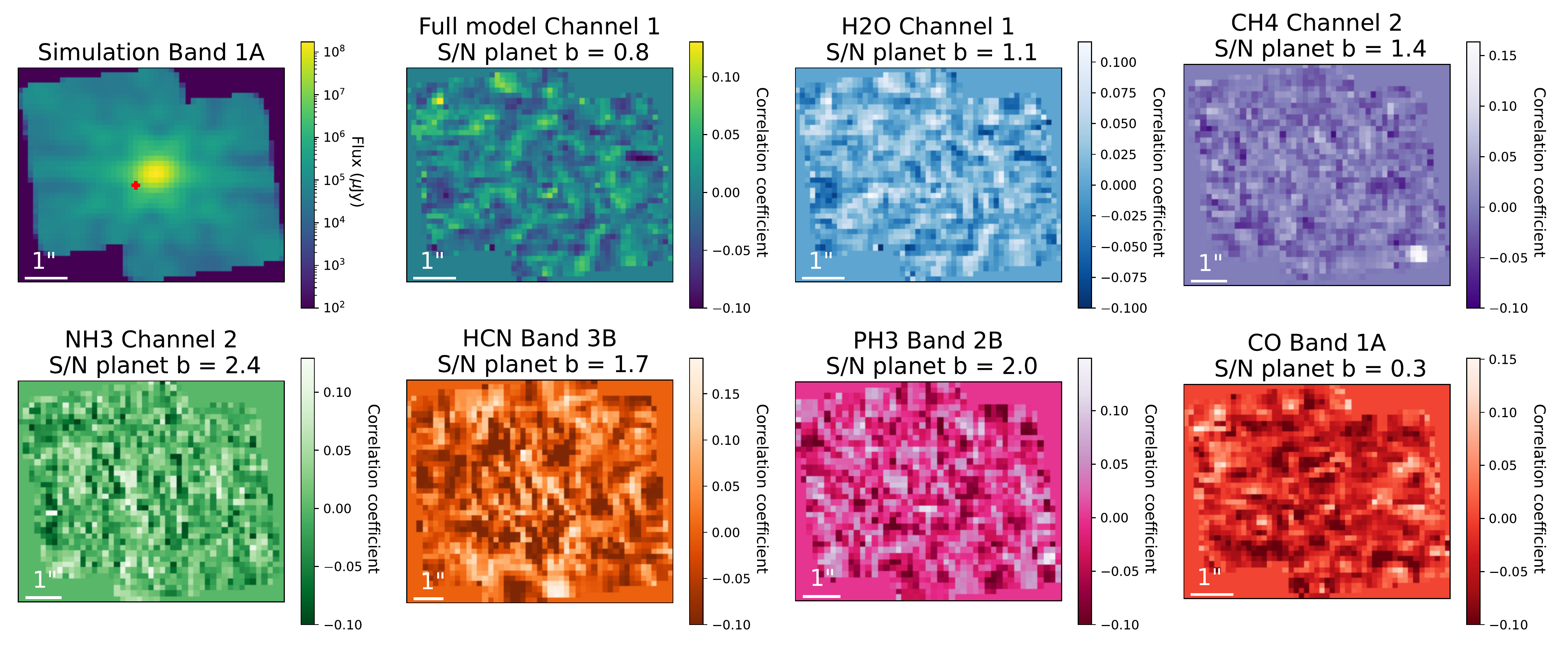}
     \caption{MIRISim simulation and correlation maps of the simulated system 51 Eri.}
     \label{fig:cc_map_51_Eri}
\end{figure*}

\begin{figure*}[h]
     \centering
     \includegraphics[width=190mm]{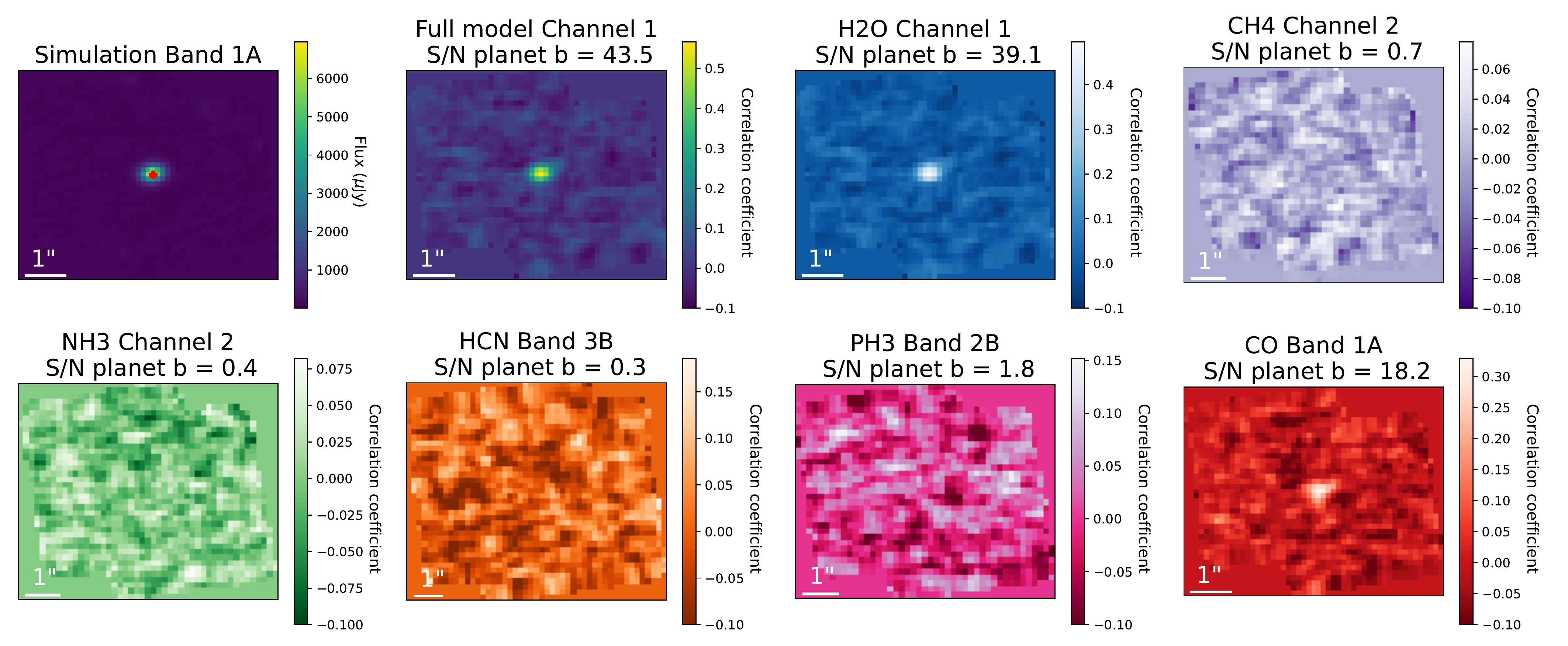}
     \caption{MIRISim simulation and correlation maps of the simulated system HD\,106906, the simulation plot shows directly the planet (in the center).}
     \label{fig:cc_maps_HD106906}
\end{figure*}

\begin{figure*}[h]
     \centering
     \includegraphics[width=190mm]{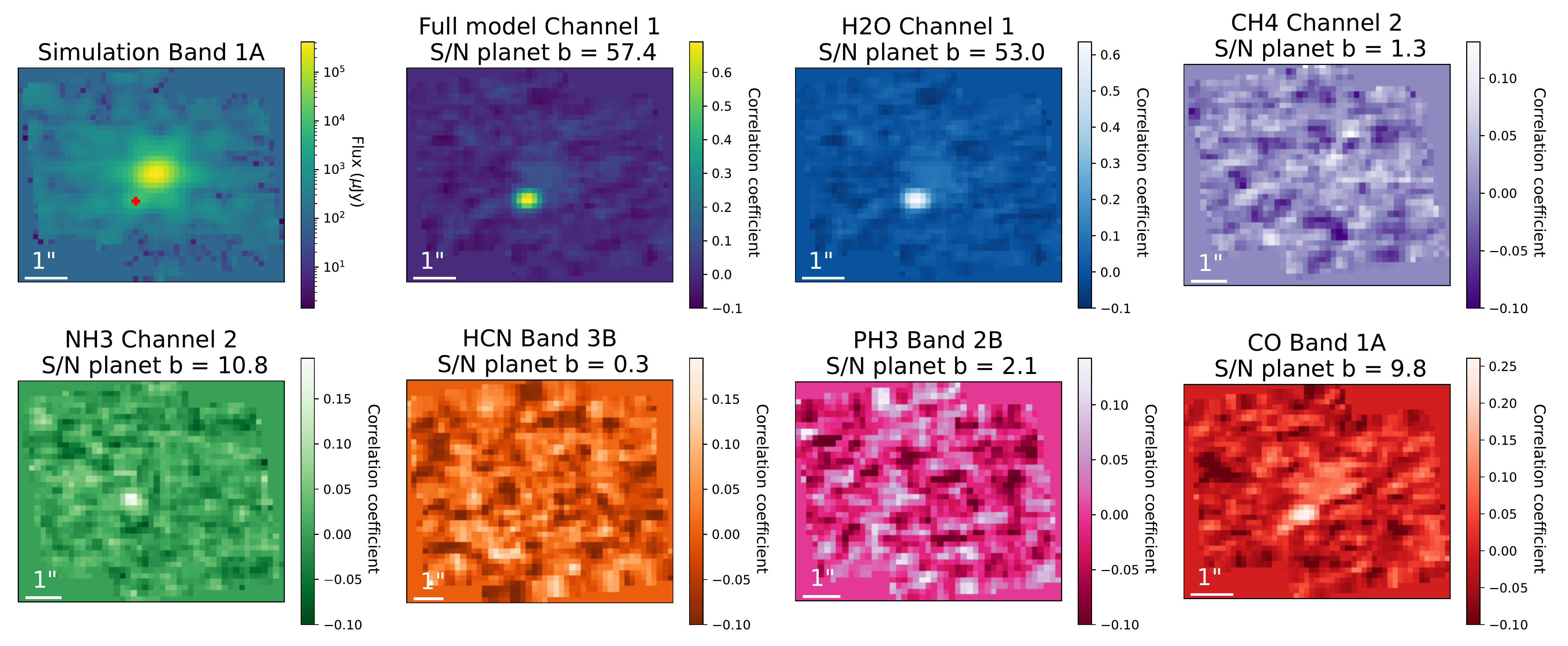}
     \caption{MIRISim simulation and correlation maps of the simulated system 2M\,1207 at 1000\,K.}
     \label{fig:cc_maps_2M1207_1000K}
\end{figure*}

\begin{figure*}[h]
     \centering
     \includegraphics[width=190mm]{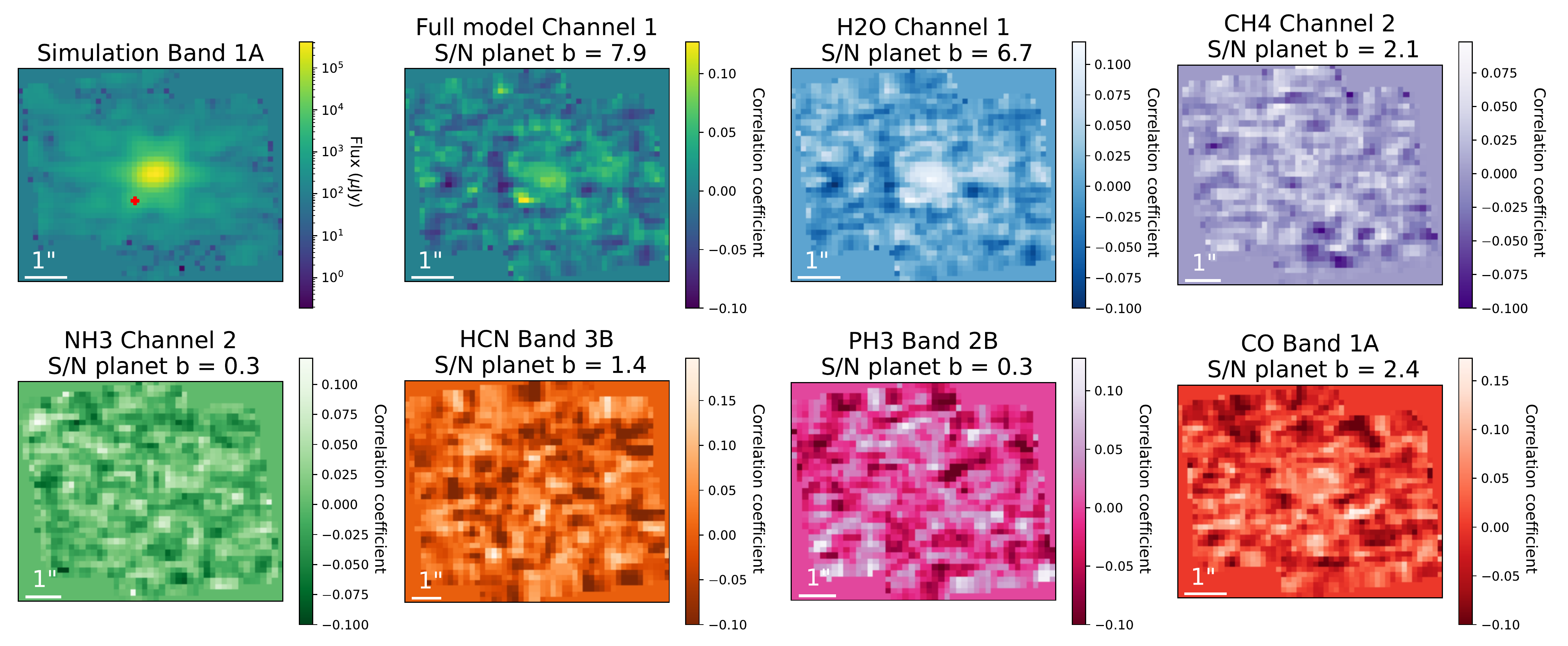}
     \caption{MIRISim simulation and correlation maps of the simulated system 2M\,1207 at 1600\,K.}
     \label{fig:cc_maps_2M1207_1600K}
\end{figure*}

\begin{figure*}[h]
     \centering
     \includegraphics[width=190mm]{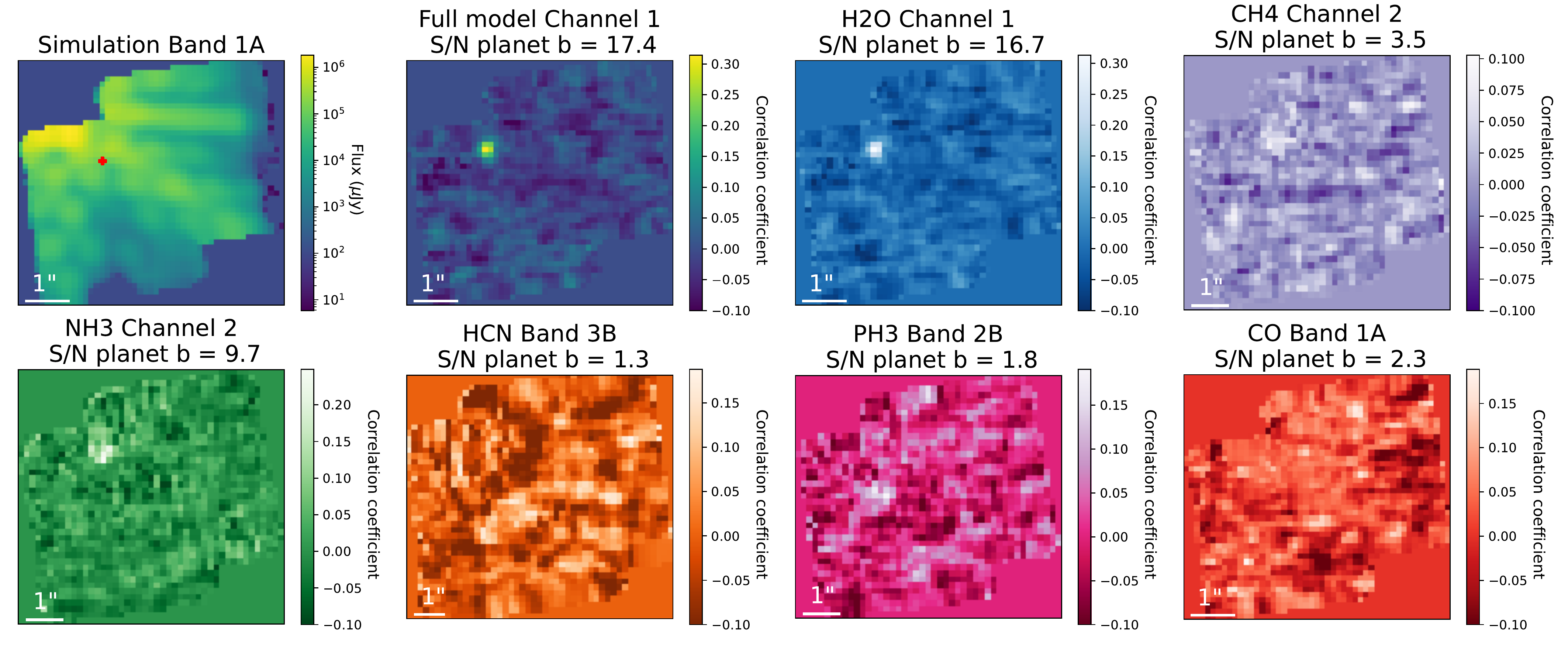}
     \caption{MIRISim simulation and correlation maps of the simulated system GJ\,758.}
     \label{fig:cc_maps_GJ758}
\end{figure*}

\begin{figure*}[h]
     \centering
     \includegraphics[width=180mm]{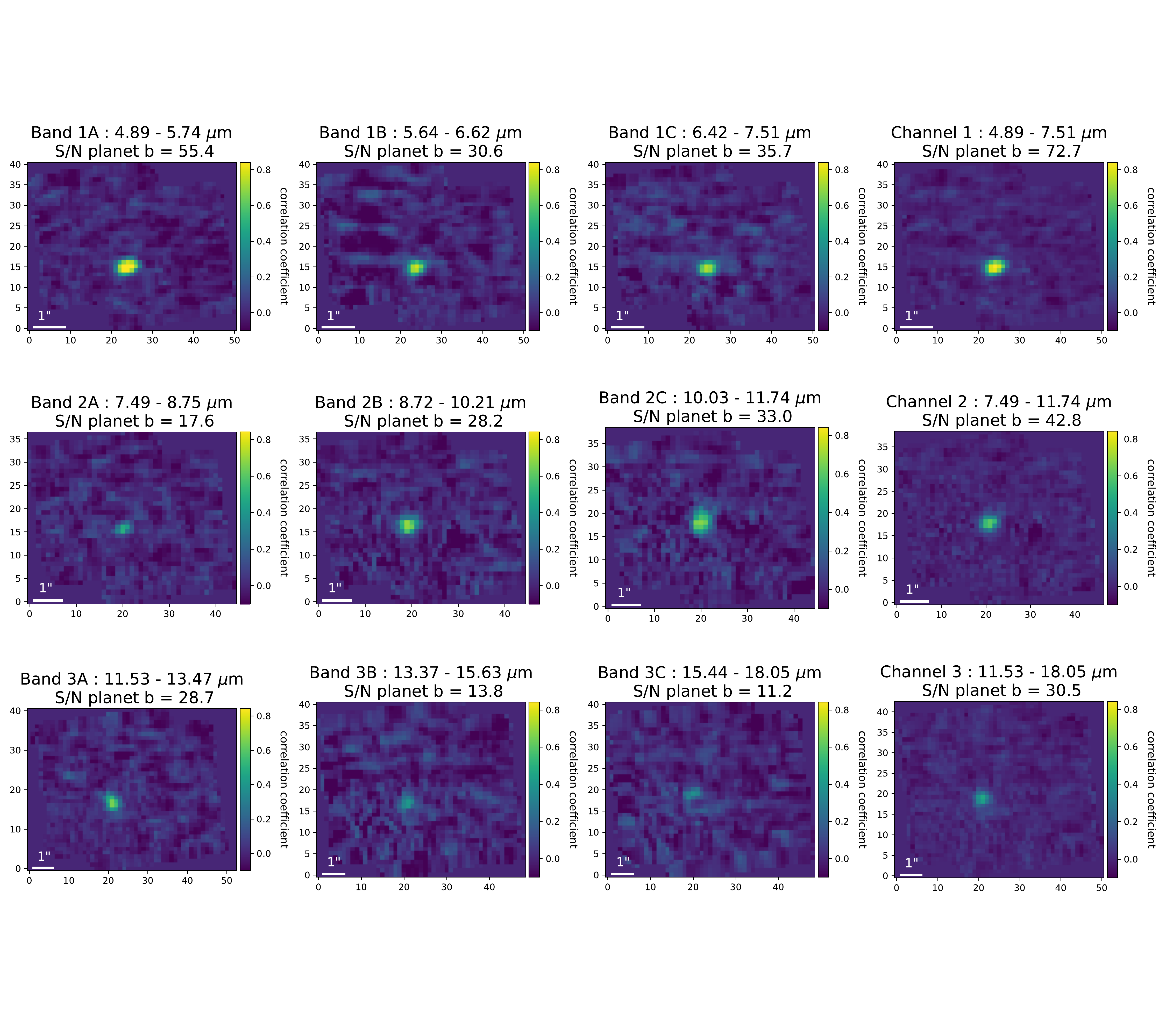}
     \caption{MIRISim simulation and correlation maps of the simulated system GJ\,504 for the 3 first channel. The scale is the same in each bands and channels.}
     \label{fig:cc_maps_GJ504_all}
\end{figure*}

\end{appendix}

\end{document}